\documentclass[iop,a4paper,fleqn,usenatbib]{mnras}%
\usepackage{newtxtext,newtxmath}  
\usepackage[T1]{fontenc}  
\usepackage{ae,aecompl}   
\usepackage{times}
\usepackage{mathtools}
\usepackage{epsf}
\usepackage{graphicx}
\usepackage{amssymb}
\usepackage{enumitem}
\usepackage{footnote}
\usepackage{tablefootnote}
\usepackage{relsize}
\usepackage{xfrac}
\usepackage{aasmacros}
\usepackage[flushleft]{threeparttable}
\usepackage{color}
\usepackage{setspace}
\usepackage[squaren, Gray, cdot]{SIunits}
\definecolor{lblue}{rgb}{0.1,0.7,1.}
\definecolor{Orange}{rgb}{1.0,0.05,0.15}
\definecolor{Green}{rgb}{0.15,0.45,0.25}
\definecolor{Blue}{rgb}{0.0,0.08,0.65}
\definecolor{Brown}{rgb}{0.7,0.25,0.0}
\definecolor{Pink}{rgb}{1.0,0.05,0.5}

\usepackage{calrsfs}
\usepackage{hyperref} 
\hypersetup{colorlinks=true, linkcolor=blue, citecolor=blue, filecolor=blue, urlcolor=black}
\DeclareMathAlphabet{\pazocal}{OMS}{zplm}{m}{n}
\newcommand{\La}{\mathcal{L}}

\title[SED-fitting performance and forecasts for future surveys]{{Horizon-AGN} virtual observatory -- 1. \\
SED-fitting performance and forecasts for future imaging surveys}

\author[Laigle, Davidzon, Ilbert et al.]{
\parbox[t]{\textwidth}{
C.~Laigle$^{1}$\thanks{E-mail: clotilde.laigle@physics.ox.ac.uk},
I.~Davidzon$^{2}$, O.~Ilbert$^{3}$, J.~Devriendt$^{1}$, D.~Kashino$^{4}$, 
C.~Pichon$^{5,6,7}$, 
P.~Capak$^{2}$, 
S.~Arnouts$^{3}$, S.~de~la~Torre$^{3}$, Y.~Dubois$^{5}$, G.~Gozaliasl$^{8,9,10}$, D.~Le~Borgne$^{5}$, S.~Lilly$^{4}$, H.~J.~McCracken$^{5}$, M. Salvato$^{11}$, A. Slyz$^{1}$}
\vspace*{6pt} \\ 
$^{1}$ Sub-department of Astrophysics, University of Oxford, Keble Road, Oxford OX1 3RH\\
$^{2}$ IPAC, Mail Code 314-6, California Institute of Technology, 1200 East California Boulevard, Pasadena, CA 91125, USA \\
$^{3}$ Aix Marseille Univ, CNRS, CNES, LAM, Marseille, France \\
$^{4}$ Institute for Astronomy, Department of Physics, ETH Zürich, Wolfgang-Pauli-strasse 27, CH-8093 Zürich, Switzerland \\
$^{5}$ Sorbonne Universit{\'e}s, CNRS, UMR 7095,  Institut d\'Astrophysique de Paris, 98 bis bd Arago, 75014 Paris \\
$^{6}$ Institute for Astronomy, University of Edinburgh, Royal Observatory, Blackford Hill, Edinburgh, EH9 3HJ, United Kingdom\\
$^{7}$ Korea Institute for Advanced Study (KIAS), 85 Hoegiro, Dongdaemun-gu, Seoul, 02455, Republic of Korea\\
$^{8}$ Finnish centre for Astronomy with ESO (FINCA), Quantum, Vesilinnantie 5, University of Turku, FI-20014, Turku, Finland;\\
$^{9}$ Department of Physics, University of Helsinki, PO Box 64, FI-00014 Helsinki, Finland \\
$^{10}$ Helsinki Institute of Physics, University of Helsinki, P.O. Box 64, FI-00014, Helsinki, Finland\\
$^{11}$ Max-Planck-Institut für extraterrestrische Physik, Garching, Germany}
\date{Accepted XXX. Received YYY; in original form ZZZ} 
\pubyear{2016}  
\begin{document}
\label{firstpage}
\pagerange{\pageref{firstpage}--\pageref{lastpage}}
\maketitle

\begin{abstract}

    Using the ligthcone from the cosmological hydrodynamical simulation {\sc Horizon-AGN}, we produced a photometric catalogue over $0<z<4$ with apparent magnitudes in COSMOS, DES, LSST-like, and \emph{Euclid}-like filters at depths comparable to these surveys. The virtual photometry  accounts for the complex star formation history and metal enrichment of {\sc Horizon-AGN} galaxies, and consistently includes magnitude errors, dust attenuation and absorption by inter-galactic medium. 
    The COSMOS-like photometry is fitted in the same configuration as the COSMOS2015 catalogue. We then quantify random and systematic errors of photometric redshifts, stellar masses, and star-formation rates (SFR). 
    Photometric redshifts and redshift errors capture the same dependencies on magnitude and redshift as found in COSMOS2015, excluding the impact of source extraction. 
    COSMOS-like stellar masses are well recovered with a dispersion typically lower than 0.1 dex. 
    The simple star formation histories and metallicities of the  templates induce a systematic underestimation of stellar masses at $z<1.5$ by at most 0.12 dex. SFR estimates exhibit  a dust-induced bimodality combined with a larger scatter (typically between 0.2 and 0.6 dex). 
We also use our mock catalogue to predict photometric redshifts and stellar masses in future imaging surveys. 
 We stress  that adding {\emph{Euclid}} near-infrared photometry to the LSST-like baseline  improves redshift accuracy especially at the faint end and decreases the outlier fraction by a factor $\sim$2. It also considerably improves stellar masses, reducing the scatter up to a factor 3. It would therefore be mutually beneficial for LSST and \emph{Euclid} to work in synergy.
    
\end{abstract}


\section{Introduction}
\label{sec-preamble}

Our understanding of galaxy formation, evolution, and {of their distribution in the large-scale structure} has taken a giant step forward in the last decade, owing to large multi-wavelength datasets. Properties of different galaxy populations, and their evolution across cosmic time,  can be constrained by measuring one-point statistics, such that the luminosity  and stellar mass functions   \citep[e.g.][]{ilbert06,ilbert13,davidzon17,bundy17}.  
Two-point statistics, i.e. measuring the spatial correlation of galaxies, make it possible to investigate the role of the local environment  \citep[e.g.][]{abbas06,delatorre10,hatfield17} and to infer  halo properties, via simplifying assumptions such as the so-called halo model \citep[e.g.][]{mccracken14,coupon15,legrand18}. More generally, higher order statistics  \citep[see][]{moresco17} as well as topological tools such as filament tracers, can help  disentangle complex    environmental effects, distinct from the isotropic influence of local density peaks \citep[e.g.][]{malavasi17,laigle18,kraljic18}. When implementing such statistics, one must asses the impact of 
observational biases on inferring the underlying properties of the population.

In particular, investigations focusing on galaxy stellar mass assembly rely on three fundamental quantities: photometric redshifts, stellar masses and star-formation rates. Large-area surveys can significantly reduce statistical errors in these kinds of measurements, and probe a wide variety of galaxy types and environments. Therefore, the dominant source of uncertainties in state-of-the-art studies became the selection biases of the surveys, the source extraction techniques, and the physical models assumed in the analysis (when needed).
Even  when a high-resolution galaxy spectral energy distribution (SED) is available, inferring physical properties from it is an ill-conditioned problem \citep{moultaka&pelat00,moultaka04}, which prevents a complete inversion approach to be successful \citep[see e.g.][]{ocvirk06}. 
Difficulties are even more severe when only apparent magnitudes in broad-band filters are available.  {In that case, SED-fitting codes are routinely used because of their versatility. These  codes fit pre-computed libraries of galaxy templates to the   photometry of observed objects \citep[see a review in ][]{walcher11,conroy13}.}
Some very promising alternative techniques are also being developed \citep[see][for a review]{salvato18}, including ``clustering redshift" \citep{newman08,menard13}, ``photo-web'' \citep{aragoncalvo15} and more recently machine-learning \citep[see e.g.][]{masters15,beck17,pasquet18,gomes18,hemmati18}. However these alternative techniques generally require large and representative spectroscopic samples, which is not the case of SED-fitting algorithms. 
In order to build a template library for the SED-fitting procedure, one relies nonetheless on several assumptions, mainly concerning star formation histories (SFHs), metal enrichment, and dust extinction and spatial distribution \citep[][]{conroy13}. These priors inevitably introduce systematics in the recovered physical quantities,  which in turn may impair the statistical measurements and bias  conclusions on galaxy mass assembly scenarii. 
For instance \citet{bundy17} find that depending on the  assumed SFH in the SED-fitting estimates, massive ($>3\times10^{11}\,M_\odot$) galaxies between $z=0$ and 0.8 may show either a mild stellar mass growth or a lack of evolution; this systematic uncertainty is  dominant, considering that their extremely large sample of galaxies ($>\!41,000$), collected across $\sim\!140$\,deg$^2$, makes shot noise and cosmic variance almost negligible.  

Furthermore, in order to understand the physical processes regulating galaxy mass assembly, it is important to compare observational measurements to  semi-analytical and hydrodynamical simulations, where different theoretical models of galaxy evolution have been implemented \citep[e.g.][]{delucia&blaizot07,vogelsberger13,dubois14,schaye15}. 
At present, such a task is not straightforward: a fair comparison should take into account biases and uncertainties affecting the observational analysis before comparing to simulated galaxies.

Therefore it is of pivotal importance to assess the performances of photometric extraction and SED-fitting codes when recovering redshift and stellar mass in order to understand their impact on the statistical analyses of the galaxy population. Broadly speaking, observational biases 
can occur because of image confusion (i.e., blending between two nearby galaxies), the choice of algorithm used to  extract galaxy flux, and the assumptions made in the SED-fitting procedure. 
Previous works have already explored some of these effects. As an example, \cite{mobasher15} have quantified the global performances of an exhaustive list of existing SED-fitting codes, while relying on a large observed and semi-analytical mock catalogue \citep[see also][]{hildebrandt10}.  Focusing on mass and age estimates, \citet{pforr12} and \citet{pacifici12} investigated the impact of the chosen template SFH, while the effect of dust and metallicity has been studied in  \citet{mitchell13} and \citet{hayward15}.\\
Beyond the impact of simplistic SFHs \citep[like the $\tau$-model defined in][]{bruzual1983}, metallicity or dust distribution \citep[see also][]{guidi16}, the performance of  SED fitting  is extremely sensitive to the choice of  photometric filters, the depth of the survey, and flux measurements  \citep[see][]{bernardi13}. 
Hydrodynamical simulations have already  been  widely used to test the impact of the photometry extraction, as they allow to work on -often high resolution-  mock images of realistic galaxies. Amongst the tested effects, the choice of the apertures \citep{price17} and  the lack of resolution \citep[integrated photometry versus pixel-by-pixel fitting]{sorba&sawicki15,sorba18} have been found to systematically underestimate stellar masses \citep[see also][]{sanderson17} or to impair morphological estimators \citep{bottrell17}. 
All these past investigations underline the importance of understanding and quantifying  biases when recovering physical parameters from surveys, which could be as large as two orders of magnitude in some particular mass and redshift ranges {\citep[see e.g. the effect of dust on stellar mass computation][]{mitchell13}}. However, most of the literature is based either on simple phenomenological prescriptions or semi-analytical models, or when the sample is based on hydrodynamical simulation, it consists in no more than a handful of galaxies \citep[e.g.][]{guidi16}.  
Hence we still lack a  study relying on a sample that combines highly realistic baryon physics with a large cosmological volume (in order to minimize statistical uncertainties), capturing both galaxies' internal properties and  environment. Moreover, this study must be an end-to-end analysis, i.e. including the same limitations introduced by the observational strategy and data reduction pipeline in current or future surveys.
The present work aims to remedy this gap. 

To this end, we exploit the ligthcone from the {\sc Horizon-AGN} cosmological hydrodynamical simulation \citep{dubois14}. From this simulation  a mock catalogue of about 750,000 galaxies was extracted between $z=0$ and $4$, down to $M_{*}= 10^9 {\rm M}_{\odot}$. Hence our sample combines large number statistics over a wide redshift range with a wealth of information on galaxy properties.  
{Our aim is to carefully understand possible systematics arising when fitting the complex photometry of the galaxies with simplified templates. For this purpose galaxy photometry has to be as realistic as possible. One advantage of using hydrodynamical simulations over SAMs for this work is to better resolve galaxies in space and time. Fluxes spatially vary across the simulated galaxies (in the limit of the resolution of the simulation) depending on metallicity enrichment and dust attenuation, and therefore the integrated photometry will present a complexity similar to the real galaxies. 
In addition, SFHs in the hydrodynamical simulation   vary on a fine time grid and depend not only on the merger history of their host halo but also on stellar and AGN feedback, and on the detail of the gas accretion history. As emphasized in e.g. \citet{mitchell18}, several quantities (e.g., the gas return time-scale) are naturally constrained by gravitational forces and hydrodynamics, whereas they would conversely need to be globally tuned in SAMs. }  
Finally, the lightcone geometry mimics that of observed surveys, and allows us for instance  to implement the attenuation by the intergalactic medium (IGM) for each galaxy  by drawing individually lines of sight  through the foreground gas distribution.

The  goal of this first study is to assess the  photometric redshift ($z_\mathrm{phot}$), stellar mass ($M_\ast$) and star-formation rate (SFR) uncertainties caused by the choice of the filters, the signal-to-noise ratio ($S/N$) of the photometry and the SED fitting recipe used for analyzing the real galaxies. \\
For this purpose, observed-frame photometry is  post-processed with COSMOS-like $S/N$ for each galaxy of the {\sc Horizon-AGN} ligthcone  (as described in Section~\ref{sec:data-simulation}).  Then photometric redshifts and physical properties ($M_\ast$ and SFR) of mock galaxies are measured by applying  the same pipeline  used in the COSMOS field \citep[][hereafter L16]{laigle16}. {This procedure allows us to identify which source of uncertainty dominate the error budget (Section~\ref{sec:resCOS}).}
After  validation on the COSMOS2015 data, we mimic (in Section~\ref{sec:forecasts}) the expected photometry  for the  \emph{Euclid} mission \citep{laureijs11}, along with the Dark Energy Survey \citep[DES,][]{abbott18} and the Large Synoptic Survey Telescope \citep[LSST,][]{LSSTsciencebook}, to predict the expected $z_\mathrm{phot}$ and $M_\ast$ accuracy they should provide at completion. The possible synergy between these surveys is also explored. 
We then summarise our analysis and draw conclusions in Section \ref{sec:conclusion}. Additional material can be found in the Appendices, where we provides more details about how the  virtual photometry
has been computed  (Appendix~\ref{appendix:photometry}); we further discuss dust and IGM absorption (Appendix \ref{appendix:photoz}), zero-point magnitude offsets (Appendix \ref{App:Lephare}),  and  redshift errors (Appendix \ref{App:zerrors}). These virtual catalogs are going to be made publicly available at \href{https://www.horizon-simulation.org/data.html}{https://www.horizon-simulation.org/data.html}.

Throughout this study, we use a flat $\Lambda$CDM cosmology with 
$H_{0}=70.4$\,km\,s$^{-1}$\,Mpc$^{-1}$,  
$\Omega_{m}=0.272$,  $\Omega_{\Lambda}=0.728$,  and $n_s=0.967$ \citep[][WMAP-7]{komatsu11}. 
All magnitudes are  in the AB \citep{Oke:1974p12716} system. The initial mass function (IMF) follows \citet{chabrier03}. Quantities are said ``observed" when they include observational noise (for magnitudes) or when they are measured through SED fitting (redshift, stellar mass and SFR). If directly derived from the simulation, they are defined as ``intrinsic''.  


\section{Data and methods}
\label{sec:data-simulation}

 \begin{table*}
 \begin{center}
 \def\arraystretch{1.2}
 \begin{tabular}{|c c c c c|}
   \textbf{Name} & \textbf{Bands} & \textbf{\textit{i}-band depth} & \textbf{NIR depth} & \textbf{References} \\ \hline
   COSMOS-like & 26 bands from $u$ to $4.5\mu {\rm m}$ & 26.2$ \pm$ 0.1 (3$\sigma$) & 24.7 $\pm$ 0.1 (${K_{\rm s}}$, 3$\sigma$) & \cite{laigle16} \\
   LSST-like & $u$,$g$,$r$,$i$,$z$,$y$ & 27.0  (5$\sigma$) & NA & \citet{LSSTsciencebook} \\ %
   \emph{Euclid}+DES & $g$,$r$,$i$,$z$,$riz$,$Y$,$J$,$H$ & 24.5  ($riz$,10 $\sigma$) and 24.3 ($i$, 10$\sigma$) & 24.0 ($H$ band, 5$\sigma$) & \cite{abbott18,laureijs11} \\
   \emph{Euclid}+LSST & $u$,$g$,$r$,$i$,$z$,$y$,$riz$,$Y$,$J$,$H$ & 27.0 (5$\sigma$) & 24.0 ($H$ band, 5$\sigma$) & \citet{rhodes17}\\
   
 \end{tabular}
 \end{center}
 \caption{A summary of the  configurations envisaged in this study. Depths are quoted in AB magnitudes. The depths in all bands are summarised in Table~\ref{tab:maglim}. {A complete list of the COSMOS bands is provided in Table~1 of L16.}
 A {\sc Horizon-AGN} photometric catalogue is built for each configuration.}
 \label{Tab:conf}
 \end{table*}

\subsection{Description of the observational surveys}
\label{Sec:obssurveys}
The virtual photometric catalogue from the {\sc Horizon-AGN} simulation is built to mimic the COSMOS2015 catalogue. It also includes the  photometry  expected from  the \emph{Euclid} space-based telescope\footnote{
\href{https://www.euclid-ec.org}{https://www.euclid-ec.org}} \citep{abbott18}, the Large Synoptic Survey Telescope \citep[LSST,][]{LSSTsciencebook} and  the Dark Energy Survey \citep[DES,][]{laureijs11} in terms of filter passbands and depths. We briefly describe hereafter the different configurations investigated in this work to quantify the performances of galaxy redshift and physical property computation.  Table~\ref{Tab:conf} provides a summary of these surveys. 
\subsubsection{The COSMOS field}
\label{subsec:data}

The COSMOS deep optical and near-infrared catalogue (COSMOS2015) 
described in \citet[][hereafter L16]{laigle16} is used as a reference  to test the performances of our estimation of galaxy properties.
The catalogue includes more than 1 million objects detected within the 2\,deg$^2$ of the COSMOS field, observed in 30 bands from UV to IR ($0.25\!-\!8\,\mu$m). Here   the analysis is restricted to the  ``ultra-deep'' stripes, i.e.~four rectangular regions that in COSMOS2015 have been covered with higher near-Infrared (NIR) sensitivity (in the UltraVISTA-DR2 survey, $K_\mathrm{s}<24.7$, $3\sigma$) than the rest of the area. 
\\
COSMOS2015 contains far- and near-UV photometry (FUV and NUV, respectively)  from GALEX \citep{zamojski07}, but only NUV was used for the estimation of photometric redshifts and masses. 
In the optical, it includes the same $u,B,V,r,i,z$ data as previous releases from the Canada-Hawaii-France and Subaru telescopes  \citep{capak2007,ilbert09}. %
This baseline is complemented with Subaru medium- and narrow-band images between $4,000$ and $8,500$\,\emph{\AA}.
In the NIR,  $Y,J,H,K_{\rm s}$ images come from the second data release (DR2) of the UltraVISTA survey \citep[][]{mccracken12}, and the $Y$-band image from Subaru/Hyper-Suprime-Cam  \citep[HSC,][]{miyazaki2012}.  The \emph{Spitzer} Large Area Survey with HSC (SPLASH, Capak et al.~in prep.)  provides mid-IR (MIR) coverage with the four IRAC channels centred at $3.6$, $4.5$, $5.8$, and $8.0\,\mu$m. 

In order to derive photometry coherently  across different bands, the point-spread function (PSF) in each filter has been rescaled using a Moffat profile modelling. After the PSF homogenisation, fluxes are extracted within fixed apertures of $3\arcsec$ using {\sc SExtractor} \citep{bertin96} in dual image mode. Following a  reduction procedure similar to that described in \citet{mccracken12}, the detection image is a $\chi$-squared sum of the four NIR images of UltraVISTA DR2 and the $z^{++}$ band.  
\emph{Spitzer} sources are extracted by means of the code \textsc{IRACLEAN} \citep{hsieh12}. 
\\
Estimates obtained through SED fitting (photometric redshift, stellar mass, and other physical quantities) are also provided for each entry of the catalogue. The method adopted to obtain these estimates is described in Section~\ref{subsec:photoz_method}, where the same technique is applied to simulated galaxies. 
Further details about the COSMOS2015 catalogue can be found in L16. 

\subsubsection{Future surveys: \emph{Euclid} and LSST}
 
To compute galaxy properties from SED-fitting in comparable conditions to \emph{Euclid} and LSST, {\sc Horizon-AGN} galaxies are also post-processed to get the photometry in \emph{Euclid}, LSST and DES filters with depths similar to the ones expected for these surveys. We note that sometimes the expected depths from the literature are provided for point sources, and as a consequence give generally too optimistic estimators of the limiting magnitudes of extended sources. Our adopted limiting magnitudes are therefore probably not exactly the ones which will be obtained in the future, but they nonetheless reflect the relative depths of these upcoming surveys. The photometric baselines is detailed below, and the adopted limiting magnitudes in all bands are summarised in Table~\ref{tab:maglim}.
\paragraph*{\emph{Euclid}+DES configuration}
\emph{Euclid} will provide photometry in one broad-band optical ($riz$ filter) and 3 NIR filters ($Y$, $J$, $H$) with expected depths at completion of 24.5 (10$\sigma$, extended sources) in the optical and 24.0 (5$\sigma$, point sources) in the NIR bands. This broadband baseline alone is not sufficient to  compute photometric redshifts with a  high enough accuracy {(especially to constrain the Balmer break in the optical)}, therefore it has to be complemented with ground-based optical photometry \citep[see e.g.][]{sorba11}. In particular, DES  provides photometry over 5000 deg$^{2}$ in the Southern sky in $g$, $r$, $i$ and $z$ with depth of 24.33, 24.08, 23.44, 22.69 \citep[10 $\sigma$, extended sources,][]{abbott18}, which matches the \emph{Euclid} requirements \citep{laureijs11}. {DES photometry provides a finer sampling of the optical range than the \emph{Euclid} $riz$ filter alone}. Collaboration between \emph{Euclid} and DES is planned. Therefore in the current work we explore what would be the expected performance of such a configuration. 
\paragraph*{LSST configuration}
The survey conducted on LSST will provide photometry in the optical over $30000$ deg$^{2}$. LSST single visit depth should reach 24.5 in $r$ ($5\sigma$, point sources), and the co-added survey depth should reach 26.3, 27.5, 27.7, 27.0, 26.2 and 24.9 ($5\sigma$, point sources) in $u$, $g$, $r$, $i$, $z$ and $y$ bands respectively \citep{LSSTsciencebook}. For weak-lensing studies, the ``gold" sample of LSST galaxies with a high $S/N$ is defined with a magnitude cut $i<25.3$. We use this cut in the present work when studying the LSST-like configuration. \\
In the Southern sky, \emph{Euclid} and LSST will overlap over at least 7000 deg$^{2}$. It is therefore natural to explore the possible gain to combine both datasets. To this end we also analyse {\sc Horizon-AGN} galaxies in the \emph{Euclid}+LSST-like configuration. 
\subsection{The {\sc Horizon-AGN} simulation}
\label{subsec:simu}

This study relies on  \textsc{Horizon-AGN}\footnote{\href{http://www.horizon-simulation.org/}{http://www.horizon-simulation.org/}} \citep[][]{dubois14}, a cosmological hydrodynamical simulation in overall fairly good agreement with  observations, in the redshift and mass regime of the present analysis \citep[see][]{kaviraj16}. \\
The simulation box, run with the {\sc ramses} code~\citep{teyssier02},  is $L_{\rm  box}=100 \, h^{-1}\,\mathrm{Mpc}$ on a side, and the volume contains $1024^3$ dark matter (DM) particles, corresponding to a DM mass resolution of $8\times 10^7 \, M_\odot$.
The initially coarse $1024^3$ grid is adaptively refined down to $1$ physical kpc. 
The refinement procedure leads to a typical number of $6.5\times 10^9$ gas resolution elements (leaf cells) in the \textsc{Horizon-AGN}  simulation at $z=1$.
\\
Heating of the gas from a uniform UV background takes place after redshift $z_{\rm  reion} = 10$, following~\cite{haardt&madau96}. 
Gas can cool down to $10^4\, \rm K$ through H and He collision and with a contribution from metals that follows the rates tabulated in~\cite{sutherland&dopita93}. 
Star formation occurs in regions where gas number density is above $n_0=0.1\, \rm H\, cm^{-3}$,  following a Schmidt law: $\dot \rho_*= \epsilon_* {\rho_{\rm g} / t_{\rm  ff}}$,  where $\dot \rho_*$ is the star formation rate mass density, $\rho_{\rm g}$ the gas mass density, $\epsilon_*=0.02$ the constant star formation efficiency  and $t_{\rm  ff}$ the gas local free-fall time.
Feedback from stellar winds and  supernova (both type Ia and  II) are included into the simulation with mass, energy, and metal releases.
Galactic black hole formation is also implemented in \textsc{Horizon-AGN}, with accretion efficiency tuned to match the 
black hole-galaxy scaling
relations at $z=0$. Black hole energy is released in either quasar or radio mode depending on the accretion rate  \citep[see][for more details]{dubois12}.
\\
The lightcone has been extracted on-the-fly as described in \cite{pichon10}. {For the lightcone extraction, gas leaf-cells were replaced by gas particles, and treated  as the stars and dark matter particles. All particles were extracted at each coarse time step according to their proper distance to the observer at the origin. In total the lightcone contains  about 22000 portions of concentric shells.}
The lightcone projected area is 5 deg$^{2}$ below $z=1$, and 1 deg$^{2}$ above. However, we restrict ourselves to 1\,deg$^{2}$ over the whole redshift range considered in this study. The full lightcone up to $z=4$ contains about 19 replica of the {\sc Horizon-AGN} box.

\subsection{Generating a mock photometric catalogue}
\label{subsec:mock_cat}
\subsubsection{Galaxy extraction}
The \textsc{ AdaptaHOP} halo finder \citep{aubert04} is run on the lightcone over $0<z<4$ to identify galaxies from the stellar particles distribution. Local stellar particle density is computed from the 20 nearest neighbours, and structures are selected with a density threshold  equal to 178 times the average matter density at that redshift. Galaxies resulting in less than 50 particles ($\simeq 10^8 \,{\rm M}_\odot$) are not included in the catalogue.
Since the identification technique is redshift dependent, \textsc{AdaptaHOP} is run iteratively on thin lightcone slices {(about 4000 slices up to $z=4$)} of few comoving Mpc (cMpc). Slices are overlapping to avoid edge effects (i.e. cutting galaxies in the extraction) and duplicates are removed. 
\subsubsection{Galaxy SED computation and dust attenuation}
\label{Sec:simPhot}
{Although the simulation assumes a Salpeter IMF \citep{salpeter55} to model stellar mass losses, we have decided to post-process it with a Chabrier IMF \citep[][]{chabrier03}.} The choice of the IMF is significant, as it controls both the stellar mass loss prescription and the overall mass-to-light ratio. The Chabrier IMF brings the simulated galaxy counts in much better agreement with the COSMOS2015 galaxy counts (see Appendix~\ref{appendix:photometry}. Magnitudes are $\sim 0.4$ mag fainter with a Salpeter IMF  compared to Chabrier's). 
For each galaxy, each stellar particle is linked to a  single stellar population (SSP) obtained with the stellar population synthesis model of \citet[][hereafter BC03]{bruzual&charlot03}.   
Because stellar particle ages and metallicities vary on a much finer grid than the BC03 models, an  interpolation is carried out between SSPs to reproduce the desired values. In addition, each SSP is also rescaled to match the initial  stellar mass of the particle, in order to follow the same mass loss fraction for the simulated galaxies as in BC03 (see Appendix \ref{appendix:imf}). This rescaling is essential to avoid discrepancies between the intrinsic and computed galaxy properties coming from the different SSP prescriptions, which is out of topic for the present work. {It should be noted that the metallicity of stellar particles in {\sc Horizon-AGN} has been boosted by a  empirically computed factor, to match the observed mass-metallicity relation. This factor $f_{Z}$ is redshift dependent as follows: $f_{Z} = 4.08430 - 0.213574z - 0.111197z^2$ \citep[see][for more details]{kaviraj16}}.

Dust attenuation is also modelled for each star particle using the gas metal mass distribution as a proxy for the dust distribution. Gas metal mass is evaluated in a cube of $138$ comoving kpc$^3$ around each galaxy 
and a constant dust-to-metal mass ratio is adopted. Although a mesh-based code is used to run {\sc Horizon-AGN} and in particular to follow the gas distribution, gas cells have been turned in particles when extracting the lightcone. In order to get a smooth metal field around the galaxy from this gas particle distribution, a Delaunay tessellation is computed on the particles to avoid cells with null values in under-dense regions, and then interpolated on a regular grid with a resolution of $\sim 1$ ckpc. The dust column density and the optical depth along the line of sight are computed for each stellar particle in the galaxy using the ${\rm R}_{V}=3.1$ Milky Way dust grain model by \citet{weingartner01}. Further details on the dust computation and dust-to-metal mass ratio calibration are given in Appendix~\ref{appendix:photometry}. This dust attenuation model only takes into account absorption and does not include scattering. While this is not a problem in the rest-frame optical and NIR, the impact in the rest-frame UV is non negligible \citep[][]{kaviraj16}. This effect is not corrected in our catalog and as a result, our galaxies are up to 0.8 mag brighter without scattering {in the UV part of the spectrum. In fact, including scattering would have a similar impact as a steepening of the dust attenuation law in the UV.} 
A dust-free version of the catalogue is also produced in order to isolate the impact of dust attenuation on the computation of galaxy physical  properties. 

Flux contamination by nebular emission lines is not included in our virtual photometry. Consistently, emission line parametrization is also turned off in the SED-fitting computation. In real surveys, emission lines can help determining the photometric redshifts. On the other hand, the dispersion in the emission line ratios is poorly modelled by the SED-fitting code and can bias the redshift estimation. 

Finally, we stress that we do not model extinction by the Milky way. Photometry from observed survey is generally corrected from galactic extinction using galactic reddening maps \citep[e.g.][]{schlegel98}. However, the amount of absorption in a given band will depend on the source SED. As shown in  \cite{galametz17}, the band-pass extinction can vary by up to 20\% from the average correction depending on the SED. The discrepancy between the effective extinction and the average correction of the photometry can potentially lead to additional systematics which are therefore not accounted for here. 

{Note that {\sc Horizon-AGN} only reproduces  global mass assembly up to some point.  Although the simulation broadly matches the mass function evolution with redshift and the SFR main sequence \citep{kaviraj16},  at low-mass ($\log M_{*}/{\rm M}_{\odot}>9.5$)  the mass function is systematically overestimated given the present sub-grid stellar feedback and star-formation recipes. Conversely, at high redshift ($z>4$) it is also underestimated because of  limited spatial resolution. 
To be conservative,  forecasts were therefore limited within the redshift and mass range where the simulation is reliable. Appendix~\ref{Ap:limitations} lists the limitations of our modelling.}

\subsubsection{IGM absorption}
Prior to convolving the galaxy spectrum with photometric filters, attenuation by the IGM must be implemented. The knowledge of the gas distribution in the ligthcone allows us to consistently implement IGM absorption, while accounting for  variation from one sight-line to the other. Conversely, it is globally accounted  for with an analytical prescription at the SED-fitting stage. Therefore our photometric catalogue allows to test the effect of the inhomogeneous IGM attenuation on photometric redshifts and galaxy properties.  The  focus is on HI absorption in the Lyman-series (hence neglecting metal lines, such as e.g. the CIV forest). Details about the implementation of the attenuation on the line-of-sight of each galaxy is given in Appendix~\ref{appendix:photometry}. 
An IGM-free version of the catalogue is also produced in order to isolate the impact of IGM absorption. 

\subsubsection{Photometry extraction and error implementation}
Eventually the integrated galaxy spectra are computed between  91 and $16\times10^{5}$\,\emph{\AA} (with 1221 wavelength points) by adding  all the SSPs of a given galaxy together. Note that IR dust emission is not computed because not used here.  Apparent total magnitudes are obtained by convolving the spectrum (redshifted to the intrinsic redshifts of the galaxies) with the same filter set as used in COSMOS2015 ($u$, $B$, $V$, $r$, $i^{+}$, $z^{++}$, $Y$, $J$, $H$, $K_{\rm s}$, [$3.6\mu$m], [$4.5\mu$m], and the 14 intermediate and narrow bands\footnote{At lower redshift, GALEX FUV and NUV filters are  relevant, as they bring information about young stellar populations. However, it is more difficult to reproduce their flux extraction  and the related uncertainties. Therefore, NUV and FUV are excluded from the mock catalogue.}),
\emph{Euclid}, DES and LSST (see Section~\ref{Sec:obssurveys}).
\\
Photometric errors are added in each band to reproduce the $S/N$ distribution and sensitivity limit of COSMOS2015 and DES, and the ones expected for \emph{Euclid} and LSST (see Table~\ref{Tab:conf} and Appendix \ref{appendix:photometry}). 
It should be emphasized here that the photometry has been derived from the entire distribution of star particles and not through a realistic flux extraction from images \citep[as it is done in real observations via  tools like {\sc SExtractor},][]{bertin96}. Consequently, our mock catalogue does not include all the associated photometric issues including potential systematic effects like blending, object fragmentation, imperfect background sky subtraction and PSF homogenisation and offsets due to the rescaling of fixed aperture to total fluxes, which will be the topic of a future work. Photometric errors as implemented in the current catalogue simply correspond to gaussian noise with standard deviation depending on the galaxy flux and the depth of the surveys. 

\subsection{SED-fitting: method}
\label{subsec:photoz_method}

\subsubsection{Photometric redshifts}
Photometric redshifts ($z_\mathrm{phot}$) are computed using the code {\sc LePhare} \citep{arnouts02,ilbert06} with a configuration similar to \citet{ilbert13}. The SED library includes spiral and elliptical galaxies from \citet{polletta07}, along with  bluer templates of young star-forming galaxies built by means of the BC03 model. 
Dust extinction is added to the templates  according to one of the following attenuation curves: \citet{prevot84}, \citet{calzetti2000}, or a modified version of \citet{calzetti2000} with the addition of the ``graphite bump'' at $\sim2\,175$~$\angstrom$   \citep[e.g.][]{fishera11,ilbert09}. The  $E(B-V)$ values range from 0 to 0.5. IGM absorption is implemented following the analytical correction of \citet{madau95}.\\
As strong nebular emission  (such as [OII] or H$\alpha$ lines) can significantly increase the flux measured in a photometric filter, nebular emission lines were considered in the $z_\mathrm{phot}$ computation of real data (L16). On the other hand,  such options are disabled when {\sc LePhare} is run on  \textsc{Horizon-AGN}  galaxies, since nebular emission lines are not modelled for them (Section \ref{subsec:mock_cat}). 

Each template in the {\sc LePhare} library is fit to the virtual photometry of the \textsc{Horizon-AGN} galaxies. The code computes the goodness of fit ($\chi^2$) for each redshift solution and their likelihood ($\La$). 
The $z_\mathrm{phot}$ estimate for a given galaxy is defined as the median of the $\La(z)$ distribution, and the $1\sigma$ uncertainty ($\sigma_\mathrm{z}$) is the interval enclosing 68 percent of its area (see Section \ref{subsec:sigmaz}). 
{In order to improve the SED-fitting performance, a mild  luminosity prior is also applied to reduce the fraction of catastrophic outliers. This prior is based on the observed luminosity function in the rest-frame $B$ band \citep[e.g.][]{ilbert06,Zucca09,lopez-sanjuan2017}. Given the extremely low number density expected in the (extrapolated) bright end of the luminosity function, our codes excludes solutions with absolute magnitude $M_B>-24.2$. No other prior on the redshift distribution is applied in {\sc LePhare}. }

{As done in L16, fluxes rather than magnitudes are used when running {\sc LePhare}. This allows to deal robustly with faint or non-detected objects}.
\paragraph*{Systematic offsets}
With real datasets, an important aspect of the $z_\mathrm{phot}$ computation is the derivation of systematic offsets which are applied to match the predicted magnitudes and the observed ones \citep{ilbert06} based on the spectroscopic sub-sample. This calibration is designed to empirically correct both for the incomplete template library and possible systematics in the galaxy magnitude extraction. However this calibration might be biased because it relies on a spectroscopic subsample.  
As described in Appendix~\ref{App:Lephare}, we test the computation of systematic offsets in the COSMOS-like catalogue by using a sub-sample of galaxies matching the spectroscopic catalog on COSMOS. We find that there is no need for this calibration in the simulated catalogue. The offsets introduced by an imperfect knowledge of the templates are therefore negligible in {\sc Horizon-AGN}\footnote{In fact, the virtual photometry of the simulated galaxies presents less diversity than in real datasets, because it is built by the mean of BC03 SSP models for all the galaxies, which might be a reason why these offsets are negligible.}. 

\subsubsection{Stellar mass and star formation rate}
Stellar mass and star formation rate (SFR) are then derived using another template library built by using the BC03 model, in the same way as for the COSMOS2015 catalogue. In this second run, {similarly to what was done in COSMOS2015 for computational reasons},  only two extinction laws are used (\citealp{arnouts13} and \citealp{calzetti2000}). 
The initial mass function (IMF) is assumed  to be  \citet{chabrier03}'s, and the stellar metallicity of each template to be either solar ($Z_\odot$) or subsolar ($0.4\,Z_\odot$).

In our library, the SFHs used to build the SEDs are parametrised with an analytic equation. It can be exponentially declining, i.e.
${\rm SFR}(t) \propto e^{-t/\tau}\,. $
An alternative definition is the ``delayed" star formation history, more suitable to model a galaxy with gas infall:
$
{\rm SFR}(t) \propto e^{-t/\tau}  {t}/{\tau} \,. 
$
In both cases $\tau$ is the $e$-folding time, varying between one tenth and several Gyr\footnote{The following timesteps are used: $\tau=0.1,0.3,1,3,4,30$ Gyr for the exponentially declining star-formation histories, and $\tau=1, 3$ Gyr for the delayed star formation histories.}. In the latter case, $\tau$ also represents the galaxy age (since its formation) at which the SFR peaks. In particular, such SFHs cannot reproduce multiple bursts of star formation. {The SFH models start forming stars from $t=0$. When building the library of templates, $t$ is sampled from a few hundreds Myr to the age of the Universe at a given redshift, with up to 44 steps at $z=0$}.

\section{SED-fitting performance: present surveys}
\label{sec:resCOS}
\subsection{Photometric redshifts}
\label{sec:photoz}

Let us first investigate our ability to recover photometric redshifts from the photometry, using the simulated COSMOS-like catalogue.

\begin{figure*}
\includegraphics[height=8cm]{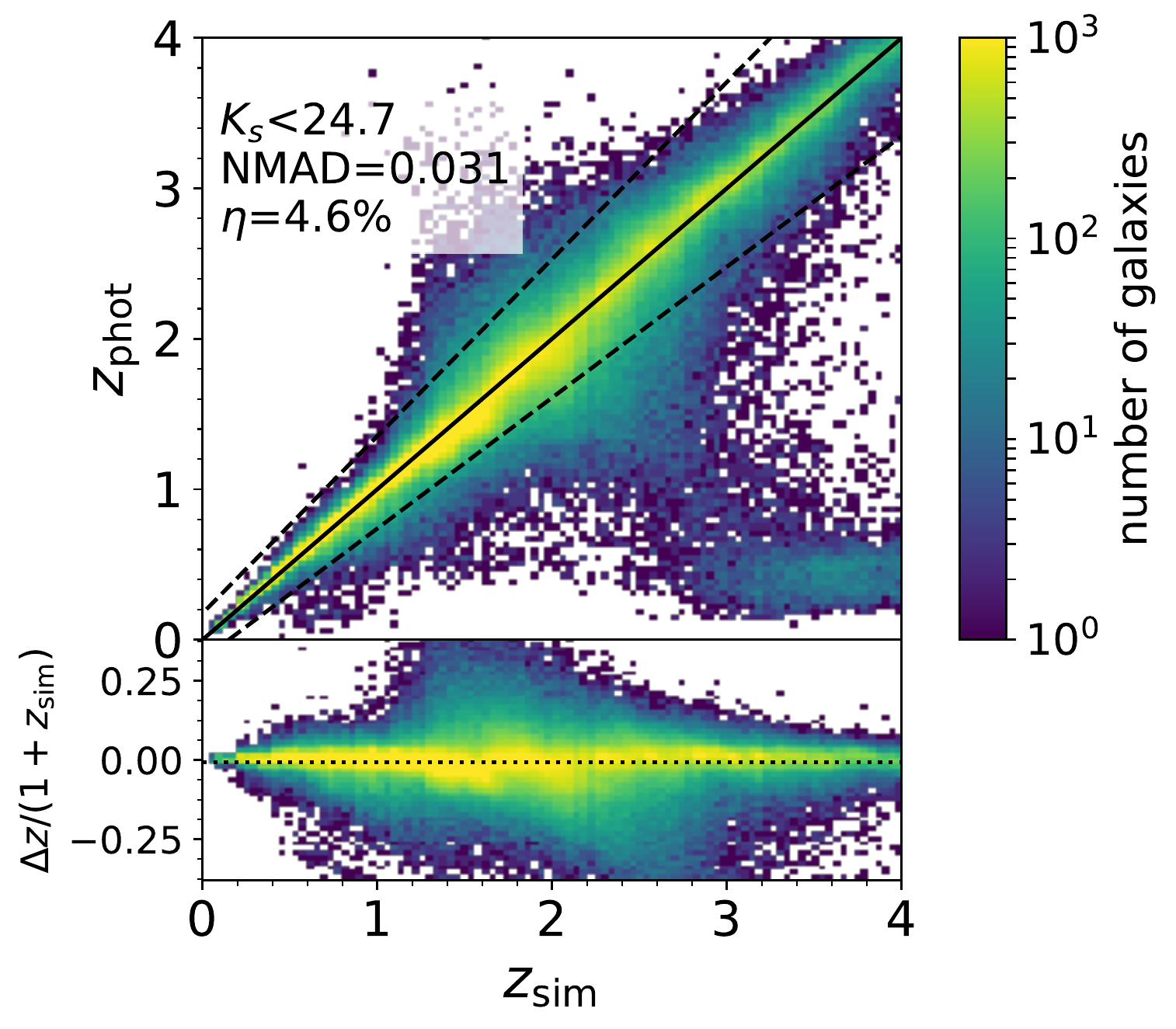}
\hspace{8pt}
\includegraphics[height=8cm]{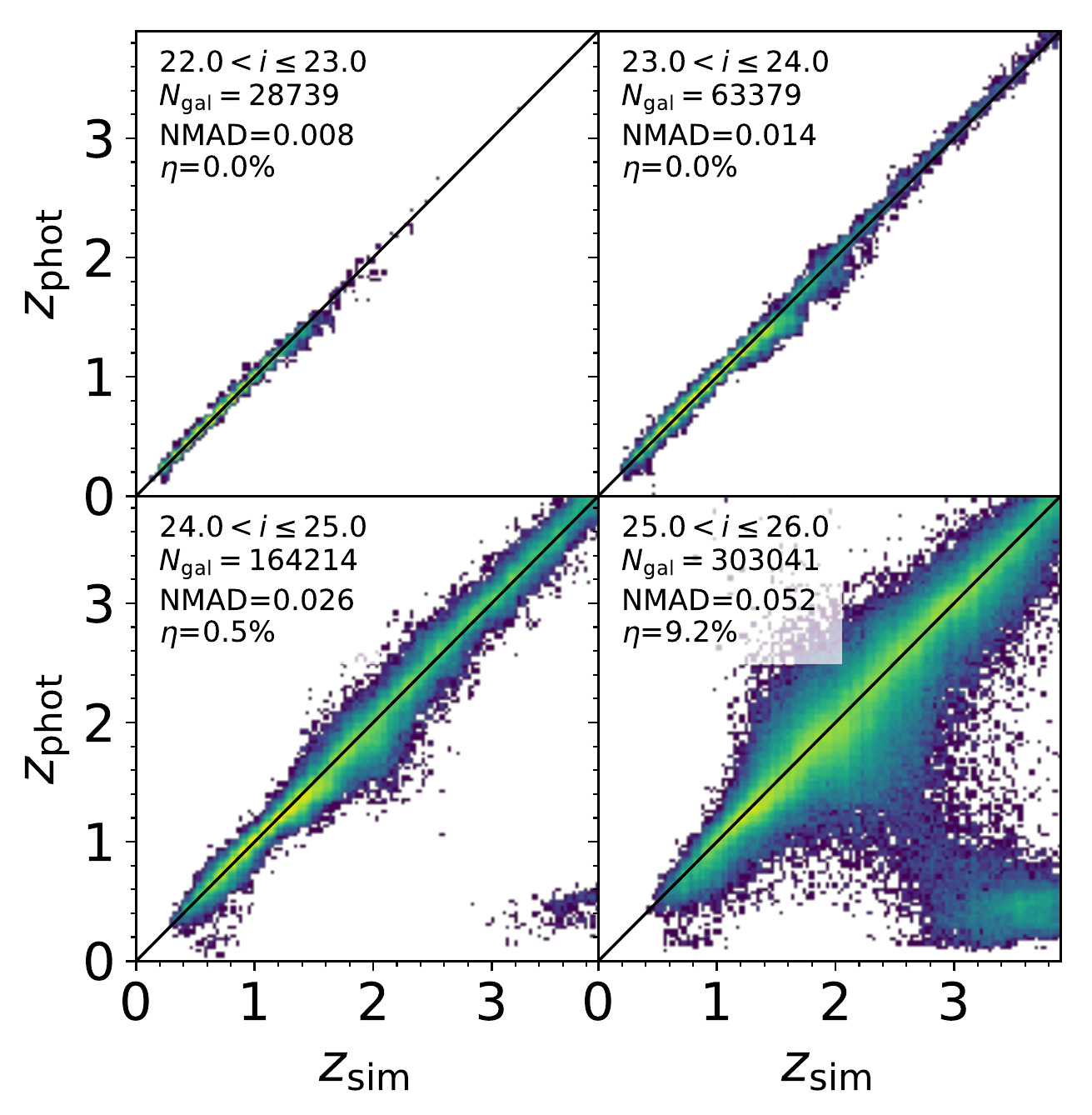}

\caption{ \textit{Left:} Comparison between galaxy redshifts in the {\sc Horizon-AGN} lightcone ($z_\mathrm{sim}$) and the photometric redshifts recovered by {\sc LePhare} ($z_\mathrm{phot}$) for the whole sample of 506,361 objects selected at $K_\mathrm{s}<24.7$.  The $z_\mathrm{phot}$ uncertainty computed as NMAD is shown along with the catastrophic error fraction ($\eta$, see Sect.~\ref{sec:photoz}). Solid line is the 1:1 bisector while dashed lines mark the $\pm0.15\,(1+z)$ threshold used to compute $\eta$. In the bottom panel $\Delta z$ is defined as $z_\mathrm{phot}-z_\mathrm{spec}$. \textit{Right:} comparison between $z_\mathrm{sim}$ and  $z_\mathrm{phot}$ as a function of apparent magnitude in the $i^*$ band (same color scale as in the left panel). The number of galaxies per magnitude bin is shown in each panel along with NMAD uncertainty and catastrophic error fraction, and is also reported in Table~\ref{tab:comp1}.}
\label{fig:zphotzspec}

\end{figure*}

\subsubsection{Comparison between $\protect z_\mathrm{sim}$ and $\protect z_\mathrm{phot}$ }
\label{subsec:zphotzspec}

The accuracy of our SED fitting method can be first tested by comparing galaxy redshifts in the lightcone ($z_\mathrm{sim}$) to those obtained by  {\sc LePhare}\footnote{
$z_{\rm sim}$ includes galaxy peculiar velocity.}. 
Fig.~\ref{fig:zphotzspec} (left panel) presents such a  comparison for galaxies with $K_\mathrm{s}<24.7$. 
In addition to the magnitude cut,   pathological cases are also excluded with reduced chi-square  values $\chi^2_\mathrm{red}>10$, which represent $<0.1$ per cent of the whole  sample (namely, 387 out of 541,555  $K_\mathrm{s}$-selected galaxies). This kind of $\chi^2$ selection is also applied in real surveys and removes similar fractions of problematic objects  \citep{davidzon17,caputi15}. 

Overall, we do not find significant systematics  affecting {\sc LePhare} redshift estimates. 
Despite the simplistic implementation of dust extinction, the limited number of templates and the fact that they have been calibrated to represent the \textit{observed} universe, our recipe 
captures the main features of the simulated galaxy SEDs, and recovers their  redshifts with a precision comparable to what is achieved with the COSMOS2015 catalogue. In our simulation, the normalized median absolute deviation \citep[NMAD,][]{hoaglin83} for the entire sample is $1.48\times\mathrm{median}(\vert \Delta z\vert)/(1+z_\mathrm{sim}) = 0.031$.  
The fraction of outliers, defined as objects with $\vert \Delta z\vert>0.15\,(1+z_\mathrm{sim})$, is $\eta=4.6$ per cent (Fig.~\ref{fig:zphotzspec}, left panel). 
In the real survey, with the same cut at $K_\mathrm{s}<24.7$, a comparison between photometric and spectroscopic redshifts results yields $\mathrm{NMAD}=0.013$ and $\eta=3.1$ per cent respectively. Such difference between simulation and observation is explained because the spectroscopic samples are not representative (as detailed below).   

Specific sub-samples of galaxies may show a smaller scatter than the global one. Galaxies' $z_\mathrm{phot}$ precision depends on their $S/N$, which in turn correlates with apparent magnitude, therefore it is useful to estimate the 
NMAD as a function of the latter.  Fig.~\ref{fig:zphotzspec} (right panel) presents the $z_\mathrm{phot}$ vs $z_\mathrm{spec}$ diagram after dividing {\sc Horizon-AGN} galaxies in four $i^+$-band magnitude bins.   NMAD and $\eta$  of each sub-sample are reported in Table~\ref{tab:comp1}, along with the corresponding metrics obtained in L16 for COSMOS2015. 
The precision  of {\sc Horizon-AGN} $z_\mathrm{phot}$ is comparable to that found in the real survey. 
However the fraction of  catastrophic failures is  systematically larger in COSMOS2015.  Such a discrepancy can be explained by observational  uncertainties related to image source extraction (e.g., confusion noise) not considered in Horizon-AGN. In addition, it should be remembered that the observed $\eta$ and NMAD are provided only for a  sub-sample of high-confidence spectroscopic galaxies, which are generally biased towards bright objects.

\begin{figure*}
\centering
\includegraphics[width=0.99\columnwidth]{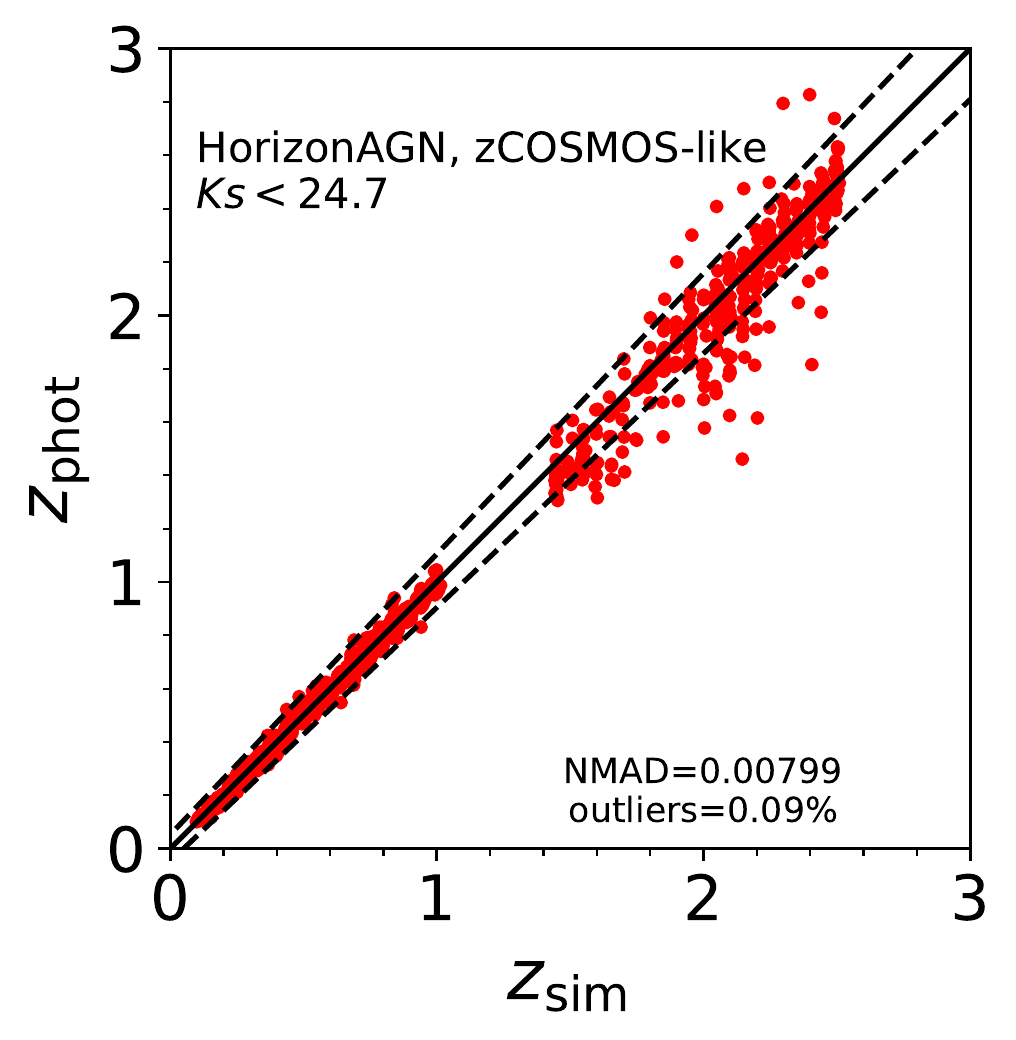}\vspace{11pt}
\includegraphics[width=0.99\columnwidth]{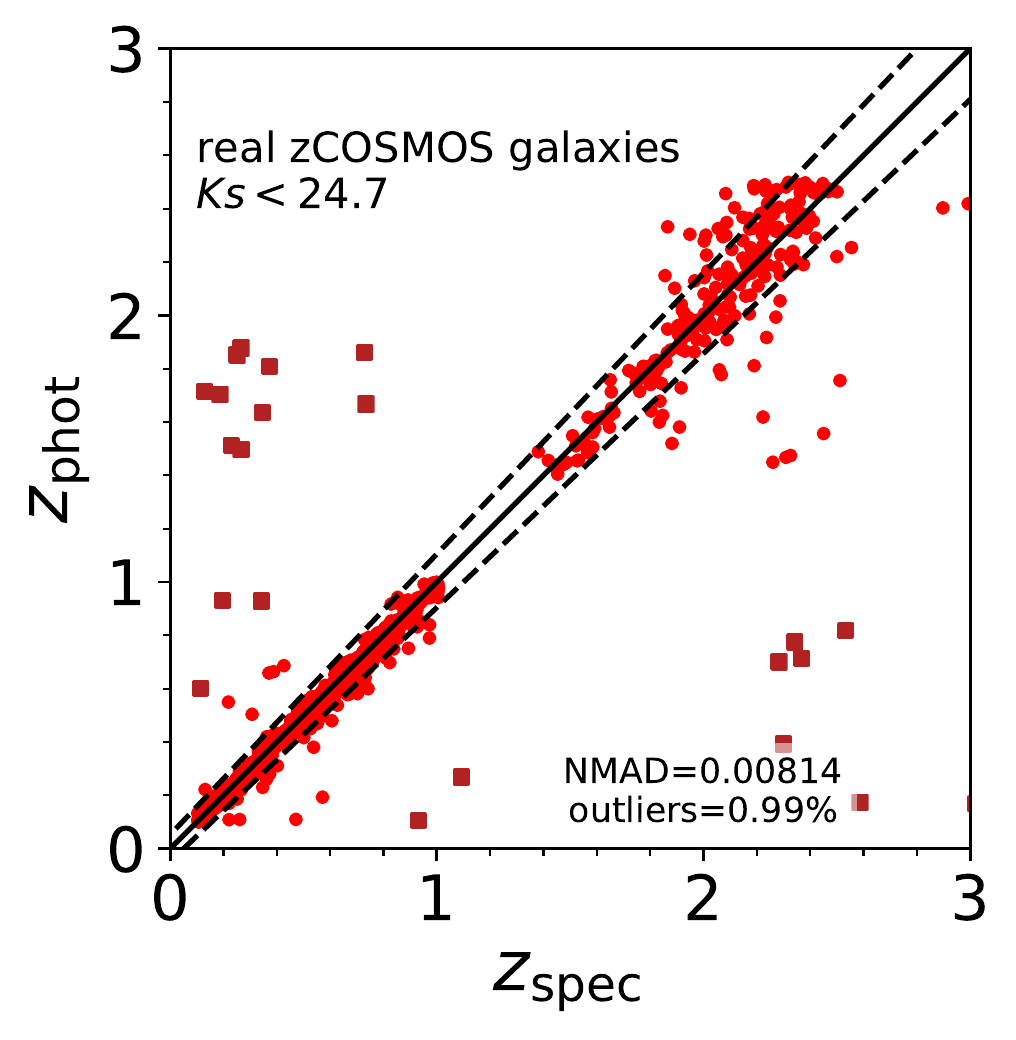}
\caption{\textit{Left:} Comparison between redshift of {\sc Horizon-AGN} galaxies (red dots) versus photometric redshifts from {\sc LePhare}. Solid and dashed lines show 1:1 relation and $\pm0.05(1+z)$ deviation. The gap at $z\sim1.5$ is due to the different selection functions of zCOSMOS-Bright and Deep. 
\textit{Right:} Same as \textit{left} panel, but for real COSMOS2015 galaxies having photometric redshift computed by {\sc LePhare} and spectroscopic redshifts from zCOSMOS-Bright and Deep survey (red circles and squares). Squares indicate catastrophic failures with $\vert \Delta z\vert>0.3\,(1+z_\mathrm{spec})$, a class of objects that is not present in the comparison using the mock sample (see Sect.~\ref{subsec:zphotzspec}). }
\label{fig:zphotzspec_cosmos}
\end{figure*}

\paragraph*{Comparison with a $z$COSMOS-like sub-sample}
The  COSMOS2015 $z_\mathrm{phot}$ precision  was assessed by using data collected during various spectroscopic campaigns (see Table 5 of L16). One of the most important  is the zCOSMOS survey, accounting for almost half of high-quality galaxy spectra at $z<3$ \citep[][]{lilly07}\footnote{The full description of the zCOSMOS-Deep sample characteristics and the evaluation of the redshift estimation performance will be published in a future paper (Lilly et al. in prep).}. 
To reproduce a similar subset in Horizon-AGN,   a sub-sample of galaxies in the lightcone is identified using selection criteria similar to zCOSMOS. We randomly extract mock galaxies at $i^+<22.5$ until we match both the magnitude and  redshift distributions of the zCOSMOS-Bright sample \citep{lilly07}, and do the same at $B<25$ to mimick the zCOSMOS-Deep (Lilly et al. in prep). The zCOSMOS-Deep selection aimed at enforcing a target selection at $z>1.5$, but some faint galaxies at lower redshift were also observed. Yet,  those interlopers are not included in the zCOSMOS-Deep mock sample by applying a sharp $1.5<z<2.5$ cut. It has to be noted that a large fraction of the catastrophic failures seen at $z_\mathrm{spec}<1$ in the real data  correspond indeed to the  selection that  is not replicated in the simulation. 
The result is a set of about 5,000 galaxies for which  pseudo-spectroscopic measurements are created by perturbing $z_\mathrm{sim}$ with a Gaussian random error having  $\sigma=0.0004(1+z)$.\footnote{ 
The standard deviation of the Gaussian random error corresponds to the 1$\sigma$ uncertainty of zCOSMOS-Bright galaxy estimated by repeated measurements \citep{lilly07}. Given the order of magnitude of $z_\mathrm{phot}$ error, that pseudo-spectroscopic perturbation is negligible in the following analysis.}    

Figure \ref{fig:zphotzspec_cosmos} shows the $z_\mathrm{phot}$ versus $z_\mathrm{spec}$ comparison in {\sc Horizon-AGN} (upper panel) and COSMOS2015 (lower panel). Since stars are not included in our simulation  the analysis is restricted to $z>0.1$ to avoid both stellar interlopers present in the real survey and  bright COSMOS2015 galaxies erroneously classified as stars. 
With such a zCOSMOS-like selection, the NMAD measured in {\sc Horizon-AGN} is in excellent agreement with COSMOS2015  ($0.0080$ and $0.0081$ respectively). 
On the other hand, catastrophic errors are more numerous in the real sample, as also found in the previous test (see Table \ref{tab:comp1}). 
Appendix~\ref{App:outliers},  focuses on the 22 outliers with $\vert \Delta z\vert>0.3\,(1+z_\mathrm{spec})$ to understand this difference. {In most of the case, the failure arises either because of uncertain photometry (fragmented or blended objects) or spectroscopic misidentification.}

\paragraph*{Impact of IGM absorption}

The {\sc Horizon-AGN} lightcone allows us to quantify the importance of correctly accounting for IGM absorption, by comparing the $z_\mathrm{phot}$ estimate computed from the IGM and IGM-free (i.e. turning off IGM both in the photometry computation and at the SED-fitting stage) versions of the catalogue. This test is carried out with the dust-free version of the catalogue. 
In {\sc Horizon-AGN}, IGM absorption is implemented along the line-of-sight of each galaxy knowing the foreground HI distribution, while in {\sc LePhare}, absorption due to the intervening IGM between the galaxy and the observer is taken into account by applying an  \textit{average} correction as a function of redshift based on an analytical relation \citep{madau95} \footnote{As explained in Appendix~\ref{appendix:photometry}, there is a slight discrepancy between the average IGM absorption in {\sc Horizon-AGN} and the correction implemented in {\sc LePhare}, the latter being  stronger than the former. However the present work does not aim  at correcting this discrepancy as it is likely to also happen  when fitting the photometry of real galaxies.}. Overestimating the IGM correction (or neglecting the line-of-sight variability of IGM opacity)  might impact the performance of  $z_\mathrm{phot}$ estimate for distant galaxies \citep[as suggested in][]{Thomas2017}. 
IGM absorption plays a role mostly at $z>2$, with a more dramatic attenuation of galaxy photometry from $z\sim 3$ as illustrated by Fig.~\ref{fig:IGM}.  

{
The global redshift accuracy  estimated by the NMAD  is impacted at a below per cent level, only at the very faint end of the galaxy population. Interestingly, implementing IGM absorption slightly helps constraining the redshift of faint galaxies, when averaging it over the entire redshift range (In the bin $25<i<26$, NMAD$=0.049$ and $\eta=8.7$ \% in the IGM-free version, while NMAD$=0.045$ and $\eta=7.6$ \% with IGM). However,
at $z>3$, the fraction of outliers populating the clump located at $\left[ z_{\rm sim}>2.5 \right]  \cap \left[ z_{\rm obs}<1.5 \right]$ strongly increases when IGM is included: at $24<i<25$, only 7\% of the existing outliers populate this region in the no-IGM case, while they are 58\% with IGM. At $25<i<26$, in this region, they are 16\% and 47\% in the no-IGM and IGM cases respectively. 
This population of outliers also occurs when fitting observed population of galaxies \citep[see Fig.~11 in][]{laigle16} and our work  suggests  therefore that the way IGM is accounted for is important to mitigate it. }

\paragraph*{Impact of medium-bands}  Let us now quantify the improvement of $z_\mathrm{phot}$ estimates  due to medium-band photometry. 
The inclusion of these bands in a deep extragalactic survey is {very useful to better constrain the redshift from spectral features occuring in the optical wavelength range (Lyman and Balmer breaks depending on the redshift, nebular emission lines in real datasets) but}  expensive in exposure time. It is therefore important to check whether  it is worthwhile. 
Even though the major advantage of those filters is to find contribution from nebular emission lines (not implemented in our virtual magnitudes), they should also in principle help to constrain the galaxy continuum.  Indeed, when  medium-band filters are removed, the $z_\mathrm{phot}$ precision degrades considerably (see Table~\ref{tab:comp1}). For instance, the NMAD of bright objects ($22<i<23$) is degraded by almost a factor 3, going from $\mathrm{NMAD}=0.008$ (when medium-bands are included) to 0.023.
At the fainter magnitudes ($24<i<26$) the difference is less remarkable,  but nonetheless the absence of medium-bands results in a $z_\mathrm{phot}$ scatter larger by $30\!-\!40$ per cent. The outlier fraction, especially at $i>25$, also increases (see Table \ref{Tab:conf}). 

\begin{table}
\begin{center}
  \caption{Statistical errors (NMAD) and percentage of catastrophic errors ($\eta$) in different $i^+$ magnitude bins. Results for COSMOS2015 galaxies (L16) are compared to the outcome of our simulation.  \label{tab:comp1}}
\begin{tabular}{ c | c c c c c c} \hline
     $i^{+}$   & \multicolumn{2}{c}{COSMOS2015} &    \multicolumn{2}{c}{Hz-AGN} &    \multicolumn{2}{c}{Hz-AGN}   \\
       & \multicolumn{2}{c}{} &   \multicolumn{2}{c}{with IB} &   \multicolumn{2}{c}{without IB}  \\
             mag            &  NMAD & $\eta$ (\%)   &  NMAD & $\eta$  (\%) &  NMAD & $\eta$  (\%) \\ \hline
$($22,23$\rbrack$ & 0.010 & 1.7  &0.008 & 0.0 & 0.023 & 0.0 \\
$($23,24$\rbrack$ & 0.022 & 6.7  &0.014 & 0.0 & 0.028 & 0.1 \\
$($24,25$\rbrack$ & 0.034 & 10.2 &0.026 & 0.5 & 0.037 & 0.8 \\
$($25,26$\rbrack$ & 0.057 & 22.0 &0.052 & 9.2 & 0.065 & 11.7 \\
  \hline
\end{tabular}
\end{center}
\end{table}

%

\begin{figure}
\includegraphics[width=0.99\columnwidth]{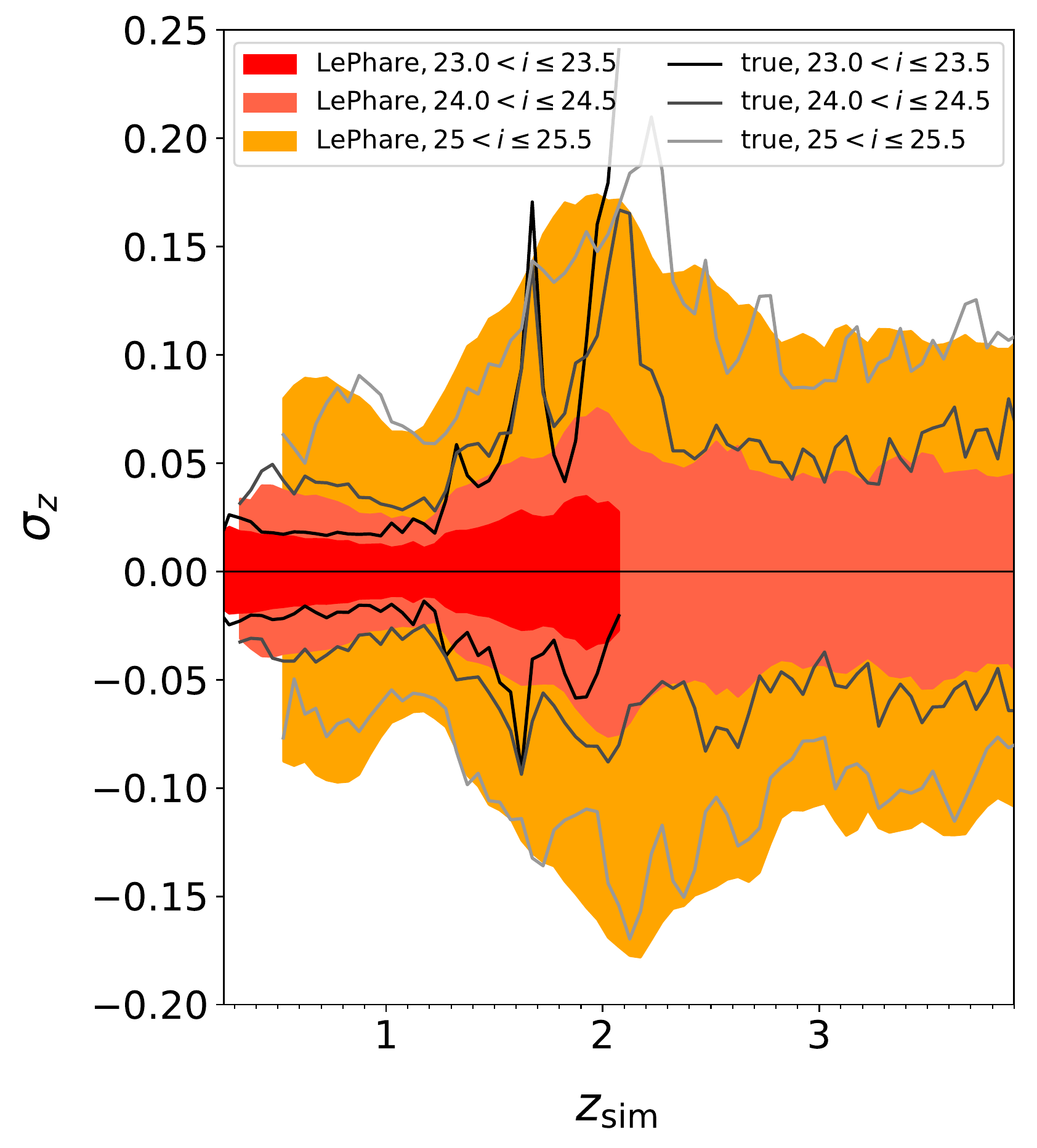} 
\caption{Photometric redshift 1$\sigma$ uncertainties in the {\sc Horizon-AGN} simulation, as a function of $z_\mathrm{sim}$ and divided in three bins of $i^+$-band magnitude. For each of these bins, the shaded area is the average $\sigma_\mathrm{z}(z)$ error interval as it results from the SED fitting likelihood analysis in {\sc LePhare} ($\sigma_\mathrm{z,fit}$ described in Sect.~\ref{subsec:sigmaz}).  Solid lines shows an alternate estimate of $\sigma_\mathrm{z}$, directly retrieved from the scatter between photometric and true redshifts ($\sigma_\mathrm{z,true}$). Both computations stop at redshifts where the statistics becomes too low ($<20$ galaxies). }
\label{fig:cfzphoterr}
\end{figure}

%
%

\begin{figure}
\centering
\includegraphics[width=0.99\columnwidth]{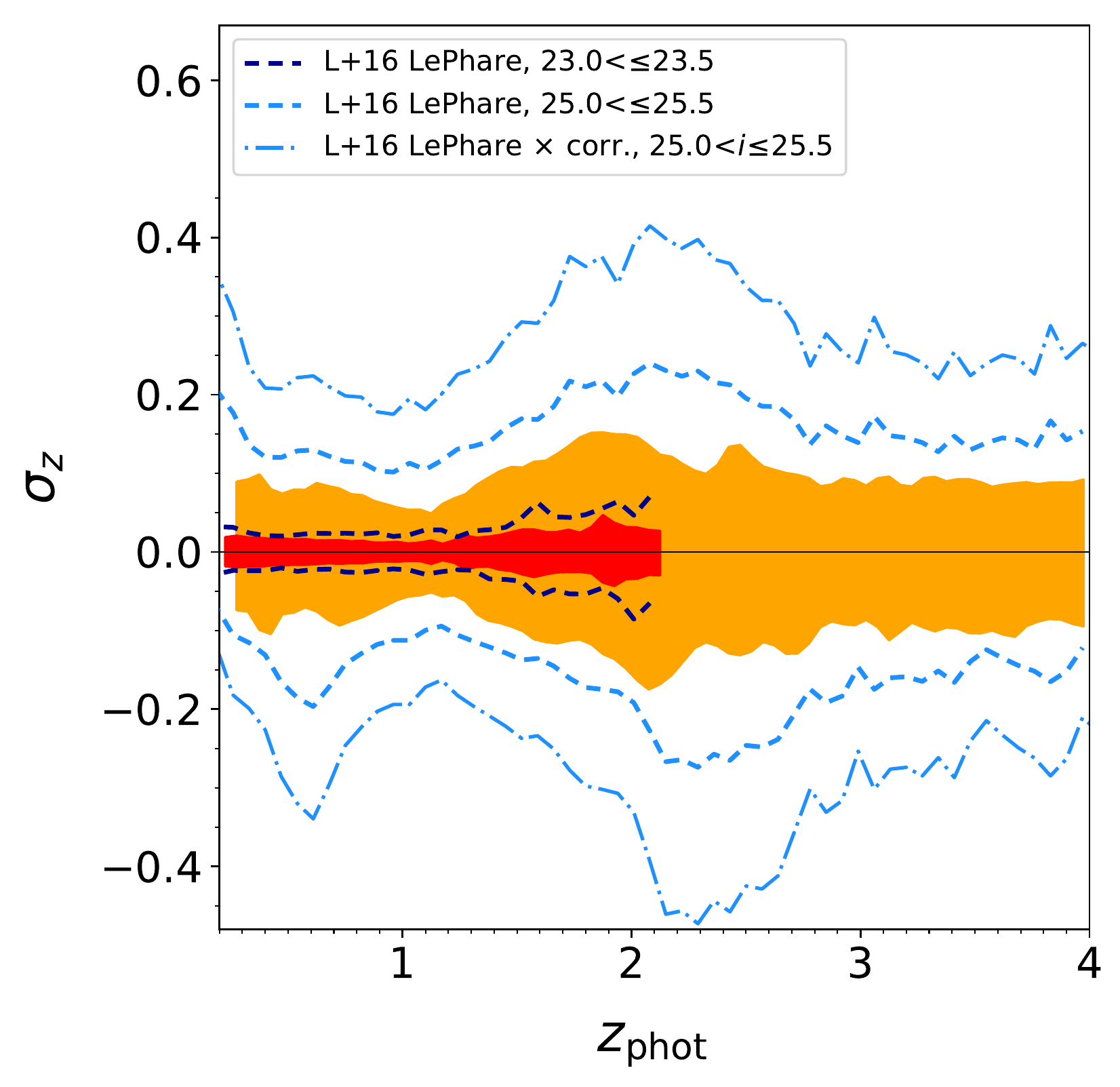} 
\caption{ Median photometric redshift 1$\sigma$ uncertainty of \textsc{Horizon-AGN} galaxies at $23<i^+\leq23.5$ and  $25<i^+\leq25.5$, as a function of $z_\mathrm{phot}$ (\textit{shaded} areas, colours as in Fig.~\ref{fig:cfzphoterr}). \textit{Dashed} lines  show the median $\sigma_\mathrm{z,L16}$ of real COSMOS2015 galaxies in the same magnitude bins (\textit{dark blue}: $23<i^+\leq23.5$, \textit{light blue}: $25<i^+\leq25.5$).
Both \textsc{Horizon-AGN} and {COSMOS2015}  errors are computed by {\sc LePhare} using the $z_\mathrm{phot}$ likelihood function. The \textit{dot-dashed} lines embrace the median of COSMOS2015 \textit{enhanced} errors at $25<i^+\leq25.5$, i.e.~the original $\sigma_\mathrm{z,L16}$ values have been increased by a multiplicative ``boosting factor'' as prescribed in L16. See Section~\ref{subsec:sigmaz} for more details. Note that the $y$-axis range is different from Fig.~\ref{fig:cfzphoterr}.} 
\label{fig:cfzphoterr_cosmos}
%
%
\end{figure}

\subsubsection{Photometric redshift errors from  marginalized likelihood}
\label{subsec:sigmaz}

When working on extragalactic surveys, an accurate  knowledge of the redshift probability distribution function (PDF) is instrumental to remove observational uncertainties in galaxy statistics. For instance, the PDFs of a certain set of  galaxies can be used to correct their luminosity function for the so-called Eddington bias \citep[see][]{schmidt14b}. 
Therefore, it is crucial to verify whether the SED fitting produces a reliable PDF($z$) for a given galaxy. 
In a Bayesian framework this PDF is the \textit{posterior} probability distribution, proportional to the product of the \textit{prior distribution} and the marginalized \textit{likelihood} $\La(z)$. 

From $\La(z)$\footnote{
$\La\propto \exp(-\sfrac{1}{2} {\chi^2})$, \\
with ${\chi^2(z)}=
{\sum}_{\rm filters\,i} 
{(F_{{\rm obs}\,i}-F_{{\rm SED}\, i}(z,T))^2}\big/{\sigma^{2}_{i, \rm obs}}$, where $F_{\rm SED\, i}(z, T)$ is the flux predicted for a template $T$ in the filter $i$ at $z$, $F_{{\rm obs}\,i}$ is the observed flux in the filter $i$ and $\sigma_{\rm obs}$ is the associated uncertainty.}, {\sc LePhare} computes $\sigma_\mathrm{z, fit}$, namely the photometric redshift 1$\sigma$ error. This is  defined as the redshift interval, centred at $z_\mathrm{phot}$, that encloses $68.27$ per cent of the $\La(z)$ area. 

\paragraph*{Reliability of redshift 1$\sigma$ errors from SED-fitting} 
The simulation  provides the true $z_\mathrm{phot}$ uncertainties directly from the difference between SED-fitting estimates and $z_\mathrm{sim}$. Hence it  can be checked whether the error bars provided by {\sc LePhare} actually represent the redshift 1$\sigma$ uncertainty. Previous tests with spectroscopic data suggest that they could be under-estimated \citep[see e.g.][L16]{Dahlen2013}. 

Fig.~\ref{fig:cfzphoterr} shows the evolution of the median of $\sigma_\mathrm{z, fit}$ as a function of $z_\mathrm{sim}$, for galaxies in three different bins of $i^+$-band magnitude. 
From these bins galaxies with degenerate redshift solutions are excluded, i.e.\ whose $\La(z)$ function shows two peaks.\footnote{
{\sc LePhare} automatically identifies a  galaxy with two acceptable redshift solutions when the secondary peak includes $>2$ per cent of the integrated $\La(z)$ distribution.} 
In such a case the integrated 68.27 per cent of the $\La(z)$  area is strongly skewed towards the secondary solution and the  $\sigma_\mathrm{z, fit}$ does not represent the pure statistical error but also includes systematics. Nevertheless,  if those galaxies are re-introduced in the sample, results shown in Fig.~\ref{fig:cfzphoterr} remain the same at $i^+<25$ and change less than 20 per cent in the faintest bin. 

We first find that $\sigma_\mathrm{z, fit}$ values increases with magnitude, as shown in Fig.~\ref{fig:cfzphoterr}. Such behaviour is expected since fainter galaxies have a lower $S/N$ ratio and a lower constraint on the SED fit. 
We also find a redshift dependency of the uncertainties, mainly due to the different efficiency of optical and NIR photometry: the former being deeper, including medium-bands and can tightly constrain the Balmer break at $z<1.3$ ; whereas at $z>1.5$ the break is entirely shifted in the NIR regime, which is sampled with fewer, less sensitive bands. 
At $z>2.5$ the $\sigma_\mathrm{z, fit}$ amplitude slightly decreases as optical blue bands start to constrain the Lyman break position.  
Overall, {\sc LePhare} predicts a symmetric scatter, with upper and lower errors such that $\sigma_\mathrm{z, fit}^+\simeq -\sigma_\mathrm{z, fit}^-$.

To establish whether the uncertainties derived by {\sc LePhare} are reliable, they are compared  to the   1$\sigma$ errors  directly retrieved from the simulation ($\sigma_\mathrm{z,true}$). 
Using the same  $i^+$-selected galaxies for which   $\sigma_\mathrm{z,fit}$ was computed, we measure $\sigma_\mathrm{z,true}$ by means of their $\Delta z=z_\mathrm{sim}-z_\mathrm{phot}$ 
distribution, finding the interval that includes 68.27 per cent of it. 
Despite some noise, relatively good agreement is found between $\sigma_\mathrm{z, fit}$ and $\sigma_\mathrm{z,true}$, an indication that the uncertainty  provided by {\sc LePhare} is generally a good proxy of the actual  1$\sigma$ error dispersion (Fig.~\ref{fig:cfzphoterr}). 
{However at bright magnitudes ($i^{+}<24.5$) and for $1<z_{\rm sim}<2.5$, $\sigma_\mathrm{z, fit}$ is generally underestimated compared to $\sigma_\mathrm{z,true}$. As shown in Appendix~\ref{App:sigz}, this underestimation might be due either to the underestimation of photometric errors, or to a lack of representativeness of the set of templates for this galaxy population, making $\La(z)$ too spiky around the median $z$. A similar trend was already discussed in L16. From a comparison with spectroscopic redshifts (their figure 13), L16 suggested indeed that the 1$\sigma$ uncertainties produced by {\sc LePhare} were under-estimated. As a consequence the authors proposed a magnitude-dependent boosting factor ($f_\sigma$)  that would enlarge  $\sigma_\mathrm{z, L16}$ so that $\sim\!68$ percent of the COSMOS2015 $z_\mathrm{spec}$ would fall within $\sigma_\mathrm{z,L16}\times f_\sigma$ from $z_\mathrm{phot}$. This boosting factor\footnote{
For galaxies with $i^+>20$,  $f_\sigma = 0.1\times i^+-0.8$;  $f_\sigma =1.2$  otherwise. } was however constant with redshift and increasing with magnitudes. From our analysis, it appears that this factor would generally over-correct the $z_\mathrm{phot}$ errors at faint magnitudes, but particular care should be given to the errors of bright galaxies at $1<z<2.5$. }

Finally, note that the global behaviour of $\sigma_\mathrm{z, fit}$ as a function of magnitude and redshift depends on the photometric baseline available, and does not necessarily hold for different configurations (see e.g. Section~\ref{sec:forecasts}). %

\paragraph*{z$_{\rm phot}$ error comparison between COSMOS-like and COSMOS2015}
Let us now compare  the \textsc{Horizon-AGN} redshift uncertainties with those   of the COSMOS2015 galaxies ($\sigma_\mathrm{z,L16}$) as calculated in L16 by means of {\sc LePhare}. The method is the same as that applied to simulated galaxies, based on the marginalized $\La(z)$.   Fig.~\ref{fig:cfzphoterr_cosmos} shows the median $\sigma_\mathrm{z,L16}$ as a function of redshift, for galaxies with $23.0<i^+<23.5$ and $25.0<i^+<25.5$. In the Figure $\sigma_\mathrm{z,fit}$ is also reported, in bins of $z_\mathrm{phot}$ instead of $z_\mathrm{sim}$ to allow the comparison with real data.  

The trend of $\sigma_\mathrm{z,L16}$ and $\sigma_\mathrm{z,fit}$ are remarkably similar, both showing an increase between $z=1.5$ and 2, as  discussed above. However, COSMOS2015 galaxies 
have a median $\sigma_\mathrm{z}$  that is about 50 
per cent larger than \textsc{Horizon-AGN}.  
{This difference is likely to be driven either by simplifications in the modelling of the photometry itself, or by failures in the photometry extraction of real data which are not modelled in the simulated catalog. On the one hand, the simulated photometry includes indeed less variety than realistic galaxies: we use a single and constant IMF, a constant dust-to-metal ratio, single SSP model for stellar mass losses, and emission lines are not modelled. These simplifications can naturally reduce the scatter of the $z_{\rm phot}$ estimate compared to the observed catalogue. On the other hand, although magnitude errors in the simulated catalogue are implemented consistently with COSMOS2015, ``catastrophic detections" (such that blended or fragmented objects) and systematics (astrometry calibration issues, background removal, lack of modelling of the PSF variation within the field, mis-centering of the galaxies ...etc.) are not considered in the virtual photometry. These errors will propagate in the $z_\mathrm{phot}$ errors.
Exploring these effects is out of the scope of this paper and will be the topic of a future work.} 

\begin{figure*}
\includegraphics[width=0.99\textwidth]{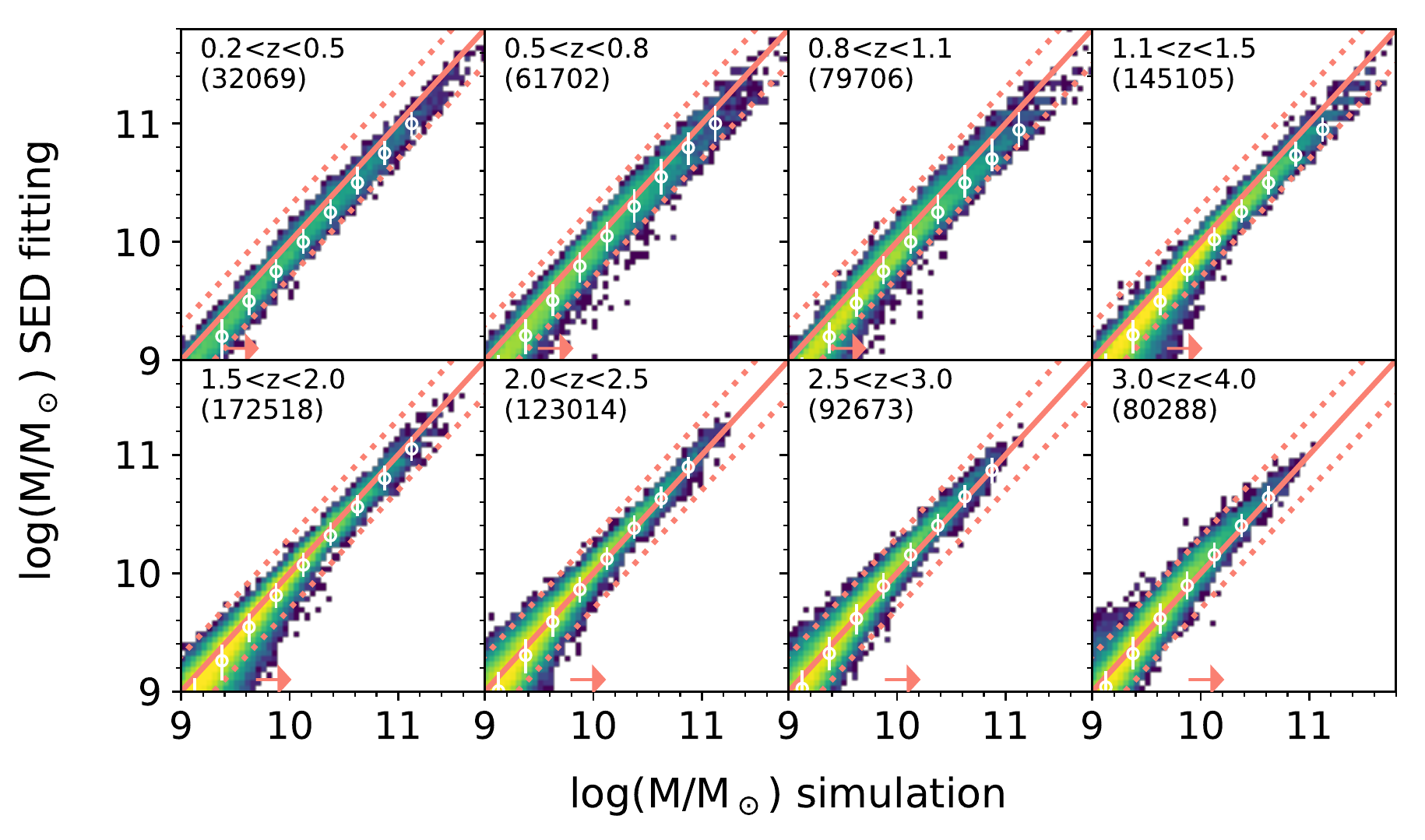}
\includegraphics[width=0.99\textwidth]{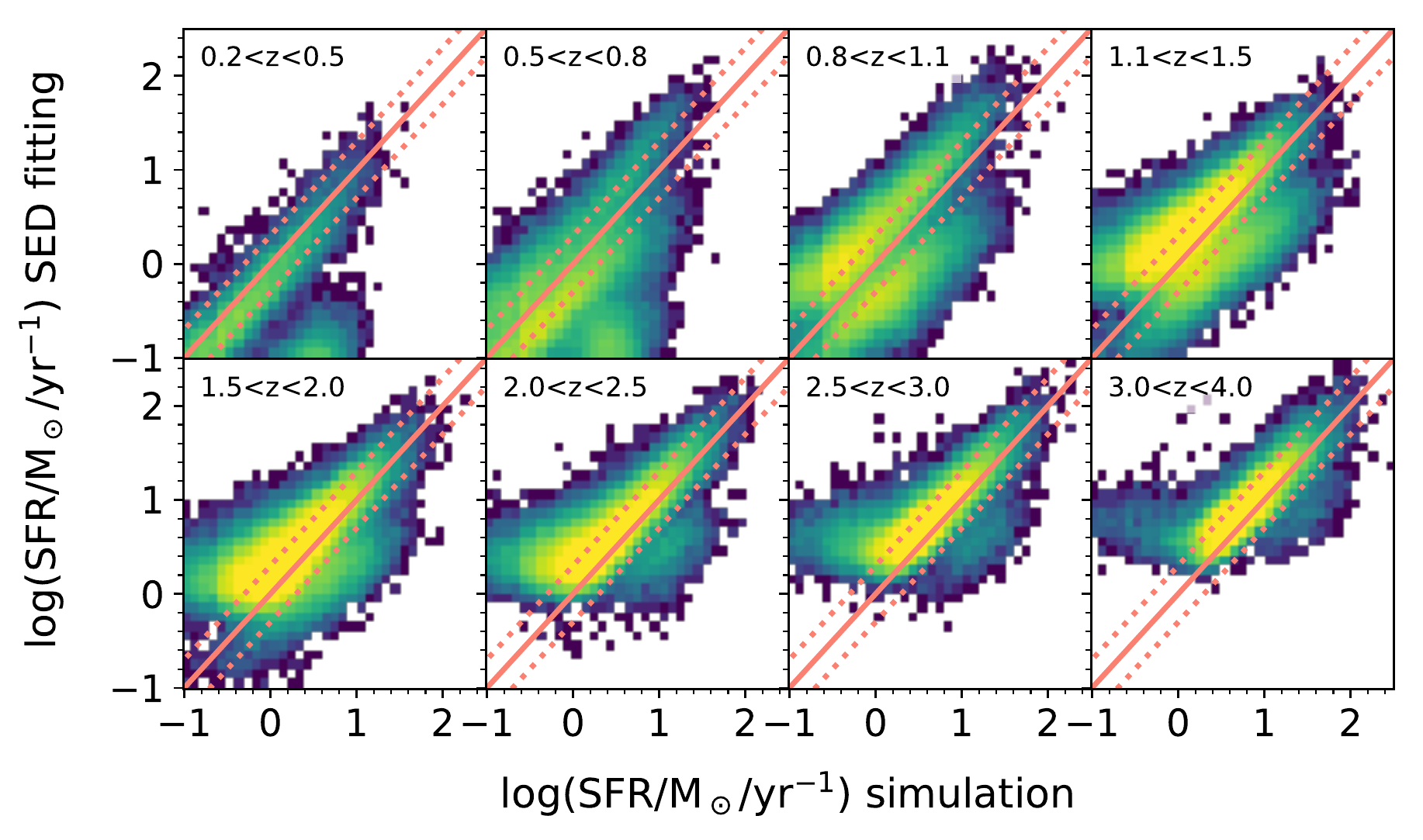}
\caption{\textit{Top:} Comparison between stellar mass  estimates obtained through SED-fitting and masses directly derived from the sum of stellar particles,  for {\sc Horizon-AGN} galaxies with $K_\mathrm{s}<24.7$ and $0.2<z_{\rm phot}<4$. Dust and IGM attenuation are included in the photometry. {The redshift is taken as being $z_{\rm phot}$ in the mass and SFR computation}. Redshift bins and the number of objects is indicated in the upper-left corner of each panel. \textit{Solid} line is the 1:1 relation and \textit{dotted} lines show $\pm0.3$\,dex offset from it. \textit{White} circles are the median of the SED-fitting estimates in running bins and the error on it. The density map has the same color scale as in Fig.~\ref{fig:zphotzspec}. \textit{Bottom:} Comparison between SFR estimates obtained through SED-fitting and the SFR directly derived by adding the mass of the stellar particles formed over the last 100 Myr,  for the same {\sc Horizon-AGN} galaxies. 
The $z_\mathrm{phot}$ range of each panel is indicated in the upper-left corner (number of galaxies in each redshift bin is the same for both mass and SFR comparison).}
\label{fig:cf_mass}
 \end{figure*}

\begin{figure*}
\includegraphics[scale=0.4]{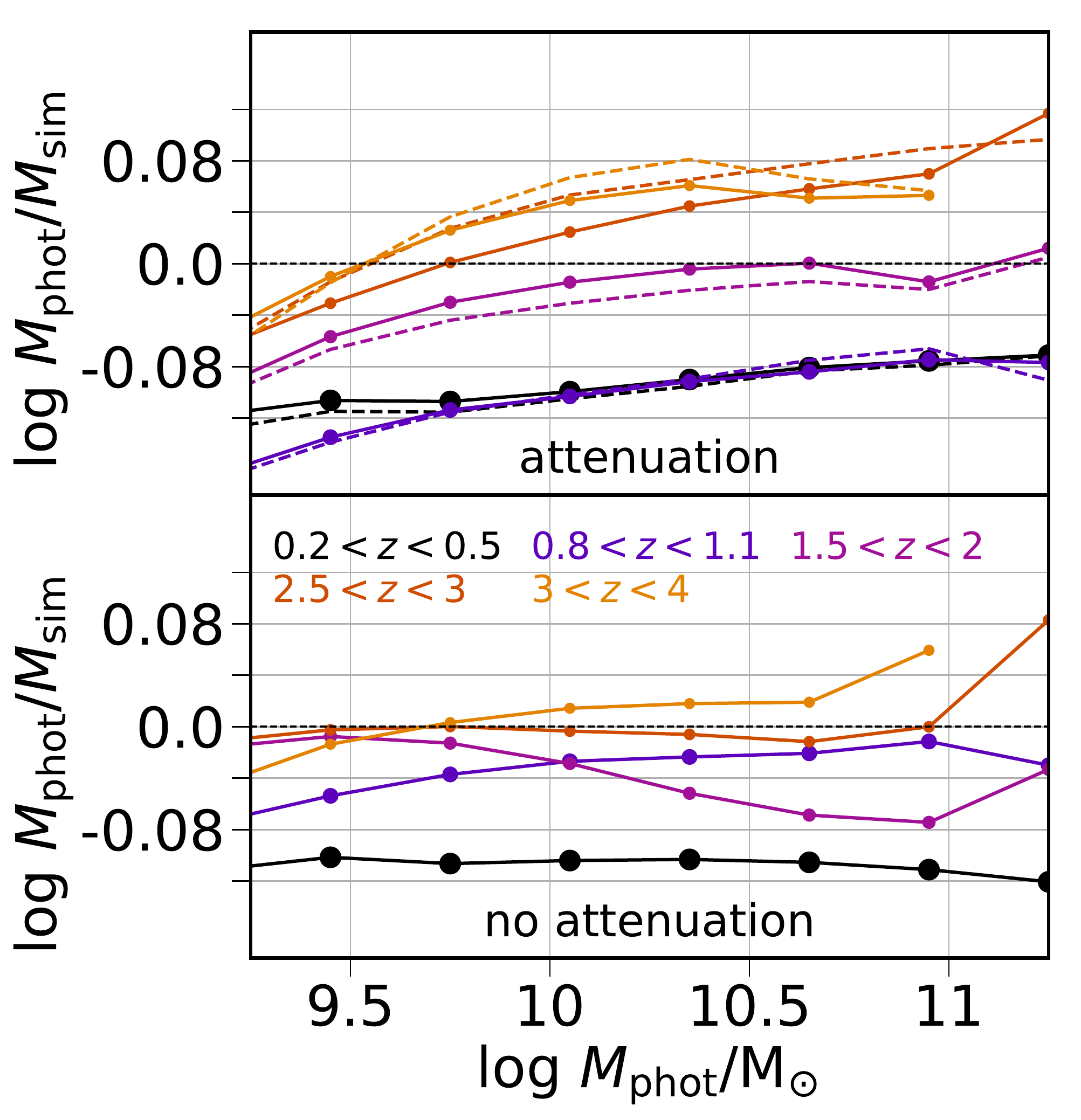}
\includegraphics[scale=0.4]{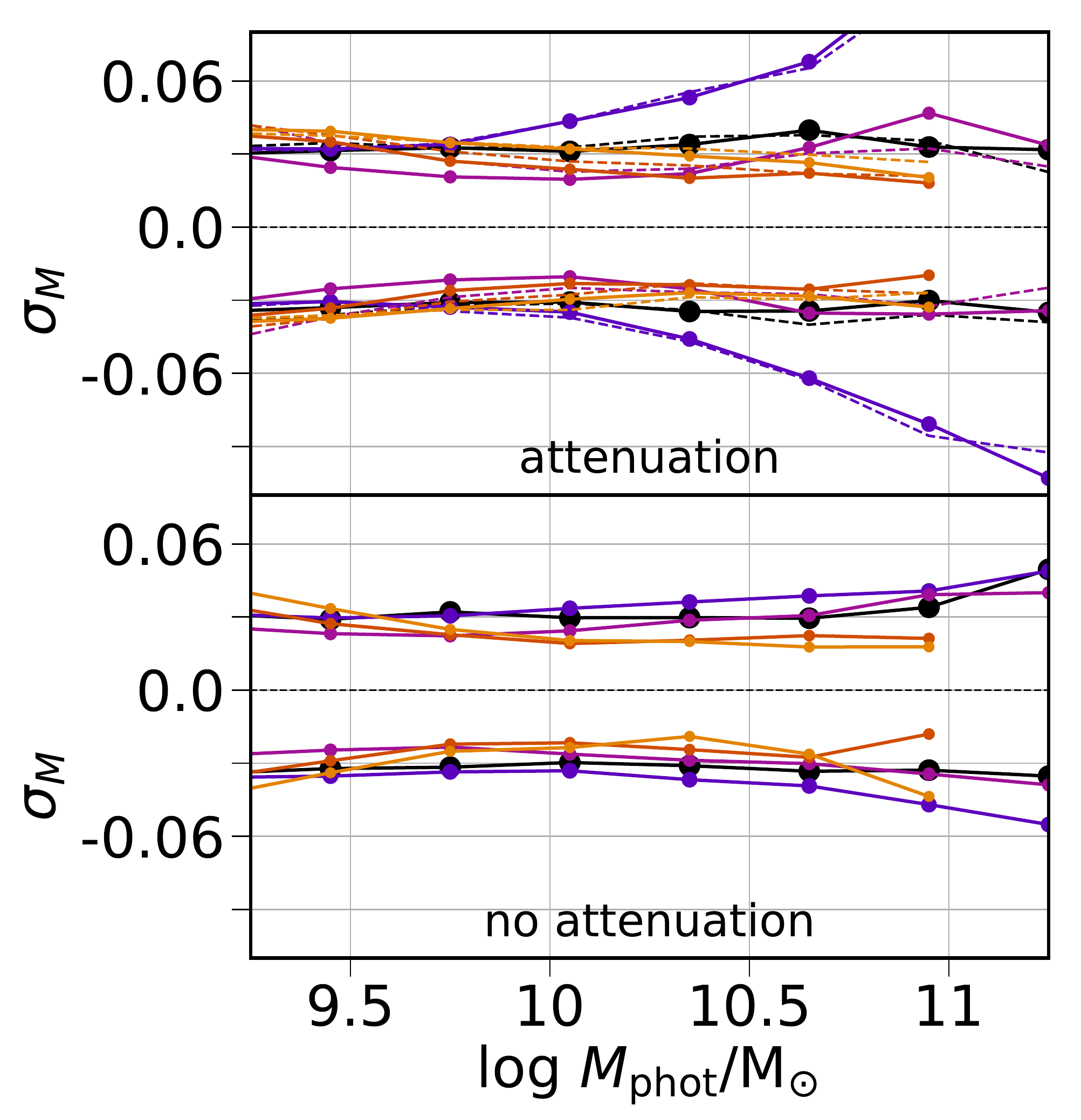}
\caption{\textit{Left}: {Median offset $\log M_{\rm phot}/M_{\rm sim}$}  as a function of $M_\mathrm{phot}$. \textit{Right}: dispersion around the median values. {The \textit{upper} and \textit{lower} lines correspond respectively to $\sigma_{M}^{+}$ and $\sigma_{M}^{-}$ and enclose 34 \% of the population above and below the median. Increasing marker sizes correspond to decreasing redshifts.} 
\textit{Top} panels are the results for \textsc{Horizon-AGN} galaxies including attenuation by dust and the IGM in their virtual photometry {while fixing the redshift at $ z_{\rm phot}$ (\textit{dashed} line) or $ z_{\rm sim}$  (\textit{solid} line) in the computation of $M_{\rm phot}$}. \textit{Bottom} panels correspond to the computation of $M_{\rm phot}$ from attenuation-free version of the same sample {with the redshift fixed to $z_{\rm sim}$}. For the sake of clarity only the results for half of the redshift bins are displayed.} 
\label{fig:cf_mass_med}
\end{figure*}

\begin{figure*}
\includegraphics[scale=0.4]{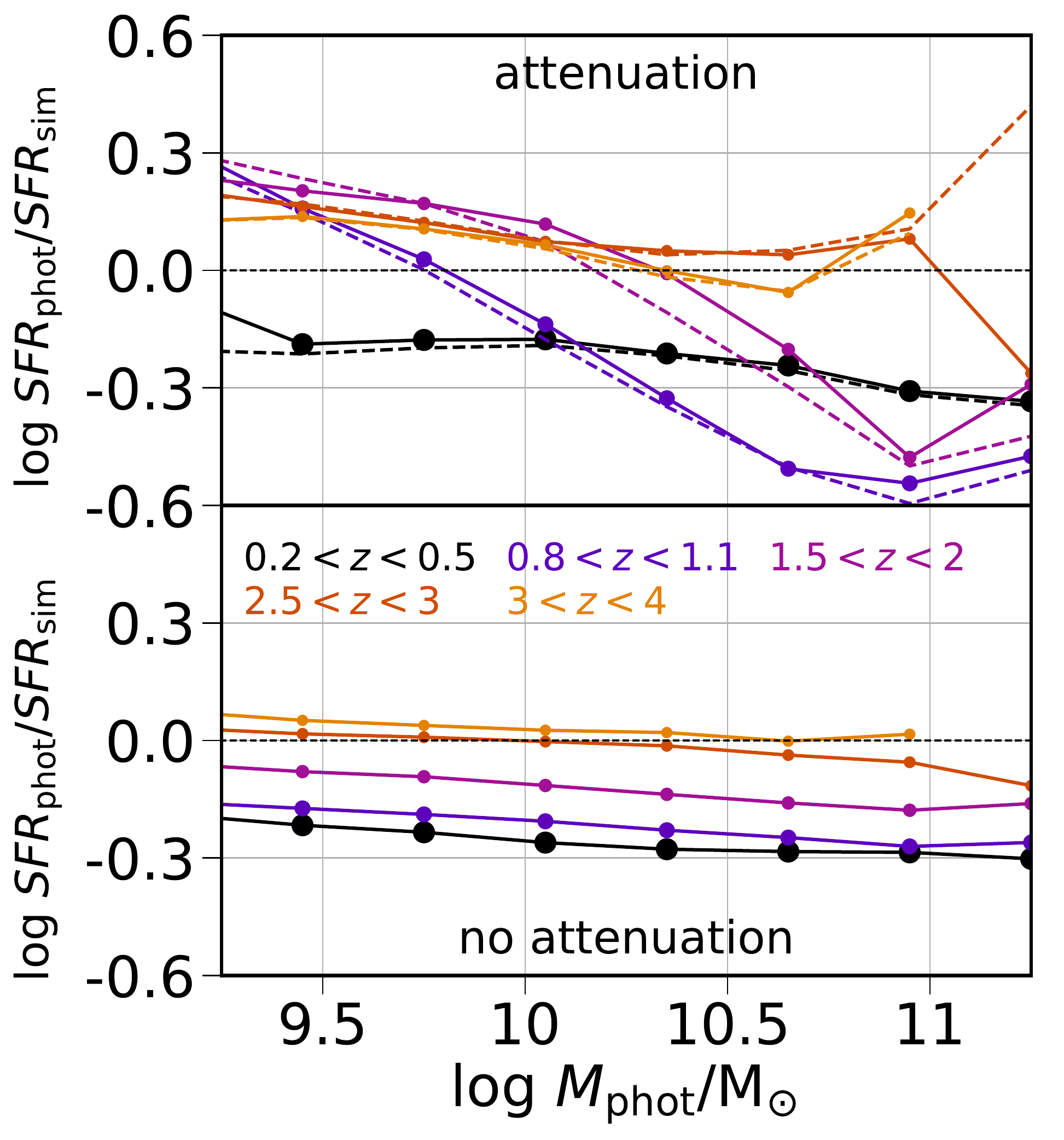}
\includegraphics[scale=0.4]{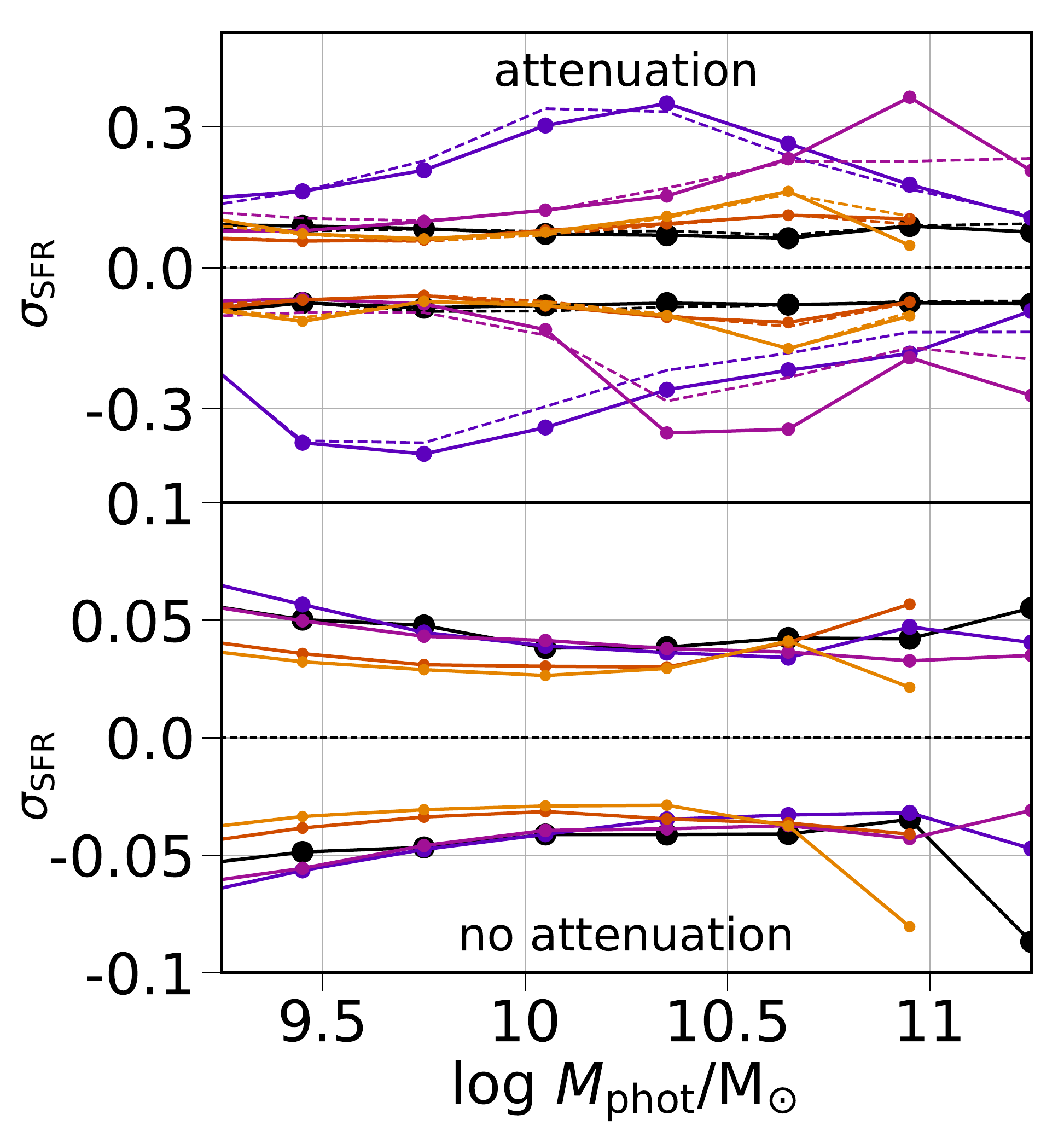}
\caption{\textit{Left}:  {Median offset $\log {\rm SFR}_{\rm phot}/{\rm SFR}_{\rm sim}$}, as a function of $M_\mathrm{phot}$. \textit{Right}: dispersion around the median values. {The \textit{upper} and \textit{lower} lines correspond respectively to $\sigma_{\rm SFR}^{+}$ and $\sigma_{\rm SFR}^{-}$ and enclose 34 \% of the population above and below the median. Increasing marker sizes correspond to decreasing redshifts}. 
\textit{Top} panels are the results for \textsc{Horizon-AGN} galaxies including attenuation by dust and IGM in their virtual photometry {while fixing the redshift at $z_{\rm phot}$ (\textit{dashed} line) or $z_{\rm sim}$  (\textit{solid} line) in the computation of ${\rm SFR}_{\rm phot}$}. \textit{Bottom} panels correspond to the attenuation-free version of the same sample {with the redshift fixed to $z_{\rm sim}$}. Note that on the \textit{left} panel, the $y$-axis range is not the same on the \textit{top} and on the \textit{bottom} panels.
}
\label{fig:cf_sfr_med}
\end{figure*}
\subsection{Physical quantities: mass and star-formation rate}
\label{sec:physprop}


\subsubsection{Stellar mass estimate}

The overall comparison between the intrisic stellar masses ($M_{\rm sim}$) and those retrived via SED-fitting ($M_{\rm phot}$) is shown in Fig.~\ref{fig:cf_mass} in photometric redshift bins up to $z\sim 4$. The observed stellar masses are  in very good agreement with the intrinsic ones. The left and right panels of Fig.~\ref{fig:cf_mass_med}  present respectively the median  and the dispersion around the median $\sigma_{M}$ of $\log M_{\rm phot}/M_{\rm sim}$ as a function of $\log M_{\rm phot}$ in different redshift bins. The dispersion around the median value being potentially asymetric, we measure $\sigma^{+}_{M}$ and $\sigma^{-}_{M}$ as the value which encloses 34\% of the full population respectively above and below the median. {The values $\sigma^{+}_{M}$ and $\sigma^{-}_{M}$ are displayed as the upper and lower lines on the \textit{right} panels of Fig.~\ref{fig:cf_mass_med}}. 

\paragraph*{Impact of $z_{\rm phot}$ uncertainties} In order to determine how much of the trend is driven by the propagation of uncertainties from the photometric redshift estimation, stellar mass computation through SED-fitting is reproduced in a second step while fixing the redshift at $z_{\rm sim}$ instead of $z_{\rm phot}$. 
{The \textit{dashed} lines in the \textit{top} panels of Fig.~\ref{fig:cf_mass_med} correspond to the median and dispersion of $\log M_{\rm phot}/M_{\rm sim}$ using the photometric redshifts in the computation of $M_{\rm phot}$, while the \textit{solid} lines correspond to the same quantities but using the intrinsic redshifts from the simulation in the computation of $M_{\rm phot}$. Comparing the \textit{solid} and \textit{dashed} lines therefore allows to quantify the impact of the photometric redshift uncertainty propagation in the stellar mass computation, which is very limited. Finally, it should be noted that the \textit{dashed} lines
provide a direct comparison with observations, as the galaxy population is split in bins of $z_{\rm phot}$, and the computation includes both dust and redshift uncertainties. As a complement, the \textit{top} panel of Fig.~\ref{Fig:masszsim} can also be compared with Fig.~\ref{fig:cf_mass}.} 
Overall, the propagation of $z_{\rm phot}$ uncertainties has only a small impact on retrieving stellar mass. 
The scatter is relatively stable over the redshift and mass ranges and is generally smaller than 0.1 dex. $M_{\rm phot}$ is preferentially underestimated up to $z\sim 2$ by at most $\sim$0.12 dex. At $z>2$ and $\log M_{\rm phot}>9.5$ the trend tends to reverse and $M_{\rm phot}$ ends up slightly overestimated. 

\paragraph*{Impact of dust attenuation} 
{In order to isolate the role played by attenuation in driving this behaviour, the same computation is performed on the attenuation-free catalogue, while using the intrinsic redshift from the simulation. The median and dispersion of $\log M_{\rm phot}/M_{\rm sim}$ in the dust-free case are shown in the bottom panels of Fig.~\ref{fig:cf_mass_med}.}
Without attenuation, one is left with a weak systematic underestimation of the stellar mass especially at low redshift, which can be driven either by the too simplistic (single-burst) SFHs \citep[see e.g.][]{leja18} or the discretisation of the metallicity in galaxy template \citep{mitchell13}.  At higher redshift, these assumptions  are more likely to correctly represent the actual SFHs and metallicity distributions. It should be noted that the impact of attenuation (overestimation) and of these simplified assumptions (underestimation) tends to compensate each other. For example, at $1.5<z<2$, $M_{\rm phot}$ is closer to $M_{\rm sim}$ when dust is included.

\subsubsection{Star formation rate estimate}

 It is known that SFR derived from SED-fitting has to be considered with caution, given the simplistic shape of the SFHs assumed in the templates, which for instance cannot account for recent bursts of star formation. \citet{ilbert15} predicted an overall offset of 0.25 dex and a scatter up to 0.35 dex, from a comparison of SFR derived on the one hand from SED-fittig and on the other hand from IR+UV flux. 
 
The \textit{bottom} panel of Fig.~\ref{fig:cf_mass} illustrates this point, as it presents the overall comparison  between the intrinsic star-formation rate (SFR$_{\rm sim}$) and that derived from SED-fitting (SFR$_{\rm phot}$) in photometric redshift bins up to $z\sim 4$.  Up to $z\sim 3$, SFR$_{\rm phot}$ presents a bimodal behavior, with a systematic underestimation for a large fraction of the galaxy population up to $z\sim 1.5$ and an overestimation of low-mass galaxies above. Moving towards high-$z$, the bi-modality tends to disappear but the scatter remains very large. The left and right panels of Fig.~\ref{fig:cf_sfr_med} present the median of $\log {\rm SFR}_{\rm phot}/{\rm SFR}_{\rm sim}$ and the dispersion around the median $\sigma_{\rm SFR}$. The dispersion around the median value being potentially asymetric, we measure $\sigma^{+}_{\rm SFR}$ and $\sigma^{-}_{\rm SFR}$ as the value which encloses 34\% of the population respectively above and below the median. The values $\sigma^{+}_{\rm SFR}$ and $\sigma^{-}_{\rm SFR}$ are displayed as the upper and lower lines on the \textit{right} panels of Fig.~\ref{fig:cf_sfr_med}. 
 The median evolves between -0.6 at $\log {M}_{*}/{\rm M_{\odot}}>10.5$ and $z<1.5$ and 0.3 at $\log {M}_{*}/{\rm M_{\odot}}<10.$, while the scatter varies from $\sim$ 0.6 at $0.8<z<1.1$ to $\sim$ 0.15 at high-$z$.
 
 \paragraph*{Impact of $z_{\rm phot}$ uncertainties}
{In order to determine the role of redshift uncertainties in driving the trend, the SFR is also computed while  fixing the galaxy redshift at their intrinsic values $z_{\rm sim}$ instead of $z_{\rm phot}$. 
The \textit{dashed} lines in the \textit{top} panels of Fig.~\ref{fig:cf_sfr_med} correspond to the median and dispersion of $\log {\rm SFR}_{\rm phot}/{\rm SFR}_{\rm sim}$ using the photometric redshifts in the computation of ${\rm SFR}_{\rm phot}$, while the \textit{solid} lines correspond to the same quantities but using the intrinsic redshifts from the simulation in the computation of ${\rm SFR}_{\rm phot}$. Comparing the \textit{solid} and \textit{dashed} lines therefore allows to quantify the impact of the photometric redshift uncertainty propagation in the SFR computation.} As a complement, the \textit{top} panel of Fig.~\ref{Fig:sfrzsim} can also be compared with the \textit{bottom} panel of Fig.~\ref{fig:cf_mass}. This comparison highlights that working with the simulated redshift removes  the bimodality in the lowest redshift range $0.2<z<0.5$, which therefore is driven by redshift degeneracies. However the bimodality remains in all the other redshift bins. 

 \paragraph*{Impact of dust attenuation}
One  expects the impact of dust on the precision of SFR to be much stronger than on the mass.  Indeed it  attenuates preferentially blue bands, which are a tracer of recently formed stars, hence directly connected to the SFR. The comparison of the SFR  in the run with and without attenuation confirms this fact. {The median and dispersion of $\log {\rm SFR}_{\rm phot}/{\rm SFR}_{\rm sim}$ in the dust-free case are shown in the bottom panels of Fig.~\ref{fig:cf_sfr_med}, and the overall comparison of ${\rm SFR}_{\rm phot}$ versus ${\rm SFR}_{\rm sim}$ without dust is shown in the \textit{bottom} panel of Fig.~\ref{Fig:sfrzsim}}. At $z>2$ and in a dust-free Universe, the SFR is very well recovered without any bimodal behaviour, while it tends to be slightly underestimated at $z<2$.  As for the mass, this remaining underestimation is likely to be driven by the oversimplified SFH and metallicity underlying models, which cannot fully render the complexity of low redshift galaxy SEDs. 
 On the contrary, when dust is accounted for in the virtual Universe, a bimodal behaviour appears, due to the SFR-dust degeneracy.    
 The cause of this trend is investigated in Appendix~\ref{Ap:sfr}. In particular, there is a direct correlation between the attenuation in the rest-frame $NUV$ and the SFR. Overestimating the attenuation $A_{NUV}$ at the SED-fitting stage yields an overestimation of the SFR. None of the two extinction curves used in {\sc LePhare} are a good fit for the one used in {\sc Horizon-AGN}, and this discrepancy is likely to be the main driver of this bimodality.

\subsection{Performance of current surveys: summary}
\label{Sec:CosSum}
The virtual photometric catalogue, calibrated to mimic COSMOS2015, has allowed a fully consistent test of  the performance of  {\sc LePhare}  when computing galaxy redshifts, masses and SFR from broad-band photometry. We summarise below our main findings:
\begin{itemize}[noitemsep,nolistsep]
    \item In the same configuration as COSMOS2015, photometric redshifts are retrieved with the same overall precision (as estimated from NMAD and 1$\sigma$ uncertainties) in the virtual dataset as in the observed  one.
    When binning the datasets in apparent magnitudes, the simulation yields as  precise estimates as the observations.  
    {In particular, the 1$\sigma$ uncertainties measured from $\La(z)$ represent on overall a good estimate of the intrinsic errors (as measured from the difference between $z_{\rm sim}$ and $z_{\rm zphot}$), except for the  bright galaxies at $1.<z<2.5$  which have in general their errors underestimated. However the averaged correcting factor for redshift errors proposed in earlier works \citep[see e.g.][]{ilbert13,laigle16} would generally over-correct the errors of faint galaxies. Redshift errors are generally smaller in {\sc Horizon-AGN} as in COSMOS, as the mock catalog does not include systematics in the extraction of photometry from noisy images, and presents less diversity in terms of photometry}.
  For the same reasons, although the simulated catalogue allows us to retrieve the overall redshift distribution of the catastrophic  population of outliers, it systematically underestimates their fraction. Intermediate bands allows to improve redshift accuracy; 
    \item Stellar masses are very well recovered, despite  the use of single-burst SFH model {and discrete metallicity} in the SED-fitting templates, which do not a priori represent the complex SFHs of  simulated galaxies. Only a small underestimation of at most $\sim 0.12$ dex persists at low redshift, and an overall scatter of the order of 0.1 dex. Conversely, dust  induces  a slight overestimation of the mass at high redshift. The impact of redshift uncertainties in driving the scatter is very limited;  
    \item Unsurprinsgly, the SFR directly derived from the SFH are a quite poor proxy of the intrinsic SFR. The simplistic SFH {and metallicity enrichment} induce an underestimation, while the dust modelling (mainly the choice of the attenuation curve) induces a bimodality. As a result, the dispersion around the median values evolve between 0.2 and 0.6 dex.
\end{itemize}

\begin{figure}
\centering\includegraphics[width=0.8\columnwidth]{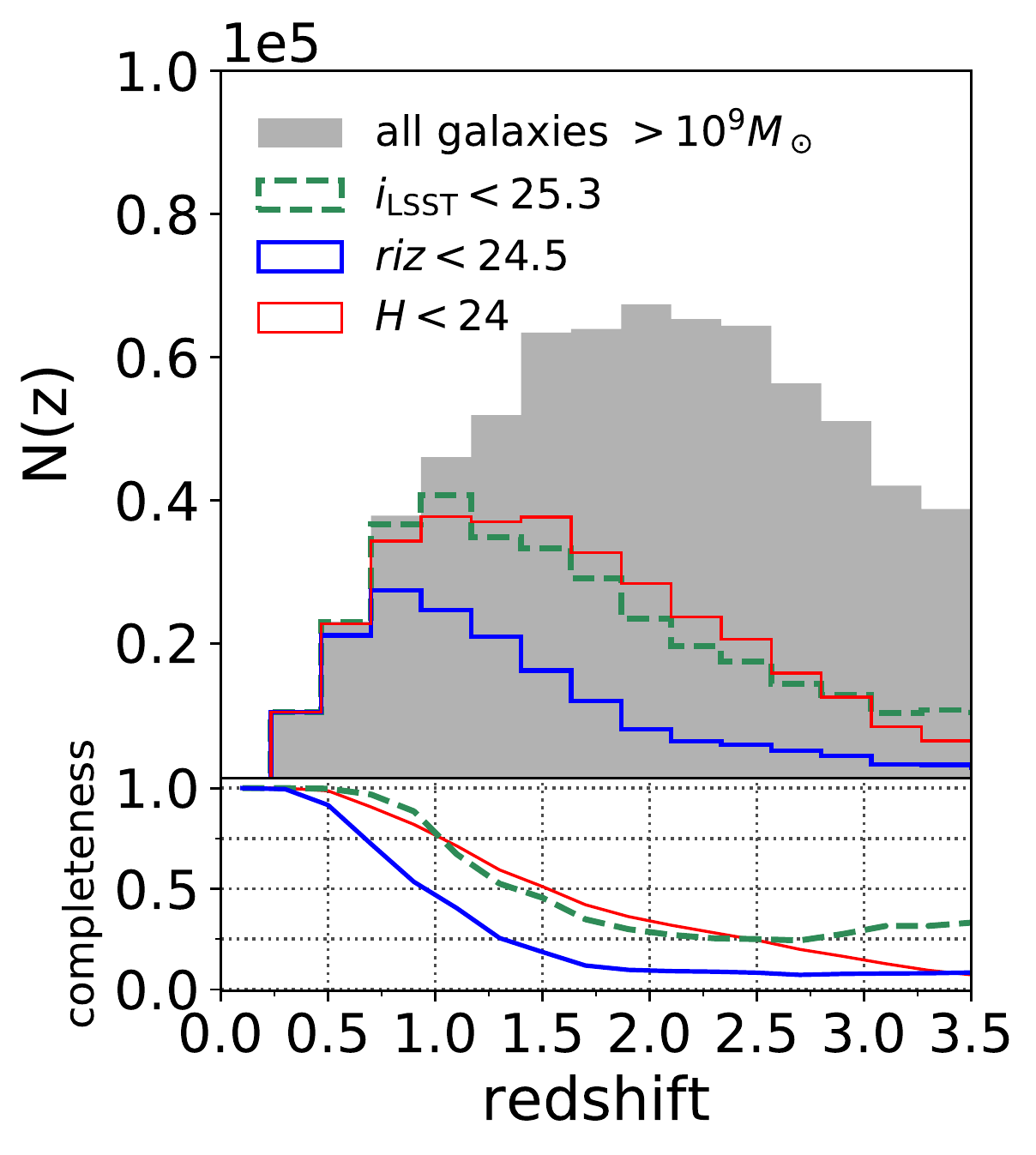}
\caption{Redshift  distribution $N(z_\mathrm{sim})$ of {\sc Horizon-AGN} simulation (gray histogram, $M>10^9\,M_\odot$) along with the $N(z)$ of different sub-samples  selected at $H<24$, $riz<24.5$, and $i<25.3$ (upper red, lower blue, green dashed histograms). For each sub-sample, the completeness fraction as a function of $z_\mathrm{sim}$ is shown in the bottom panel using same colours. }
\label{fig:euclid_nz}
\end{figure}

\begin{figure}
\includegraphics[width=0.99\columnwidth]{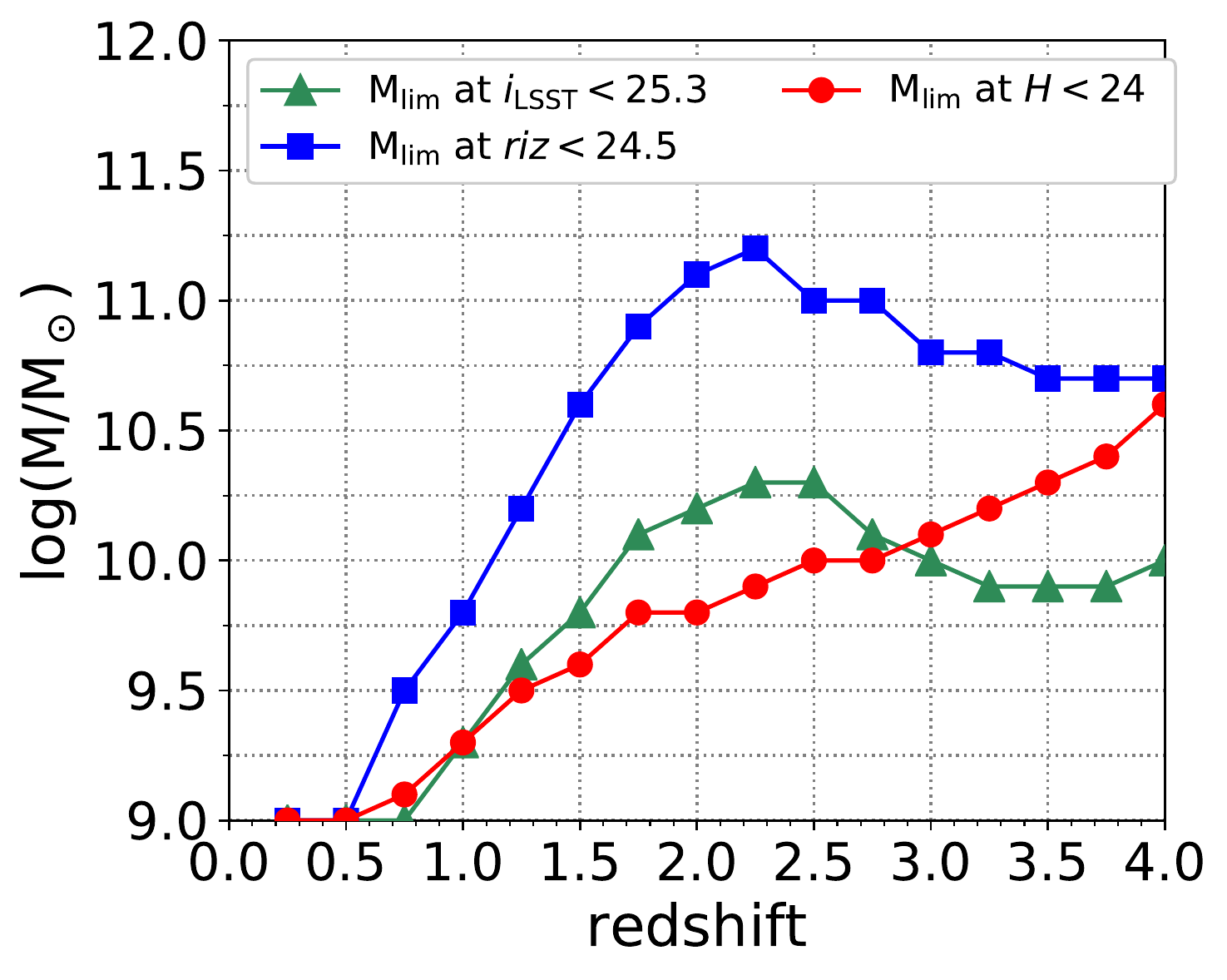}
\caption{ Redshift evolution of the stellar mass limits defined as the 90 per cent completeness threshold for  galaxy samples selected  in \emph{Euclid} $H$ (\textit{red} line and circles), $riz$ (\textit{blue} line and squares), and LSST $i$ filter (\textit{green} line and triangles).} 
\label{fig:Mcompl}
\end{figure}

\section{SED-Fitting performance: Forecasts}
\label{sec:forecasts}
In this Section, we use the mock catalogues reproducing \emph{Euclid}-like, LSST-like, and DES-like photometry in order to predict the performance of these surveys in the three configurations presented in Table~\ref{Tab:conf}.  

\subsection{Redshift and  mass completeness in  \emph{Euclid} and LSST}
Let us first  present the expected completeness of each survey,  estimated from  virtual catalogues using the intrinsic redshift,  intrinsic stellar masses and total  unperturbed magnitudes. 
The fraction of ``detected'' galaxies (i.e., those brighter than the  magnitude limit of the survey) is measured as a function of redshift and stellar mass. The magnitude cuts  correspond to those used for weak lensing galaxy selection in \emph{Euclid} ($riz<24.5$) and LSST ($i<25.3$); the  completeness at $H<24$ is also computed,  namely the 5$\sigma$ detection limit expected at the completion of the \emph{Euclid} mission.  It is argued in the following that the latter  threshold is the most  suited for galaxy evolution science cases.  

Fig.~\ref{fig:euclid_nz} compares  the intrinsic redshift distribution of all the objects in the lightcone with $M_\mathrm{sim}>10^{9} {\rm M}_{\odot}$  
to the sub-sample of galaxies detected in the \emph{Euclid}-like catalogue, in the case of either $H$-band  
or optical 
selection.  The $i$-band sample expected to be detected in LSST is also included, 
showing a redshift distribution similar to the \emph{Euclid} $H$-band selected sample. The cut applied in the $riz$ band results in a lower completeness, below 50 per cent already at $z_\mathrm{sim}=1$. Both the $H<24$ and $i<25.3$ completeness drop below 50 per cent at $z_\mathrm{sim}=1.5$. The fact that Fig.~\ref{fig:euclid_nz} does not include galaxies below  $10^9 M_\odot$  has a negligible impact  at $z>0.5$ because such a low-mass population is generally fainter than the magnitude limits considered here (see L16). Even at $z<0.5$, where all  selections are complete, one  does not expect the addition of $M_\mathrm{sim}<10^9 M_\odot$ galaxies to   significantly impact our results.

 \begin{table*}
 \caption{Mass completeness limits (90 and 50 per cent threshold) in the \emph{Euclid} ($H<24$ or $riz<24.5$) and LSST ($i<25.3$) configurations as a function of redshift, as estimated from intrinsic quantities. 
 }
 \label{Tab:compl}

 \begin{center}
 \def\arraystretch{1.2}
 \begin{tabular}{|l c cccc cccc|}\hline\hline
   $z_\mathrm{sim}$ range & $z_\mathrm{sim}$ median & \multicolumn{3}{c}{90\% mass  completeness [$\log M_\odot$]} & \multicolumn{3}{c}{50\% mass completeness [$\log M_\odot$]}\\
   & &  $ H<24$ &  $\mathit{riz}<24.5$ &  $i<25.3$  & $ H<24$ &  $\mathit{riz}<24.5$ &  $i<25.3$ & \\\hline
(0.00,0.25] & 0.22 & 9.00 & 9.00 & 9.00 & 9.00 & 9.00 & 9.00  \\
(0.25,0.50] & 0.41 & 9.00 & 9.02 & 9.00 & 9.00 & 9.00 & 9.00  \\
(0.50,0.75] & 0.65 & 9.08 & 9.43 & 9.00 & 9.00 & 9.05 & 9.00  \\
(0.75,1.00] & 0.89 & 9.31 & 9.81 & 9.27 & 9.05 & 9.42 & 9.00  \\
(1.00,1.25] & 1.14 & 9.48 & 10.12 & 9.67 & 9.26 & 9.70 & 9.30  \\
(1.25,1.50] & 1.38 & 9.64 & 10.52 & 9.77 & 9.38 & 9.92 & 9.44  \\
(1.50,1.75] & 1.63 & 9.76 & 10.65 & 10.01 & 9.51 & 10.22 & 9.60  \\
(1.75,2.00] & 1.87 & 9.83 & 10.89 & 10.13 & 9.60 & 10.31 & 9.65  \\
(2.00,2.25] & 2.11 & 9.88 & 10.78 & 10.23 & 9.66 & 10.32 & 9.70 \\
(2.25,2.50] & 2.37 & 9.94 & 10.76 & 10.14 & 9.70 & 10.25 & 9.72  \\
(2.50,2.75] & 2.62 & 10.01 & 10.72 & 10.06 & 9.75 & 10.26 & 9.66  \\
(2.75,3.00] & 2.87 & 10.04 & 10.68 & 9.98 & 9.82 & 10.22 & 9.58  \\
(3.00,3.25] & 3.12 & 10.17 & 10.53 & 9.93 & 9.87 & 10.22 & 9.54  \\
(3.25,3.50] & 3.36 & 10.21 & 10.34 & 9.81 & 9.94 & 10.15 & 9.48  \\ 
(3.50,3.75] & 3.61 & 10.27 & 10.33 & 9.82 & 9.96 & 9.98 & 9.37  \\ 
(3.75,4.00] & 3.87 & 10.28 & 10.26 & 9.71 & 9.99 & 10.07 & 9.32  \\ \hline
  
 \end{tabular}
 \end{center}
 \end{table*}

The stellar mass completeness ($M_{\rm lim}$) is shown in  Fig.~\ref{fig:Mcompl}. This is a  lower limit, as a function of redshift, above which $>$90 per cent of galaxies are detected in the selection band ($i$, $riz$, or $H$). The results are summarized in Table~\ref{Tab:compl}, along with a less conservative $M_{\rm lim}(z)$ threshold (50 per cent completeness). %

As shown in Fig.~\ref{fig:Mcompl}, the 90 per cent stellar mass completeness of LSST galaxies is below $10^{9}\,{\rm M}_{\odot}$ at $z<0.5$. It increases at $z>0.5$, because of dimming, reaching $M_{\rm lim}=1.5\times10^{10} {\rm M}_{\odot}$ at $z\sim2$.  This threshold decreases at higher redshift  as   galaxies within our mass range  in the early universe have higher  SFRs \citep[see e.g.,][]{speagle14} so they become brighter in the rest-frame UV probed by the $i$ band. Conversely, at $z\lesssim2$ the SFR starts to decline while more stellar mass is assembled, allowing an easier detection in the $H$ band \citep[see a similar discussion in][for a \emph{Spitzer}/IRAC selected  sample]{davidzon17}.  

For galaxy evolution studies relying on  the \emph{Euclid} photometry one should prefer an $H$-band selection, instead of the nominal $riz<24.5$, if the scientific goals require {a sample highly complete in stellar mass}. Given an average galaxy SED, the \emph{Euclid} optical selection would correspond to a cut at $H\sim23$, while the survey will go deeper in the NIR by about 1 mag. Therefore, in the following sections, the analysis is carried on with the $H<24$ selected sample. 
Modulo observational uncertainties, the nominal mass completeness for the \emph{Euclid} $H<24$ sample is well fit by the function 
$M_\mathrm{lim}(z) = 4.5\times10^8 (1+z)^{2.4}\,M_\odot$, 
reaching a maximum of $2\times10^{10}M_\odot$ at $z=4$ {in the studied redshift range} (see Fig.~\ref{fig:Mcompl}). Although the limit based on $riz$ detections is generally higher (e.g., $8\times10^{10} M_\odot$ at $z\sim2$) it is enlightening to note that it starts to decline at $z>2$, as already discussed for the optical selection in LSST\footnote{Recall that the star forming main sequence in {\sc Horizon-AGN} reproduces well the observed one at $z\sim 4$, but at lower redshifts the simulation underestimates it by $\lesssim 0.3$ dex  \citep[see figure 3 in ][]{kaviraj16}. Consequently, observed galaxies at $2<z<4$ are expected to be brighter in the rest-frame UV, which would therefore enhance their $riz$ magnitude by at most $\sim$0.5 mag. 
Such a magnitude offset corresponds to  a stellar mass limit approximately  $\sim$0.2 dex lower than that displayed in  Fig.~\ref{fig:Mcompl}.  A  $H<24$ selection would still provide a higher completeness. 
 The underestimation of the SFR in the virtual catalogue  being  independent of mass, this remark is also valid for the LSST catalogue. }.

\subsection{Forecasts for \emph{Euclid} and LSST photometric redshifts}
\label{sec:photoz_euclid}
\begin{figure*}
\includegraphics[scale=1]{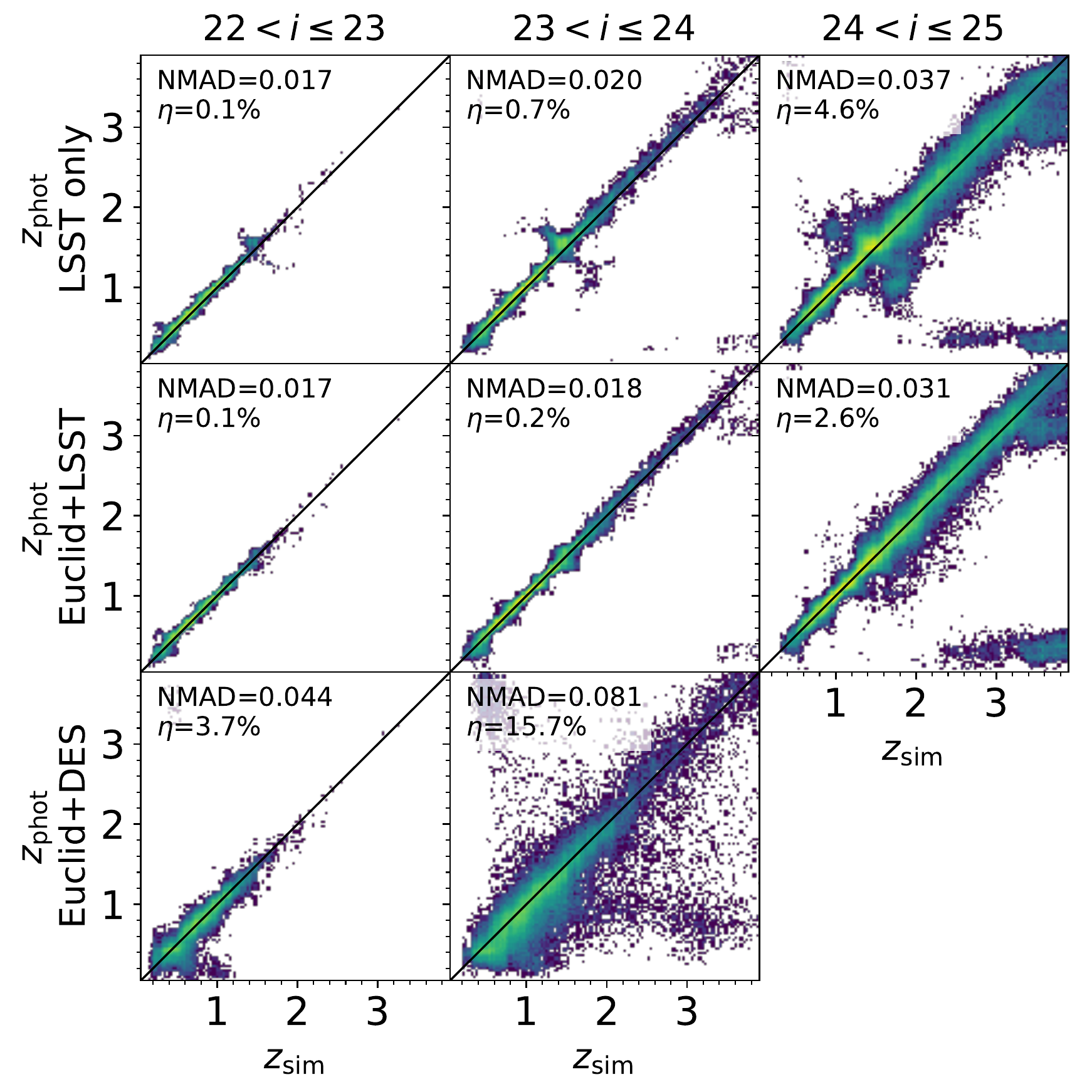}
\caption{Predictions of $z_\mathrm{phot}$ quality in future galaxy  surveys (see Section \ref{sec:photoz_euclid});  symbols and colours  as in Fig.~\ref{fig:zphotzspec}. 
Results for three different  baselines are shown: LSST (\textit{top}), \emph{Euclid}+LSST (\textit{middle}), and \emph{Euclid}+DES (\textit{bottom}); Table \ref{tab:comp2} summarizes these results. Each column shows galaxies in a different $i$-band magnitude bin, with the $i$-band virtual observations coming from either LSST or DES. In the latter case, the comparison is limited to $i<24$ because of the shallower DES sensitivity. 
}
\label{fig:zphot_all}
\end{figure*}

Photometric redshifts are computed in the same way as in the COSMOS-like case (Section~\ref{subsec:photoz_method}). 
The performance of SED-fitting in the \emph{Euclid}+DES, LSST-only and \emph{Euclid}+LSST configurations for $z_{\rm phot}$ estimation is presented in Fig.~\ref{fig:zphot_all}. The sample is split in different $i$-band bins ($i$ taken from either LSST or DES photometry). 
The results in terms of NMAD and catastrophic outliers are presented in Table~\ref{tab:comp2}, for both $H$ and $i$ magnitude bins.

In all configurations, the usual population of outliers at high redshift and faint magnitudes $i>24$ is found (see the discussion in  Section~\ref{sec:photoz} about the impact of IGM absorption). 
\emph{Euclid} (combined with optical DES photometry) performs relatively well at bright magnitudes ($i<23$). However, 
because of the lack of blue optical bands to constrain the Balmer break, the accuracy at very low redshift ($z<0.5$) is lower than in COSMOS, even for bright galaxies. At fainter magnitudes, the main limitation of this configuration is the shallow depth of the survey. 

In the LSST configuration, without NIR photometry, a large fraction of catastrophic outliers is present at $1.2<z<2.5$ at all magnitudes with a relatively symmetric patterns. The reason is the same as for the increase of $z_{\rm phot}$ uncertainties in this redshift range (see e.g. Fig.~\ref{fig:cfzphoterr_cosmos}). At this redshift, the Balmer break is not constrained anymore by the optical bands and enters NIR. Without NIR bands to properly constrain its position, determining the redshift is challenging \citep[see also][]{fotopoulo18,gomes18}. The situation improves from $z>2.5$ when the Lyman break enters the optical bands. 

\begin{figure}
    \centering
    \includegraphics[width=0.99\columnwidth]{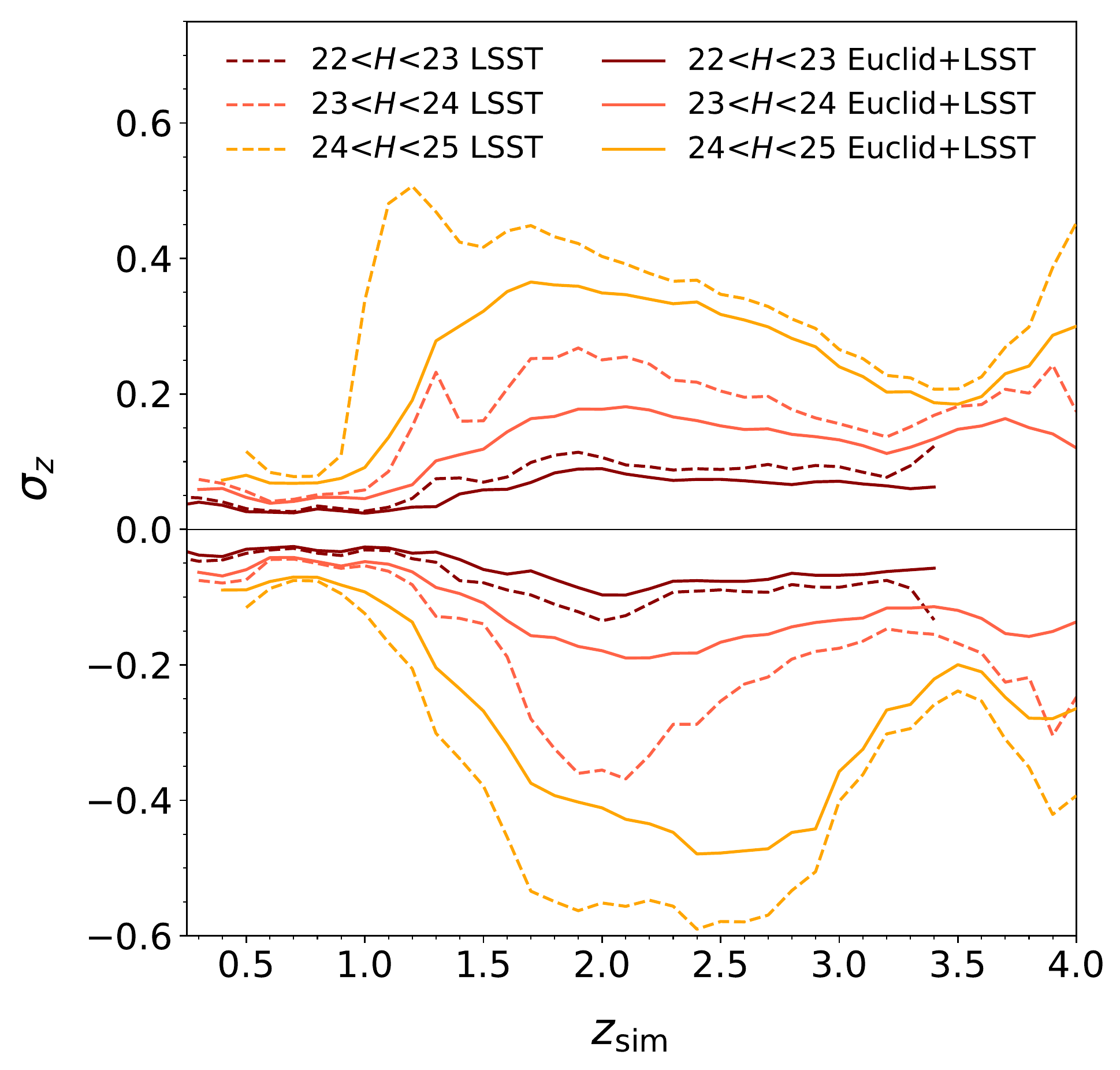}
    \caption{Photometric errors computed by {\sc LePhare} in the  LSST-only (\textit{dashed} line) and \emph{Euclid}+LSST (\textit{solid} line) configurations in three different $H$ bins.}
    \label{fig:sigEuclid}
\end{figure}

The LSST-like catalogue alone performs therefore less well than the LSST+\emph{Euclid} one. While the NMAD improves only by $\sim$ 1 percent in the faintest magnitude bin, adding the \emph{Euclid} NIR bands allows to reduce the fraction of outliers by a factor 2. 
The benefit to combine \emph{Euclid} and LSST is further illustrated in Fig.~\ref{fig:sigEuclid}. It presents the photometric errors computed by {\sc LePhare} in the LSST-only and \emph{Euclid}+LSST configurations in different $H$ bins. The $z_{\rm phot}$ errors dramatically improve in the redshift range $1.5<z<2.5$, especially in the faintest magnitude bins. This is true as long as galaxies are bright enough in the NIR. At fainter magnitudes $H\gtrsim 25$,  \emph{Euclid} is not deep enough to properly constrain the Balmer break. 
{From the \emph{Euclid} perspective, adding the LSST optical bands to the \emph{Euclid}+DES baseline considerably decreases redshift uncertainties and fraction of outliers (see Table~\ref{tab:comp2}) especially at faint magnitudes, at is provides deeper photometry in the $g$, $r$, $i$ and $z$ band. Furthermore the addition of the $u$-band is considerably useful from $z\sim 2.5$, when the Lyman-break enters the $u$-band.}

\begin{table}
\begin{center}
  \caption{Statistical errors (defined as normalized median absolute deviation, NMAD) and percentage of catastrophic errors ($\eta$) in different $i$ and $H$ magnitude bins, in the different configurations studied here.}\label{tab:comp2}
\begin{tabular}{ c | c c c c c c} \hline\hline
     $i$ band   & \multicolumn{2}{c}{Euclid+DES} &    \multicolumn{2}{c}{Euclid+LSST} & \multicolumn{2}{c}{LSST only}  \\
             mag            &  NMAD & $\eta$ (\%)   &  NMAD & $\eta$  (\%) &  NMAD & $\eta$  (\%)  \\ \hline
$($22,23$\rbrack$ & 0.044 & 3.7  & 0.017 & 0.1 & 0.017 &  0.1 \\
$($23,24$\rbrack$ & 0.081 & 15.7  & 0.018 & 0.2 & 0.020 & 0.7 \\
$($24,25$\rbrack$ & -- & $>$40.0 & 0.031 & 2.6 & 0.037 & 4.6 \\
  \hline
\end{tabular}\\
\vspace{3mm}
\begin{tabular}{ c | c c c c c c} \hline\hline
     $riz$ band  & \multicolumn{2}{c}{Euclid+DES} &    \multicolumn{2}{c}{Euclid+LSST} & \multicolumn{2}{c}{LSST only}  \\
             mag            &  NMAD & $\eta$ (\%)   &  NMAD & $\eta$  (\%) &  NMAD & $\eta$  (\%)  \\ \hline
$($22,23$\rbrack$ & 0.039 & 3.0  & 0.017 & 0.1 & 0.017 & 0.1  \\
$($23,24$\rbrack$ & 0.065 & 10.7  & 0.018 & 0.3 & 0.020 & 0.8 \\
$($24,24.5$\rbrack$ & 0.119 & 28.8 & 0.029 & 3.2 & 0.034 & 4.6 \\
  \hline
  \end{tabular}\\
\vspace{3mm}
\begin{tabular}{ c | c c c c c c} \hline\hline
     $H$ band  & \multicolumn{2}{c}{Euclid+DES} &    \multicolumn{2}{c}{Euclid+LSST} & \multicolumn{2}{c}{LSST only}  \\
             mag            &  NMAD & $\eta$ (\%)   &  NMAD & $\eta$  (\%) &  NMAD & $\eta$  (\%)  \\ \hline
$($21,22$\rbrack$ & 0.045 & 2.6 & 0.018 & 0.0 & 0.018 & 1.2 \\
$($22,23$\rbrack$ & 0.080 & 12.4  & 0.022 & 0.1 & 0.026 & 2.4 \\
$($23,24$\rbrack$ & 0.153 & 35.4  & 0.040 & 3.6 & 0.049 & 7.1 \\
  \hline
\end{tabular}
\end{center}
\end{table}

\subsection{Forecast for \emph{Euclid} and LSST stellar masses}\label{sec:physprop_euclid}

\begin{figure*}
\includegraphics[width=0.95\textwidth] {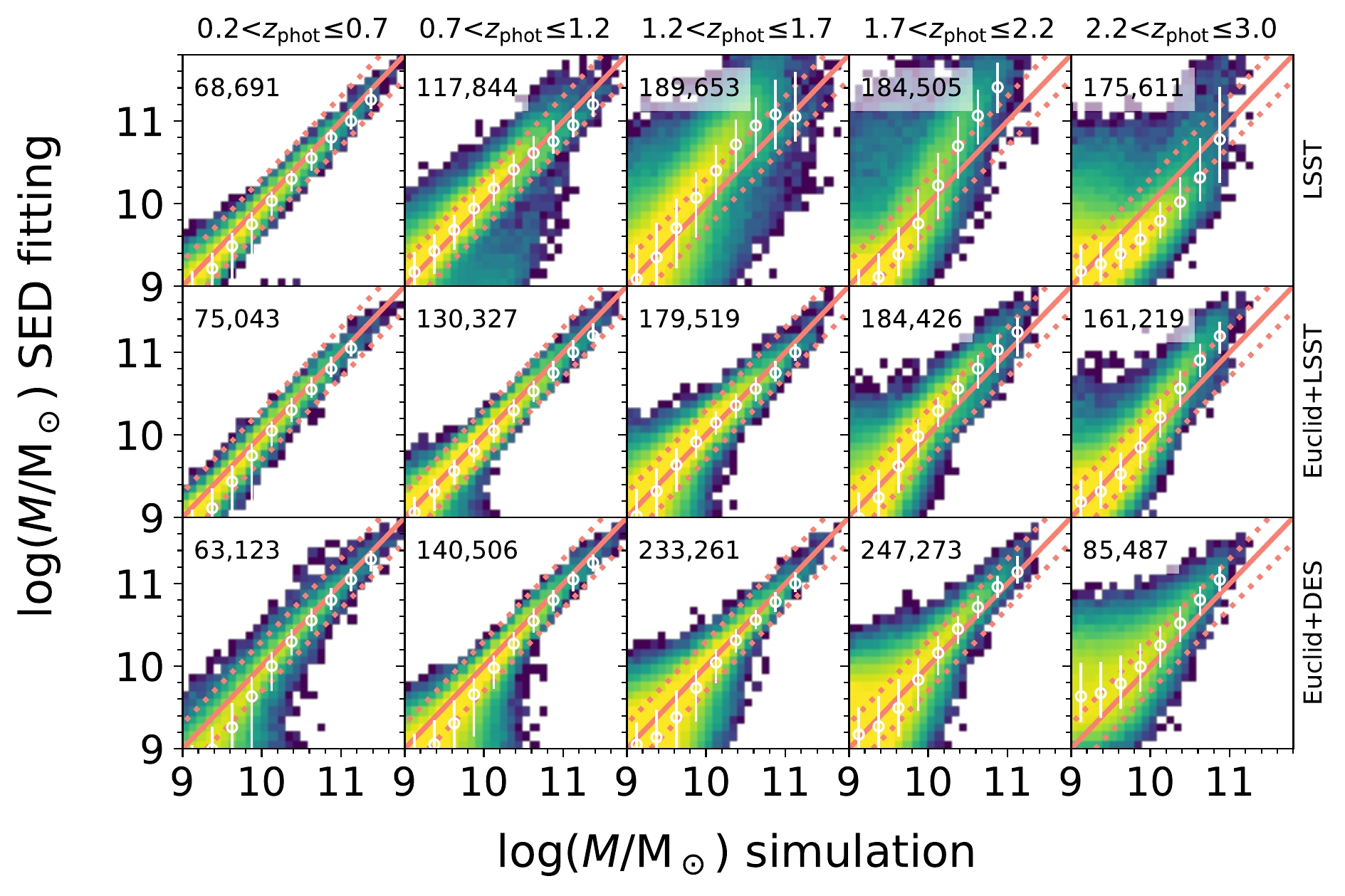}\vspace{6mm}
\caption{Stellar mass comparison for the three photometric baselines: LSST (\textit{top}), \emph{Euclid}+LSST (\textit{middle}) and \emph{Euclid}+DES (\textit{bottom}). Panels in a given column include galaxies in the photometric redshift bin indicated on the top (the number of galaxies in each bin is quoted in the upper-left corner of each panel). Colors and symbols are the same as in Fig.~\ref{fig:cf_mass}. 
}
\label{fig:cf_mass_euclid}
\end{figure*}

Figure \ref{fig:cf_mass_euclid} presents the overall comparison between intrinsic and reconstructed stellar masses in five photometric redshifts bins between 0.2 and 3, in the \emph{Euclid}+DES, LSST-only and LSST+\emph{Euclid} configurations. The number of objects in each redshift bin varies as the performance of $z_{\rm phot}$ estimation varies from one configuration to the other. 

\begin{figure*}
    \centering
    \includegraphics[width=0.99\textwidth]{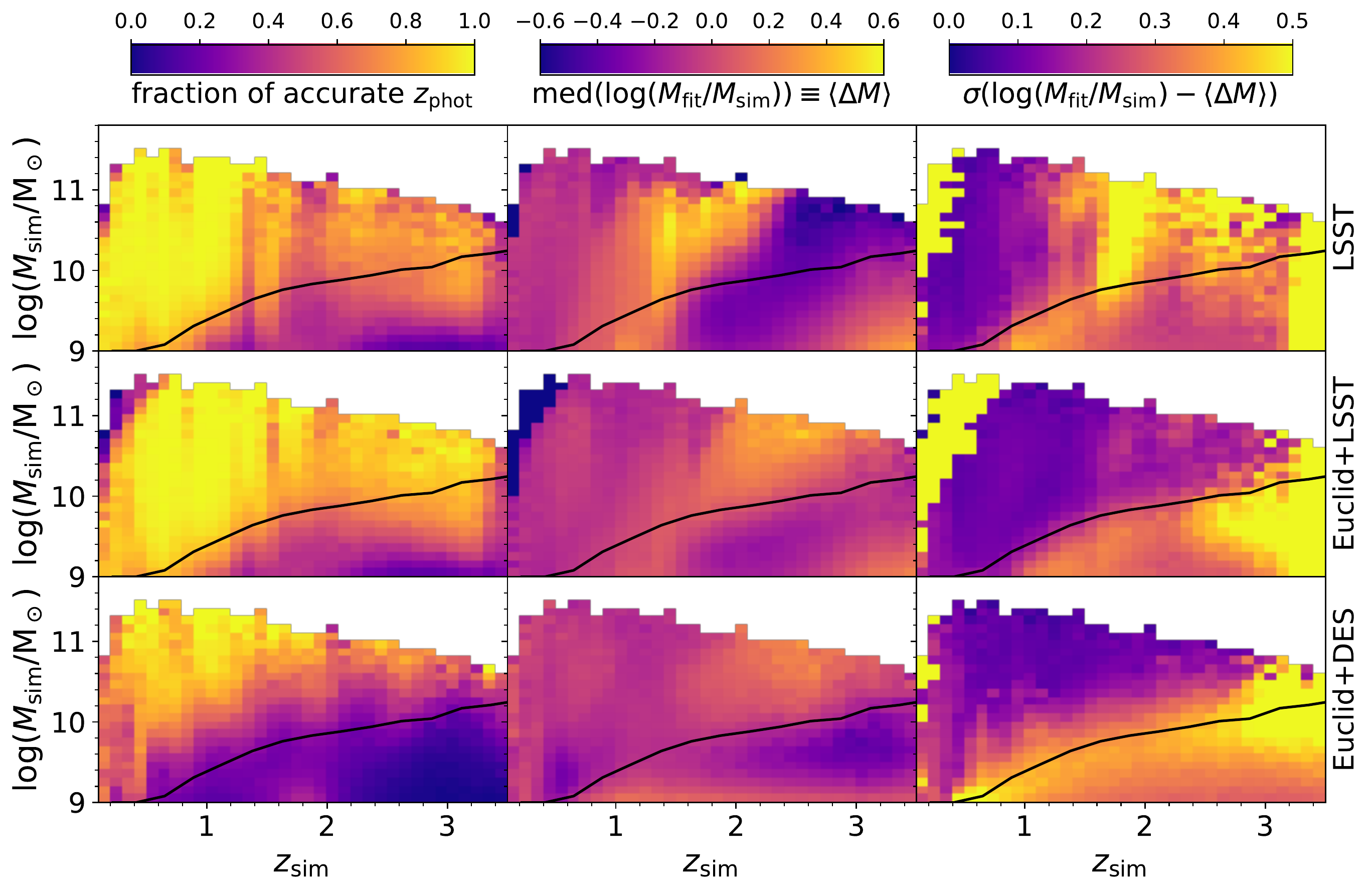} 
    \caption{SED-fitting properties of galaxies in bins of $z_\mathrm{sim}$ and $M_\mathrm{sim}$ from the LSST catalogue (\textit{top}) and  \emph{Euclid} (\textit{middle} and \textit{bottom}, combining with LSST and DES photometry respectively). \textit{Left}: fraction of accurate photometric redshift, defined as the fraction of galaxies within the given pixel having $\vert \Delta z\vert/(1+z_\mathrm{sim})<0.05$; \textit{Middle}: median logarithmic offset between intrinsic stellar masses ($M_\mathrm{sim}$) and SED-fitting estimates ($M_\mathrm{fit}$);  
    \textit{Right}: standard deviation of the scatter between $M_\mathrm{sim}$ and $M_\mathrm{fit}$ after removing the systematic offset. In all the panels a \textit{black solid} line delimits the 90 per cent stellar mass completeness expected for a galaxy sample selected at $H<24$ (see  Fig.~\ref{fig:Mcompl}). }
    \label{fig:lim}
\end{figure*}

With the LSST-like catalogue, the performance is much poorer than with the \emph{Euclid}+DES or \emph{Euclid}+LSST configurations, with a very large scatter from $z>0.7$. Indeed, without NIR photometry, the stellar mass will be determined on the basis of the photometric filters which trace the young stellar populations. {For example, the massively star-forming galaxies at high redshift (e.g. the massive galaxies in the bin $1.7<z_{\rm phot}<2.2$) will generally get their mass overestimated, which drives the very large scatter above the median}. On the contrary, passive galaxies generally get their mass underestimated (e.g. in the bin $0.7<z_{\rm phot}<1.2$), {which drives the very large scatter below the median}. The resulting scatter ({as defined from the RMS of $\log M_{\rm phot}/M_{\rm sim}$}) can be as large as 0.5 at $z>2$ (see e.g. Figure~\ref{fig:lim}).

It can also be noted that the \emph{Euclid}+LSST configuration performs in general better than the \emph{Euclid}+DES one at $z<2$: the additional LSST optical bands help to constrain the mass reconstruction. However more bands do not always yield a better fit. At $z>2$ and $\log M_{\rm sim}>10.5$,  the scatter is larger in the \emph{Euclid}+LSST  configuration than in the \emph{Euclid}+DES one (namely with NIR photometry only). Although  conter-intuitive, this discrepancy might be a consequence of the very different depth in optical and NIR. {Flux errors being much smaller in the optical, the blue part of the spectrum will provide a stronger constraint to the fit compared to the NIR. When no template can fit well both the optical and NIR photometry, the preference will be given to the optical since the errorbars are smaller. As a result, the error on the mass might be higher, because the optical part of the spectrum is a poorer proxy for stellar mass than the NIR. In the case of star-forming galaxies, it can lead to an overestimation of the mass}. Removing the LSST $u$-band is in general not sufficient to bring better agreement, and the other optical bands still contribute a lot to this discrepancy. 

As a summary, Fig.~\ref{fig:lim} presents the evolution of completeness with redshift and mass, along with the evolution of the median and RMS of $\log M_{\rm sim}-\log M_{\rm phot}$. 

 \subsection{Performance of future surveys: summary}
\label{Sec:EucSum}
We can draw the following conclusion concerning the expected performance of future surveys. 
\begin{itemize}[noitemsep,nolistsep]
    \item With the depth of the surveys for weak lensing galaxy selection in \emph{Euclid} ($H<24$) and  LSST ($i$<25.3), one can expect at $z=2$ a 90 percent completeness at $\log M_{*}/{\rm M}_{\odot}>9.9$ and $10.2$ respectively. 
    \item The \emph{Euclid}+DES $z_{\rm phot}$ accuracy is of the order of several percent even at low redshift and for bright objects, due to the absence of deep optical photometry to constrain the Balmer break position ($z<1.5$); the fraction of catastrophic outliers dramatically increases with fainter objects. The LSST $z_{\rm phot}$ accuracy is of the order of 2 percent at bright magnitudes ($i<24$). The absence of NIR photometry does not allow to properly constrain the Balmer break at $1.5<z<2.5$, leading to a significant fraction of catastrophic outliers in this redshift range. A dataset which would include \emph{Euclid} NIR photometry in addition to the LSST optical photometry would provide a better $z_{\rm phot}$ accuracy than LSST alone or \emph{Euclid} alone; it  would also decrease the fraction of outliers by 2. 
    \item Our LSST-like catalogue benefits a lot from NIR \emph{Euclid}-like photometry for stellar mass reconstruction, reducing the scatter up to a factor 3. There is therefore a mutual benefit for LSST and \emph{Euclid} to work in synergy.
\end{itemize}
 \section{General summary and conclusion}
 \label{sec:conclusion}

Using the realistic photometric catalogue extracted from the {\sc Horizon-AGN} hydrodynamical simulation, we investigated the performance of SED-fitting algorithms to compute galaxy properties. 
Compared to previous studies, the additional value of the present modelling relies on the use of an hydrodynamical lightcone from which the photometry has been consistently post-processed. This lightcone contains a large diversity of galaxies in terms of masses and star formation activity (but also orientation with respect to the line-of-sight), over a representative cosmological volume.  Galaxy photometry therefore naturally accounts for the diversity of SFHs, metallicity enrichment and dust distribution which result from the complex history of their formation, driven by the combination of pristine gas infall, stellar and AGN feedback, mergers, etc, which is consistently followed in the  simulation. This  lightcone also allowed for the self-consistent implementation of in-homogeneous IGM absorption within each galaxy spectrum in order to test its impact on $z_{\rm phot}$ estimation. 

We used the well-calibrated COSMOS2015 dataset to assess the performance of the SED-fitting software {\sc LePhare}  in extracting galaxy $z_{\rm phot}$
from our mock catalogue. We also quantified our ability  to estimate the corresponding $z_{\rm phot}$ uncertainties. 
 We then estimated the biases in galaxy masses and SFR estimation through SED-fitting  relying on our ability to turn on and off various physical processes in the mocks (see Table~\ref{tab:comp1} and Section~\ref{Sec:CosSum} for a detailed summary). 
 Finally, we quantified the expected performance for the upcoming imaging surveys \emph{Euclid} and LSST, given the available photometric baseline and expected depths (see Table~\ref{Tab:compl}, Table~\ref{tab:comp2} and Section~\ref{Sec:EucSum} for a detailed summary).\\
In addition to these findings specific to some survey configurations, this work has allowed to draw the following  general conclusions on the process of measuring galaxy properties from their photometry:
\paragraph*{Choice of the photometric baseline:}
The added value of having medium-bands in the photometric baseline to improve redshift precision is obvious when  comparing $z_{\rm phot}$ computed  with and without medium-bands (Table~\ref{tab:comp1}). At the faint end of the galaxy population, better estimating the galactic continuum with these bands improves NMAD and $\eta$ by $\sim 50$ per cent. One can expect that the gain is even larger in the real universe, when nebular line emission can be used to constrain redshift more efficiently;\\
Deriving $z_{\rm phot}$ without deep optical photometry is challenging below $z<1.5$ (bottom panel in Fig.~\ref{fig:zphot_all}).\\
NIR photometry is  mainly driving  the performance of $z_{\rm phot}$ at $1.5<z<2.5$ and of stellar mass computation (e.g. compare top and bottom lines in Fig.~\ref{fig:cf_mass_euclid}). Adding optical bands to the NIR photometry helps reducing the scatter 
(compare middle and bottom lines in Fig.~\ref{fig:cf_mass_euclid});
\paragraph*{Impact of dust and IGM attenuation on the $z_{\rm phot}$ estimates}
While the impact of dust is significant, it globally does not bias much  the $z_{\rm phot}$ reconstruction performance with the current method to include it in SED-fitting. Overestimating the IGM absorption at the SED-fitting stage also impacts the $z_{\rm phot}$ estimate and can explain a large fraction of the population of catastrophic outliers at $z_{\rm sim}>2.5$ and $z_{\rm phot}<1.5$, as seen in real data. {There are however other possible explanations for this observed population of outliers, including systematics at the stage of photometry extraction, which we do not test in the present work;}
\paragraph*{Uncertainties on $z_{\rm phot}$ estimates:}
1-$\sigma$ uncertainties estimated from the SED-fitting code {\sc LePhare} are a good representation of the intrinsic $z_{\rm phot}$ errors, {except for the bright galaxies at $1<z<2.5$ for which the errors are generally underestimated.}. The remaining discrepancies {can be understood} in the limits of our end-to-end pipeline (e.g. no emission lines, {no modelling of the possible failures in the extraction of the photometry});
\paragraph*{Uncertainties on $M_{\rm phot}$ and SFR estimates:}
The scatter and systematics in the stellar mass and SFR computation are a combination of three effects: 1) the inherently limited SFH and metallicity enrichment pattern used to build the template library, which usually drives a global underestimation of  stellar mass and SFR 2) the way dust is accounted for (in particular the choice of the dust extinction curves at the SED-fitting stage) and the degeneracy between dust and SFR in the blue bands 3) the propagation of $z_{\rm phot}$ errors in the mass and SFR estimates, which increase the scatter. The net result exhibits a complex trend, which also depends on the photometric baseline available (e.g. compare Fig.~\ref{fig:cf_mass} and Fig.~\ref{fig:cf_mass_euclid}). As mentioned before, the impact of these effects on SFR is more dramatic, with in particular a bimodal behaviour mainly driven by dust.

\vspace{0.5cm}
Amongst possible actions to improve the redshift, mass and SFR estimates, building a template library with an additional parametrizable double burst star-formation history could temper the remaining systematic offsets at low redshift. When computing masses and SFR, one could also try to build the template library in {\sc LePhare} with additional dust extinction curves, which could mitigate the bimodal behaviour in the SFR computation. {Finally, it would be worth testing if allowing the mean IGM absorption to slightly vary at $z>3$ (in order to account for the line-of-sight variability of IGM opacity or the uncertainty on the model) can reduce the fraction of catastrophic outliers.}
\vspace{0.5cm}

{
Our study does not account for systematics in the photometry extraction, and therefore our estimates of $\sigma_{z}$, $\eta_{z}$ and $\sigma_{M}$ must be understood as lower limits. However in the light of our results, we can discuss if LSST and \emph{Euclid} will, at face value, fulfill their requirements. 
\\
For LSST, the redshift errors quantified from the root-mean-square scatter ($\sigma_{z}^{\rm rms}=$rms$(z_{p}-z_{s})/(1+z)$) and fraction of outliers  $\eta_{3 \sigma}$ (i.e the fraction of objects with $(z_{\rm p}-z_{\rm s})/(1+z_{\rm s})>3\sigma_{z}^{\rm rms}$) must be respectively smaller than 0.05 (with a goal of 0.02) and  10\% at all redshifts, as specified by \cite{LSSTsciencebook}. For  $z<3$ and $i^{+}<24.5$ (resp. $i^{+}<25.0$), we found $\sigma_{z}^{\rm rms}=0.043$ (resp. 0.060) and $\eta_{3 \sigma}=1.03$\% (resp. 1.50\%). Note that $\sigma_{z}^{\rm rms}$ is quite sensitive to the presence of outliers, and computing it by excluding the outliers (as defined by $\eta_{3 \sigma}$) yields $\sigma_{z}^{\rm rms}=0.026$ (resp. 0.032). At face value, the requirements are fulfilled, but as cautioned previously, these errors might be increased because of the systematics in the photometry extraction.
When adding the \emph{Euclid} photometric baseline to LSST, the errors decreases to $\sigma_{z}^{\rm rms}=0.031$ (resp. 0.044) (and $\sigma_{z}^{\rm rms}=0.025$ (resp. 0.030) when excluding the outliers). }\\
{
As for \emph{Euclid}, the expected redshift error $\sigma_{z}^{\rm rms}$  must be smaller than 0.05 (with a goal of 0.03) and the fraction of outliers $\eta_{0.15}$  (same definitition as in this paper) is required to stay below 10\%, with a goal of 5\% \citep{laureijs11}. In the \emph{Euclid}+DES configuration, at $riz<23.5$ (resp. 24.5) we get $\sigma_{z}^{\rm rms}=0.09$ (resp. 0.17) and  $\eta_{0.15}=3.45$\% (resp. 9.67\%). Excluding the outliers in the computation yields $\sigma_{z}^{\rm rms}=0.46$ (resp. 0.057). Photometry deeper than DES in bands narrower than the actual \emph{Euclid} $riz$ filter (e.g. the photometric baseline provided by LSST) will be required to improve these performances and extend them at fainter magnitudes.
\\
Although measuring stellar mass is not pivotal for weak-lensing based cosmology, it is of great interest for galaxy evolution science, i.e. to make the best of the legacy programs of \emph{Euclid} and LSST, and therefore to fullfill their secondary science goals. In particular, the huge area of these surveys will allow to drastically decrease the statistical errors on mass functions, two-point correlation functions or any other environmental measurements (e.g. groups and clusters, cosmic web analysis). For these studies, the NIR coverage provided by \emph{Euclid} will be of prime importance to extract accurate galaxy masses. On the other hand, without deep optical photometry in narrow optical filters, \emph{Euclid} will be unable to separate galaxy populations from their colours, which is pivotal e.g. to study galaxy bimodality and the growth of the population of quiescent galaxies. To this end, combining \emph{Euclid} and LSST would be a powerful configuration, which would benefit to both surveys and allows for the first time to address some of the most pressing questions in the field of galaxy formation today. \\
In following works we will pursue this discussion by exploring how redshift and mass errors propagate into one and two-point statistics, and we will quantify the effect of imperfect photometry extraction from mock images.}


\section*{Acknowledgements}
{\sl 
CL is supported by a Beecroft Fellowship, and thanks the Korean Astronomy and Space science Institute for hospitality when this work was finalized.
ID was supported in part by NASA ROSES grant 12-EUCLID12-0004. 
OI acknowledges the funding of the French Agence Nationale de la Recherche for the project
``SAGACE''. JD and AS acknowledges funding support from Adrian Beecroft, the Oxford Martin School and the STFC. HJMCC acknowledges support from the Programme national cosmologie et galaxies (PNCG) and the Domaine d'int\'er\^et majeur en astrophysique et conditions d'apparition de la vie (DIM--ACAV+). The authors thank A.\ L.\ Serra for her suggestions to improve graphic rendering of the figures. 
 This work  relied on the HPC resources of CINES (Jade) under the allocation 2013047012 and c2014047012 made by GENCI
and on the Horizon and CANDIDE clusters hosted by Institut d'Astrophysique de Paris. We warmly thank S.~Rouberol for maintaining  these clusters on which the simulation was post-processed. 
This research is part of Spin(e) (ANR-13-BS05-0005, \href{http://cosmicorigin.org}{http://cosmicorigin.org}), ERC grant 670193 and 
{\sc horizon-UK}.   
  This research  is also partly supported by the Centre National dEtudes
Spatiales (CNES). This work is based on data products from
observations made with ESO Telescopes at the La Silla Paranal
Observatory under ESO programme ID 179.A-2005 and on data products
produced by TERAPIX and the Cambridge Astronomy Survey Unit on behalf
of the UltraVISTA consortium. 
}
\bibliographystyle{mnras}
\bibliography{papers}

\appendix

\section{Mocks additional features}
\label{appendix:photometry}
Let us provide more details about how the virtual photometry  has been computed. 

\subsection{Calibration of the dust attenuation}
\subsubsection{Dust-to-metal mass ratio}
Dust attenuation is implemented assuming that the distribution of gas metallicity  is a good proxy for the dust distribution. This computation implies to choose a value for the dust-to-metal mass ratio, i.e. to define which fraction of metals are locked into dust grains. For the sake of simplicity, the dust-to-metal mass ratio is assumed constant, though some works have shown that it could vary with redshift  or within a same galaxy as a function of metallicity \citep[e.g.][]{galametz11,mattsson12,decia13,fisher14}. Most of the time, the implementation of dust attenuation in simulations uses a dust-to-metal ratio of 0.4 \citep[e.g.][]{jonsson06}, which is the Milky-Way value \citep{dwek98}. Nevertheless, there is no evidence that this factor, derived from  high-resolution models of our Galaxy,  should be used at face value in the simulation, especially because the spatial resolution ($\sim 1$ physical kpc) of the simulation implies that dust scattering and absorption occurs at the subgrid scale. One expects therefore {that the emergent dust attenuation will depend on the smaller scale distribution of dust and metals, which are not resolved. Given the low resolution of the simulation, we do not implement a two-component dust attenuation models to account separately for dust obscuration in both birth clouds and diffuse interstellar medium, although this has been successfully implemented elsewhere \citep{trayford15}}. \\
In addition, prior to converting the metal mass into dust mass, it is important to reproduce the correct gas phase metallicity in the simulation. As discussed in \citet{kaviraj16}, the relatively low resolution reached in {\sc Horizon-AGN} implies a delayed enrichment of star-forming clouds, which  underestimates  the gas phase metallicity  compared to observations. To correct for this, a redshift-dependent boosting factor (varying from 4 at $z\sim 0$ to 2.4 at $z\sim 3$) has been computed in order to brings  the simulated mass-gas phase metallicity relation in agreement with obsevations from \citet{mannucci10} at $z=0$, 0.7, 2.5  and \citet{maiolino08} at $z=3.5$. However, the normalization of the mass-gas metallicity relation undergoes large variations depending on the chosen observable used to measure gas metallicity, up to a factor of 5 \citep[see e.g.][]{andrews13}. In particular, the renormalisation of {\sc Horizon-AGN} gas phase metallicity tends to align onto the highest values derived from observations \citep[namely using the R23 method, see e.g.][]{lian15}. As a consequence, one can expect the boosting factor derived in \citet{kaviraj16} to be an upper limit. In fact, while comparing the simulated galaxy counts with COSMOS2015 in various bands as a function of redshift after renormalizing the gas-phase metallicity, we find that a dust-to-metal mass ratio of 0.4 is excessive {as it results in too few galaxy counts compared to observations}. Therefore, we empirically choose a dust-to-metal mass ratio of 0.2, which yields an overall better agreement with the observed counts in all bands at all redshifts. 
\subsubsection{Attenuation curve}
\begin{figure}
\includegraphics[scale=0.43,trim={0.5cm 0cm 0.5cm 0},clip]{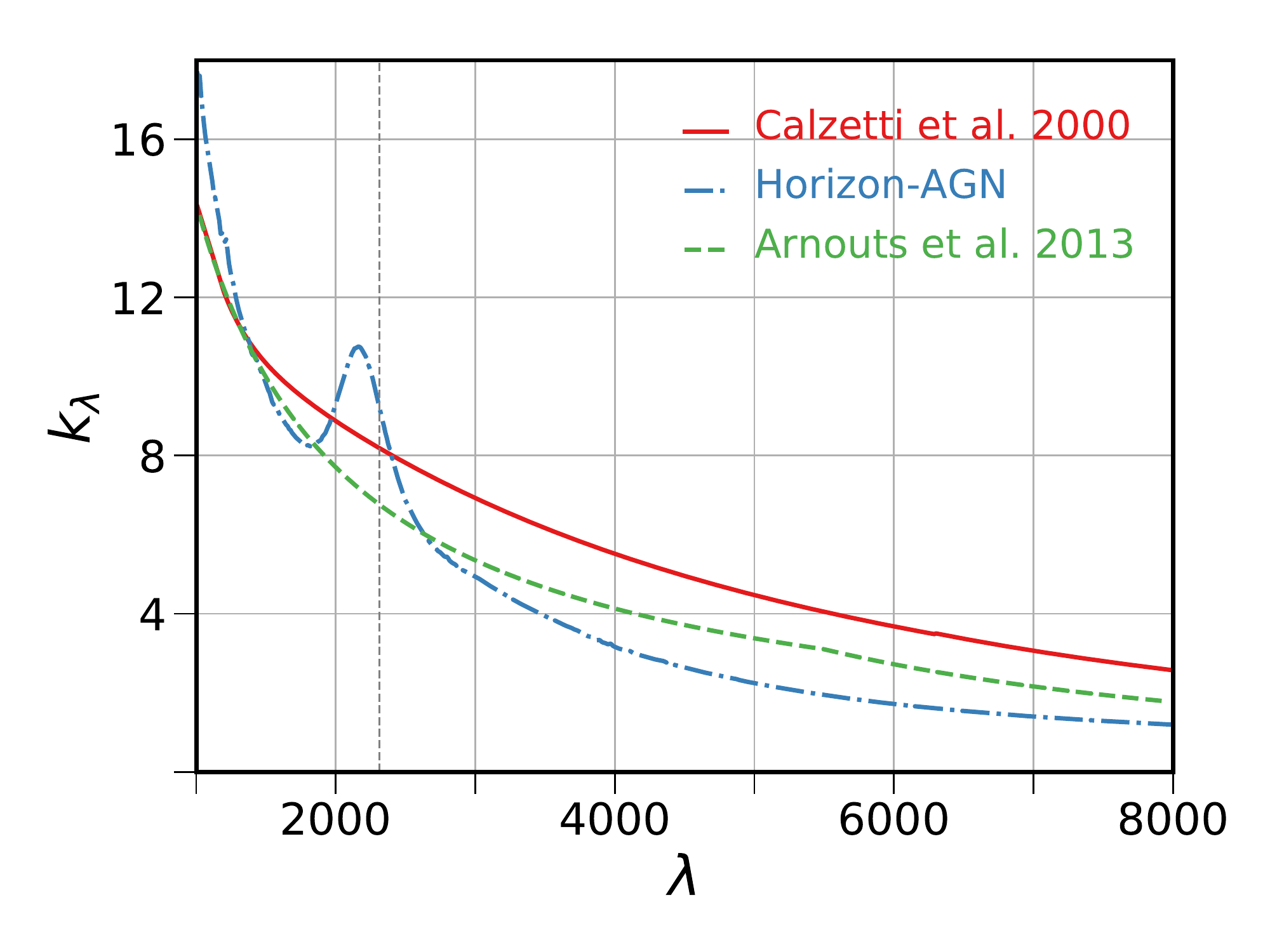}
\caption{Extinction curves used when fitting the photometry to get the physical properties (\textit{red solid line} and \textit{green dashed line}) and the average extinction for $\sim 1000$ {\sc Horizon-AGN} galaxies randomly selected (\textit{blue dashed-dotted line}). The vertical dashed line indicates the median wavelength of the \textit{NUV} filter.}
\label{fig:Att}
\end{figure}

Beyond the dust mass distribution in the galaxy, the amount of extinction at a given wavelength will be dependent upon the chosen attenuation curve. In this work, the ${\rm R}_{V}=3.1$ Milky Way dust grain model by \citet{weingartner01} is used for post-processing the simulated galaxies. This model includes in particular the prominent 2175 $\angstrom$-graphite bump. 
In the simulation, the spectrum of each stellar particle, assumed to be a SSP, is attenuated with this model and knowing the specific dust column density in front of each particle. However, when summing up the contribution of all the SSPs, we find that the overall attenuation curve of the resulting galaxy spectrum is  less steep and the bump tends to reduce. As discussed in \citet{fishera11}, turbulence  can further reduce the bump and flatten the extinction curve. 
Fig.~\ref{fig:Att} presents the extinction curve in {\sc Horizon- AGN} as averaged over one thousand galaxies randomly selected, and the two curves used in {\sc LePhare} for the computation of stellar masses and SFR. $k_{\lambda}$ is defined as $A_{\lambda}/{\rm E}(B-V)$.
None of the two curves used in {\sc LePhare} can correctly reproduce the one in {\sc Horizon-AGN}, and this discrepancy is likely to be the reason for the bimodality in the SFR, as further discussed in Appendix~\ref{appendix:photoz}. 

\subsection{IGM absorption}
In order to implement the IGM absorption, the Lyman-$\alpha$ forest is implemented on each galaxy line-of-sight from the gas density, velocity and temperature in the IGM in front of the galaxy. \\
Let us consider the line-of-sight of a  background source emitting at the observed wavelength $\lambda_{0}$. Here $\lambda_{\alpha}=1215.7$~$\angstrom$ corresponds to the transition from the ground state to the first excited state of the Hydrogen atom.  The wavelength of the photons emitted by the background source spectra are redshifted by a factor of $(1+z)$. At some point the light from the source will be redshifted at $\lambda_{\alpha}=1215.7$ $\angstrom$. At this point, it may be absorbed by HI from the IGM. The probability of transmission of the light at the observed frequency $\nu_{0}$ will be given by:\begin{equation}
F(\nu_{0})=e^{-\tau_{\alpha}(\nu_{0})}\,,
\label{Eq:flux}
\end{equation}
where $\tau_{\alpha}(\nu_{0})$, the Ly-$\alpha$ optical depth at the observed frequency $\nu_{0}$ is given by
\begin{equation}
\tau_{\alpha}(\nu_{0})=\int_{0}^{x_{s}} \mathrm{d}x \dfrac{\sigma_{\alpha}n_{\rm HI}(x,z)}{1+z}\label{Eq:depth}\,,
\end{equation}
where  $x$ is the comoving coordinate of the comoving point varying along the line of sight between the observer ($x=0$) and the source ($x=x_{s}$), $z$ is the corresponding redshift, $n_{\rm HI}$ the neutral hydrogen density at point $x$ and redshift $z$, and $\sigma_{\alpha}$ the Ly-$\alpha$ cross-section. Here $\sigma_{\alpha}$ is a function of the frequency $\nu$ of the photon with respect to the rest frame of the neutral hydrogen at position $x$. Here $\nu=\nu_{0}(1+z)(1+v/c)$, where $v$ is the peculiar velocity along the line-of-sight. So $\sigma_{\alpha}$ may be then written as
\begin{equation}
\sigma_{\alpha}=\frac{ \sigma_{\alpha,0}\,c}{b(x,z)\sqrt{\pi}}  e^{\displaystyle-\frac{\left(v(x,z)(1+z)\nu_{0} -c \nu_{\alpha} +c (1+z)\nu_{0}\right)^{2}}{\nu^2_{\alpha} b^2(x,z)}}\!\!\!, \nonumber
\end{equation}
where $b(x,z)=\sqrt{2k_{B}T(x,z)/m_{p}}$,  $\sigma_{\alpha,0}=(3\pi\sigma_{\rm T}/8)^{1/2}f\lambda_{\alpha}$, with $\sigma_{\rm T}=6.25\times10^{-25}$cm$^{2}$  the Thomson cross-section, and $f=0.4162$  the oscillator strength. 
 {As we do not save the neutral hydrogen outputs for {\sc Horizon-AGN}}, the neutral hydrogen density is computed in post-processing by considering that the fraction $x_{\rm HI}=n_{\rm HI}/n_{\rm H}$ is a balance between photo-ionizations, collisional ionizations and recombinations. At equilibrium with the cosmic UV background field, it yields:
\begin{equation}
\alpha(T)n_{\rm e}(1-x_{\rm HI})=\gamma(T)n_{\rm e}x_{\rm HI} +\Gamma x_{\rm HI}\,,
\end{equation}
where $\alpha$ and $\gamma$ are the collisional recombination and ionisation rates, $\Gamma$ is the photoionisation rate, and $n_{e}$ is the free electron number density. Considering a uniform background radiation field as implemented in the simulation, the photoionisation rate is assumed to be spatially uniform. Its overall normalization is quite uncertain and is adjusted in order to match the PDF of the transmitted flux at $z=1.5$, $z=2$, $z=2.5$ and $z=3$ \citep[see and e.g.][]{theuns98,bolton05,lukic15,bolton16}.   
The chemical composition of the IGM is close to primordial, so it can safely be assumed that $n_{e}$ receives contribution only from ionised Hydrogen and Helium (assumed entirely in its fully ionized form), which allows to determine its composition \citep{choudhury00}. Prescriptions from  \cite{black81} are used to determine the collisional recombination and ionisation rate as a function of gas temperature. 

{In order to take into account the full Lyman series absorption, we assume that the only difference between Lyman-$\alpha$ and other Lyman transition comes from different absorption cross-sections. This modelling hence neglect  different widths of Lorentz profiles from one transition to an other, but this is expected to be a secondary effect \citep{irsic14}.} 

\begin{figure}
\includegraphics[scale=0.43,trim={0.5cm 0.5cm 0.5cm 0},clip]{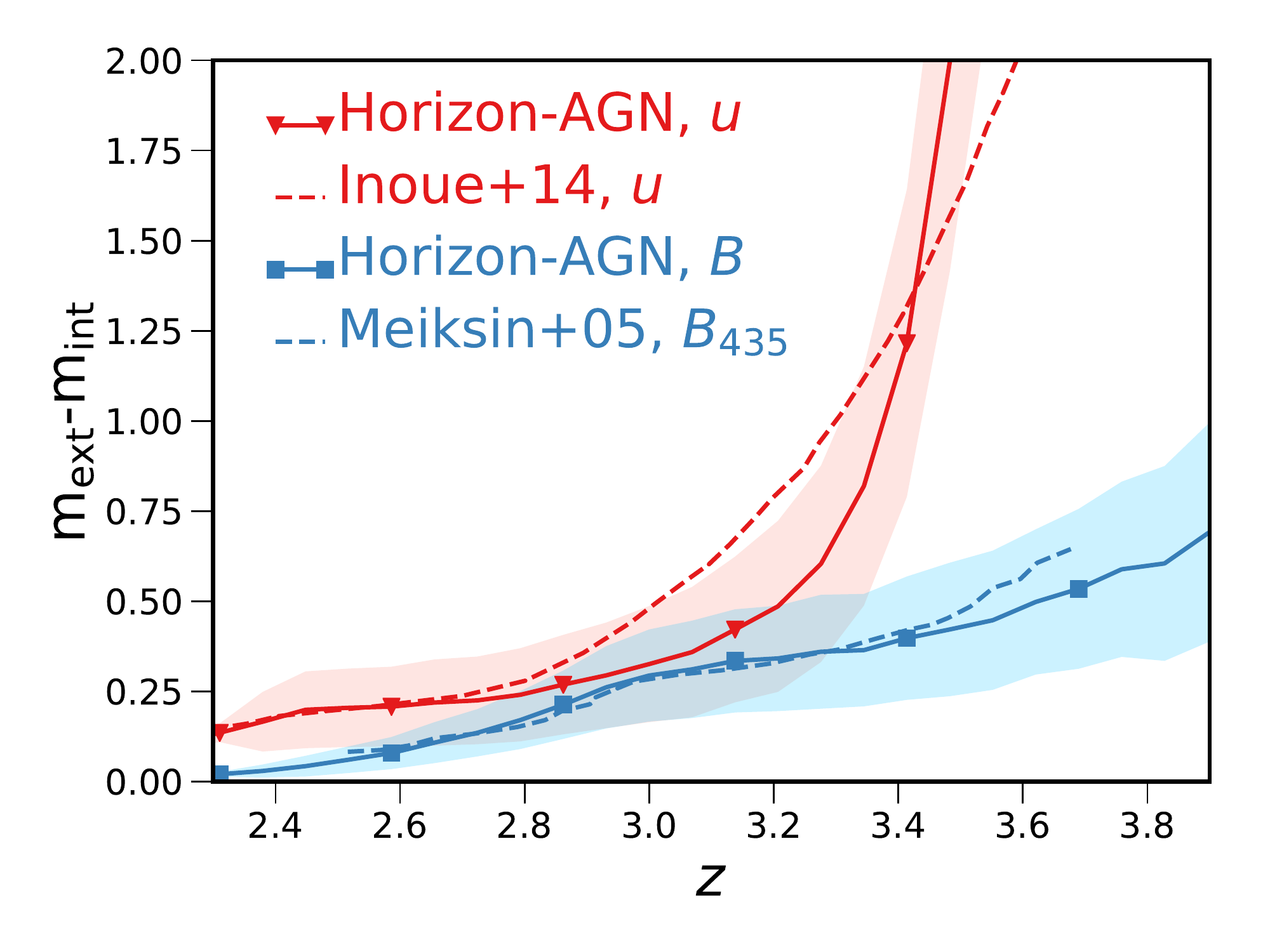}
\caption{Median absorption by the IGM in the $u$,  and $B$ bands as a function of redshift for {\sc Horizon-AGN} galaxies. Here  $m_{\rm int}$ and $m_{\rm ext}$ are respectively the intrinsic magnitudes and magnitudes after IGM absorption.}
\label{fig:IGM}
\end{figure}

{Eventually, galaxy spectra are then multiplied with the Lyman-series absorption lines. Fig.~\ref{fig:IGM} shows the median absorption by the IGM in the $u$ and $B$ bands as a function of redshift and compares with the litterature \citep{meiksin05,inoue14}. The median absorption in a given band will also depends on the hardness of the galaxy UV spectrum, and on the presence or absence of Lyman-limit systems \citep{meiksin05}. On overall, our implementation of IGM absorption matches well the litterature. As noted in e.g. \cite{inoue14}, the Madau model tends to slightly overestimate the correction at $z>3$ compared to observations, and therefore also overestimate the correction of our virtual galaxies.}

\begin{figure*}
\includegraphics[width=0.97\columnwidth]{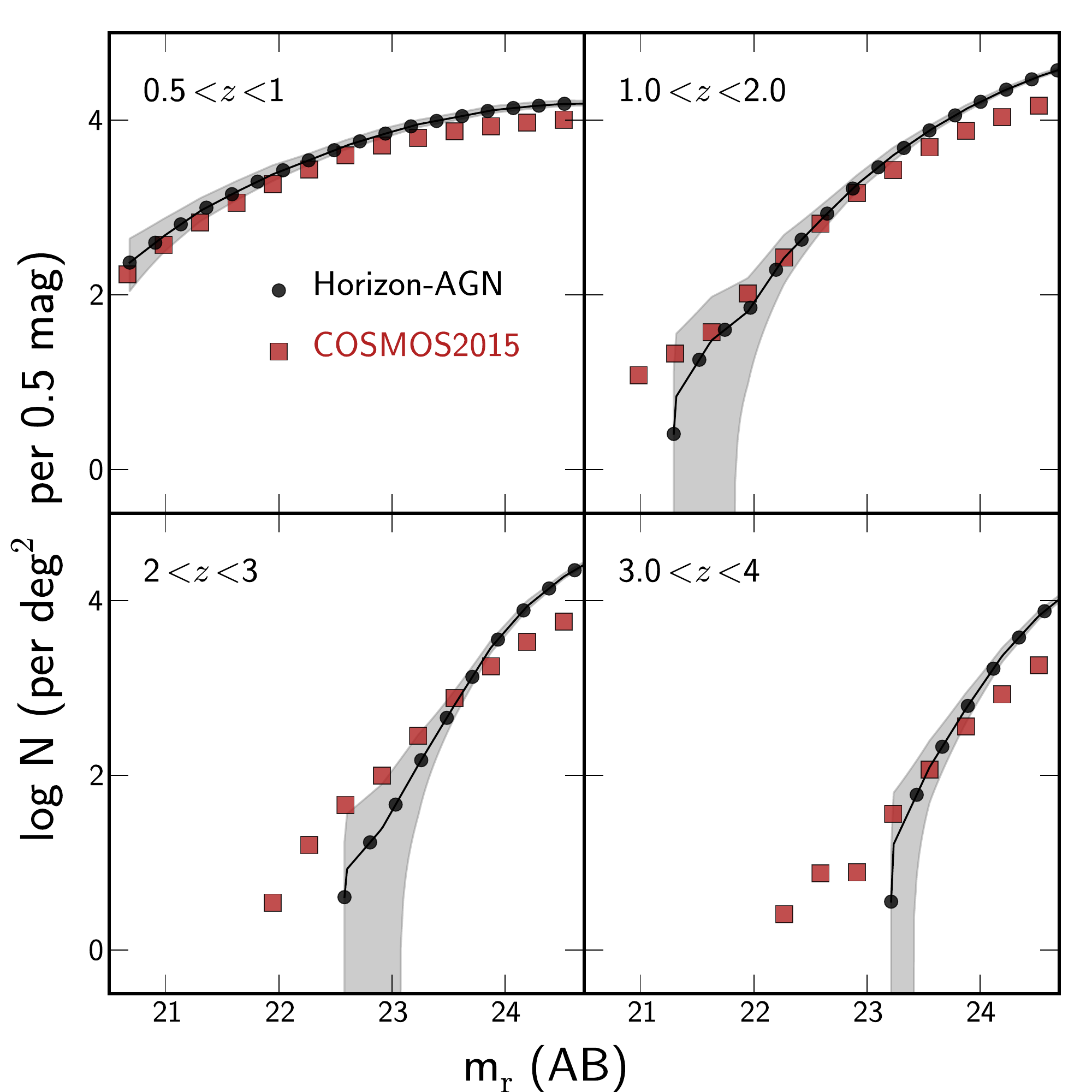}
\includegraphics[width=0.97\columnwidth]{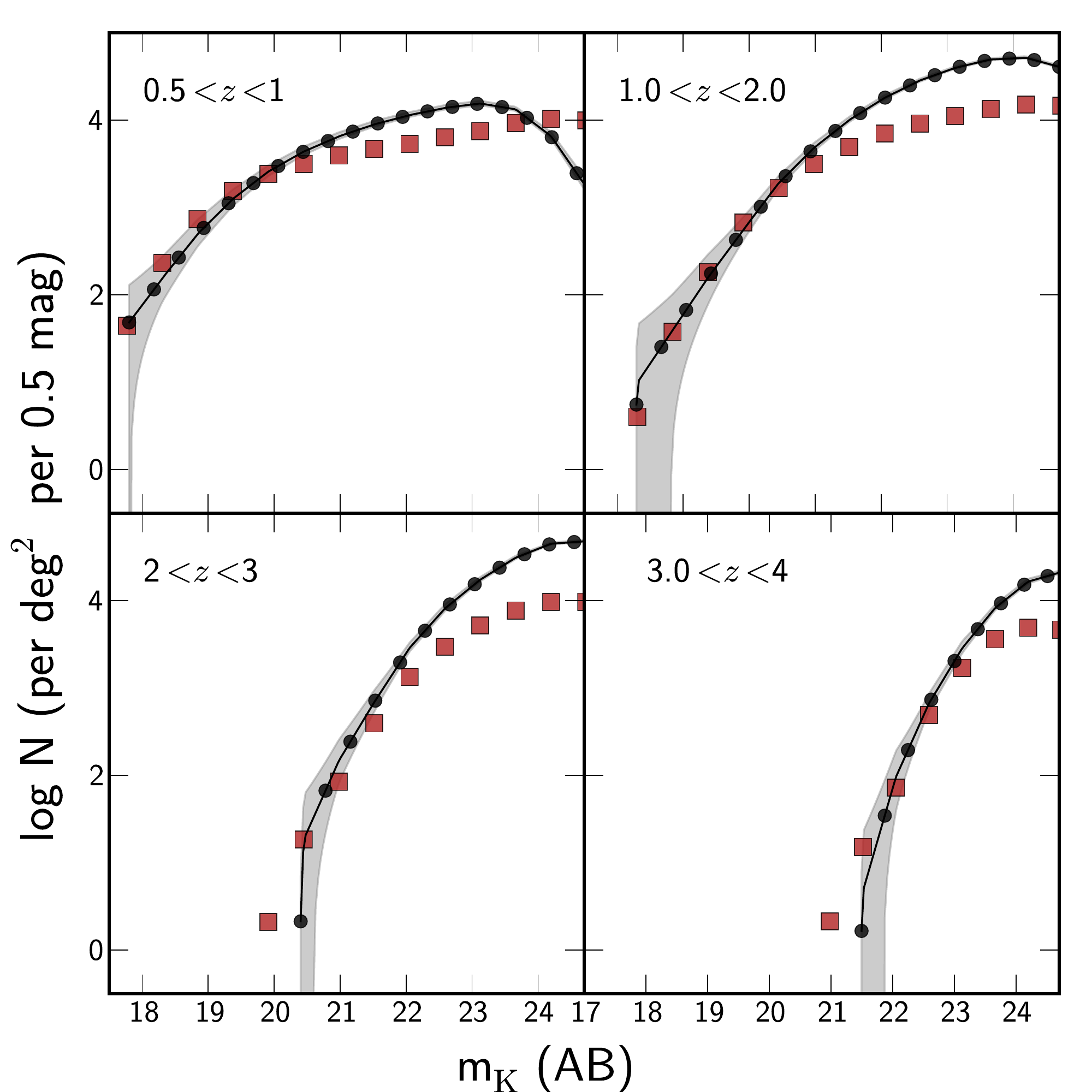}
\caption{Galaxy total magnitude count in the $r$-band (\textit{left}) and $K_{\rm s}$ band (\textit{right}) as a function of redshift in COSMOS2015 (red squares) and {\sc Horizon-AGN} (sold line). COSMOS2015 photometry is corrected for Galactic extinction. }
\label{fig:magapp}
\end{figure*}

\subsection{Flux error implementation}
\label{Ap:fluxErrors}
\begin{figure}
\includegraphics[width=0.97\columnwidth]{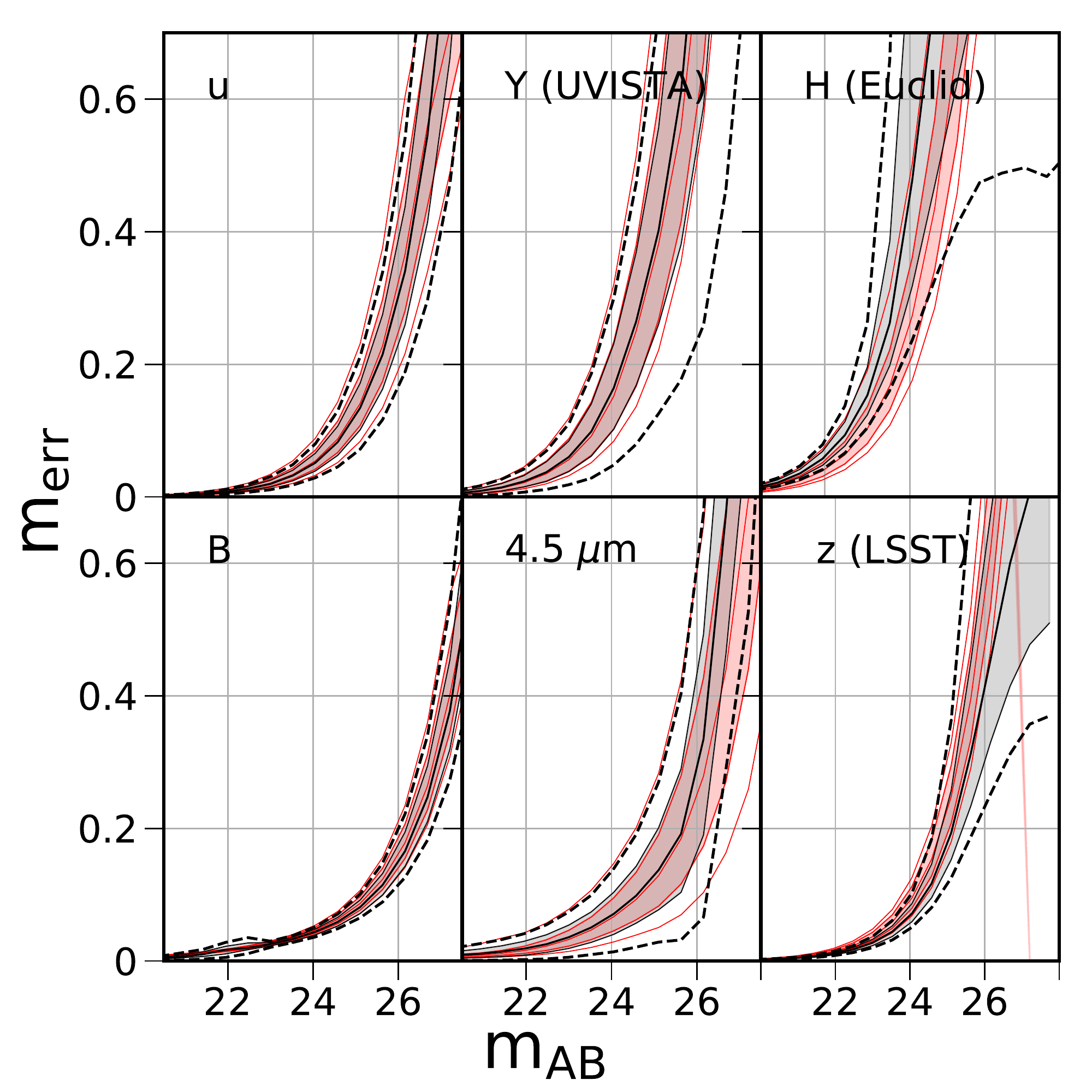}
\caption{Comparison between observed  and simulated  apparent magnitude errors (${\rm m}_{\rm err}$) in $u,B,Y$, IRAC $[4.5]$ bands from COSMOS, $H$ from \emph{Euclid} and $z$ from LSST. In each panel, \textit{grey} area shows 68 percent of the galaxies from {\sc Horizon-AGN} and the \textit{red} one shows 68 percent of the galaxies from COSMOS. \textit{Solid} lines are the median, and outer lines encompass  95  percent of the ${\rm m}_{\rm err}$  distribution. In the case of \emph{Euclid} and LSST bands, the observed errors are those in COSMOS in the corresponding filter passbands, but shifted to match the magnitude limits at completion. $H$ from \emph {Euclid} will be at completion $\sim 0.6$ dex shallower than the current $H$-band from UVISTA, and $z$-band from LSST will be $\sim 0.3$ dex deeper than the current $z^{++}$ band from Subaru.}%
\label{fig:magerr_vs_mag}
\end{figure}

Implementing realistic errors on the flux is crucial for the accuracy of our forecasts to retrieve correct redshift and masses. To this end, we compute flux errors for our mock galaxies and perturb their fluxes accordingly. 
\\
For the COSMOS-like catalogue, this implementation is done in each band while relying on the COSMOS2015 catalogue. The real dataset is split in small bins of flux\footnote{
In COSMOS2015 the total fluxes are computed from the corrected aperture magnitudes (using the offset defined in Equation~4 of L16), and then accounting for the Galactic foreground extinction (following their Equation~10).
}. In each bin the flux error distribution is fitted with a Gaussian function.  
The {\sc Horizon-AGN} catalogue is then divided in the same way. For each simulated galaxy in a given bin, and for each COSMOS filter, an error is randomly chosen according to the COSMOS2015 error distribution. Galaxy apparent fluxes, which initially corresponded ``exactly'' to the star particles' content, are now perturbed according to their $1\sigma$ error. We note that for this implementation, the intrinsic fluxes from {\sc Horizon-AGN} are confronted to the -already perturbed- observed ones. As a result, the horizontal width of the faint-end tail tends to be larger in the simulation as in observations. \\
For the {\emph Euclid}-like, LSST-like and {\emph Euclid}+LSST catalogues, errors are implemented according to the COSMOS2015 flux error distribution in the closest filters pass-bands, and shifted according to the expected depth of the surveys at similar number of $\sigma$. We emphasize that, although reasonable, these errors might no reflect the specific noise of the survey, and in any case it does not take into account possibly systematics in the photometry. \\
Fig.~\ref{fig:magerr_vs_mag} shows the distribution of $1\sigma$ errors, as a function of total magnitudes, in several bands in {\sc Horizon-AGN} (grey area) and compares them to observed data (red area) when available (for the COSMOS-like sample). The adopted depths in all bands are summarised in Table~\ref{tab:maglim}. 
 \begin{table}
 \begin{center}
 \def\arraystretch{1.2}
 \begin{tabular}{|c | c c|}
   \textbf{Survey} & \textbf{band} & \textbf{depth}  \\ \hline
    & $u$ & 26.6   \\
    & $B$ & 27.0   \\
    & $V$ & 26.2   \\
    & $r$ & 26.5   \\
    & $i^{+}$ & 26.2   \\
  COSMOS-like   & $z^{++}$ &25.9   \\
  $3\sigma$ depth,  & $Y$ & 25.3  \\
  extended sources    & $J$ & 24.9   \\
    & $H$ & 24.6   \\
    & $K_{\rm s}$ & 24.7  \\
    & IB & 25-26  \\\hline
    & $riz$ & 24.5   \\
  \emph{Euclid}-like  & $Y$ & 24.0   \\
 $5\sigma$ depth,    & $J$ & 24.0   \\
 extended sources   & $H$ & 24.0   \\\hline
     & $g$ & 24.6   \\  
  DES-like   & $r$ & 24.1   \\  
   $5\sigma$ depth,   & $i$ & 24.0   \\  
    extended sources   & $z$ & 23.9   \\  \hline
     & $u$ & 26.3   \\  
     & $g$ & 27.5   \\  
LSST-like     & $r$ & 27.7   \\  
$5\sigma$ depth,      & $i$ & 27.0   \\  
  extended sources     & $z$ & 26.2   \\  
     & $y$ & 24.9   \\  \hline
 \end{tabular}
 \end{center}
 \caption{A summary of the adopted depths in all bands. The depths of intermediate bands (IB) are detailed in \protect\cite{laigle16}.}
 \label{tab:maglim}
 \end{table}
Fig.~\ref{fig:magapp} presents the simulated galaxy count in the $r$ and $K_{\rm s}$ bands in bins of redshift and compares them to the counts from COSMOS2015. Several features can be noticed. First of all the {\sc Horizon-AGN} simulated catalogue is mass limited while COSMOS2015 is not, which explains the drop-off of the count at faint magnitudes in the $K_{\rm s}$ band at low-redshift. Part of the discrepancy at bright magnitudes is driven by the smaller area covered by {\sc Horizon-AGN} (1deg$^{2}$) compared to COSMOS2015 ($\sim$1.4deg$^{2}$). {Finally, we find that the galaxy counts are overestimated at the faint end of the distribution}. This effect has already been discussed in \cite{kaviraj16} and is mostly driven by stellar feedback not being strong enough in {\sc Horizon-AGN}.

\begin{figure}
\begin{center}
\includegraphics[scale=0.43,trim={0.5cm 0 0.5cm 0},clip]{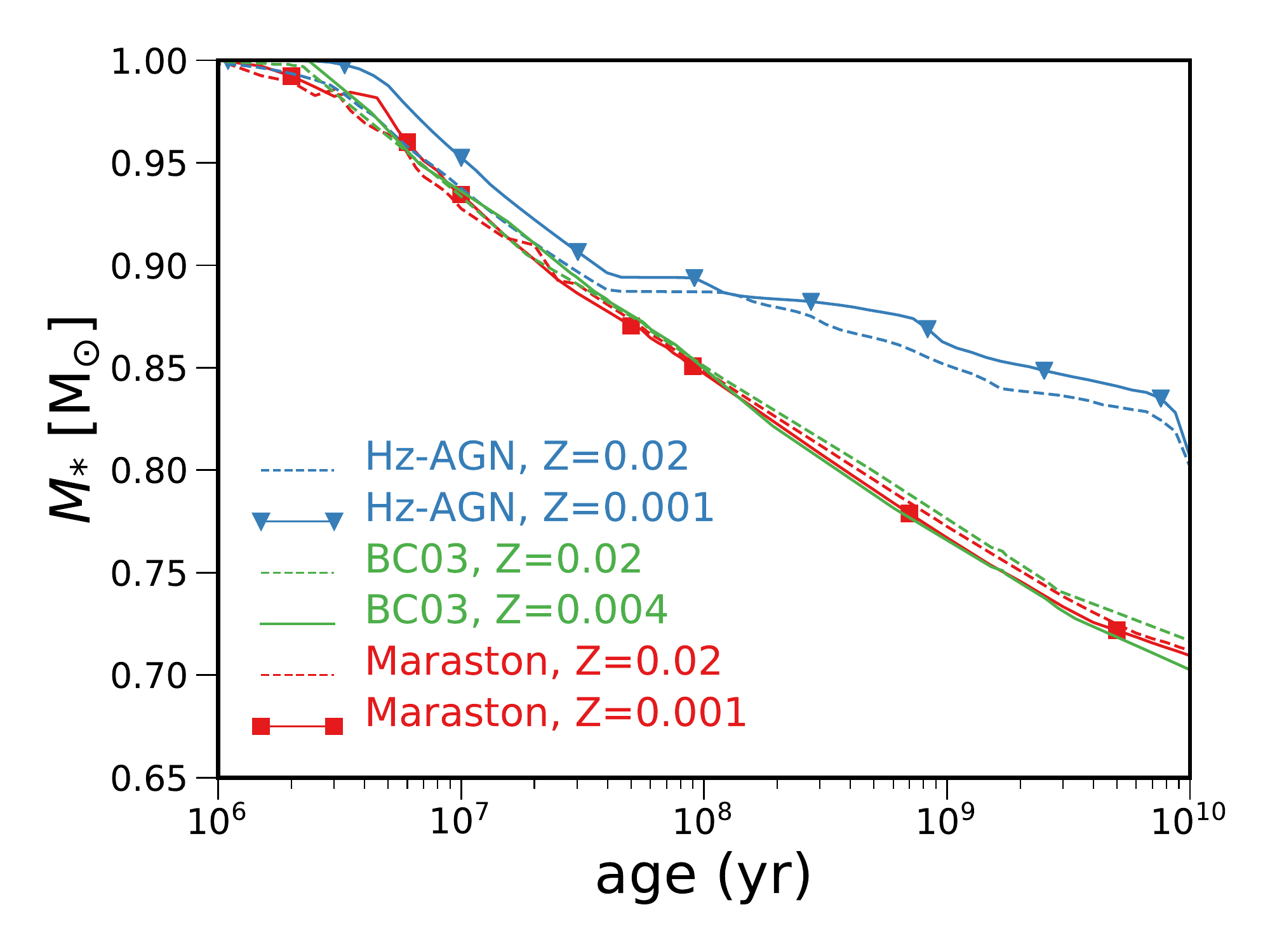}
\end{center}
\caption{Evolution of the stellar mass as a function of time for a 
stellar particle in {\sc Horizon-AGN} (\textit{blue} lines) and for the SSPs produced by  BC03 (\textit{green} lines) and \citet[][\textit{red} lines]{maraston05}  stellar population synthesis model, for solar (\textit{solid} lines) and sub-solar (\textit{dashed} lines) metallicity. Salpeter IMF is assumed here.}
\label{fig:massloss}
\end{figure}

\subsection{Stellar mass loss}
\label{appendix:imf}
Let us  discuss  the prescriptions used in both hydrodynamical simulations and stellar population synthesis models to take into account stellar mass losses  due to galactic winds, remnants and supernovae.  The impact on the comparison between simulated and observed data is not negligible if the two samples rely on different prescriptions. 

In  \textsc{Horizon-AGN}, the  stellar mass loss is modelled  as a function of time and metallicity assuming that stars are distributed with a \citet{salpeter55} IMF and  supernovae type Ia occur with the frequency described in  \citet{greggio&renzini83}, assuming a binary fraction of 5 percent. 
To compare to SED-fitting estimates, one may either rescale the $M_{*}$ values in the simulation by a factor $\sim$1/1.7 \citep[the usual conversion from Salpeter to Chabrier IMF, e.g.][]{Santini:2011p12432} or perform the SED fitting with BC03 templates that assume a Salpeter IMF. 
Neither of these solutions is sufficient to fully remove the bias because even when the IMF is the same, the resulting fraction of ejected  stellar mass may significantly differ. In other words, the mass of {\sc Horizon-AGN} stellar particle also account for remnant mass. Different SSP models will implement their formation differently, and therefore at a given time and metallicity the remnant mass will change from one model to the other.  
Fig.~\ref{fig:massloss} presents the mass evolution of a SSP as a function of time in {\sc Horizon-AGN} and using the \citet{maraston05} and BC03 models.   

To account for this discrepancy, one possible solution is to choose a SED-fitting library  based on a stellar population synthesis model whose features are in better agreement with that used in the simulation. However, we cannot modify {\sc LePhare}  as we want to compare our findings to COSMOS2015's and therefore be consistent with the set-up used for that catalogue. Instead, we prefer to correct the \textsc{Horizon-AGN} virtual photometry by matching a BC03 SSP to each hydrodynamical particle at the time of its formation. We let the SSP evolve so that at any age of the stellar particle we can compute the mass loss fraction according to the BC03 model. 
 
We stress that these details, related \textit{a-priori} to the ``sub-grid physics'', should be reconciled before  comparing the simulation to real data  \citep[e.g.,][]{henriques15}. However, although most of the studies  take into account the IMF conversion, it is difficult to find in the literature comparisons  that correct for the different stellar mass loss parametrisation  \citep[see e.g.,][]{davidzon18}.

\subsection{Limitations of our modelling}
\label{Ap:limitations}
It should be finally emphasized that our end-to-end modelling  still has shortcomings, which are listed below. 
\paragraph*{No systematics in the photometry} Although statistical photometric errors are consistently implemented in the virtual dataset (see Section~\ref{Ap:fluxErrors}), systematics arising when extracting the photometry from the images (blending of objects, clumpy objects possibly split at the extraction, lensing magnification and PSF, image artefacts) are not accounted for here and will be the topic of future work. 

\paragraph*{Spatially constant IMF and stellar mass loss prescriptions} Our modelling also ignores specific aspects of galaxy evolution which can modify the photometry. For example, nebular emission lines are not implemented in the photometry.  
Furthermore, the pipeline implicitly assumes that the IMF does not spatially vary within the galaxies, and is perfectly known at the SED-fitting stage.  A Chabrier IMF  \citep{chabrier03} is {\it de facto} used both when computing the photometry from the simulated catalogue and to build the galaxy template library for SED fitting. Fitting the galaxy photometry with a different IMF from the one chosen to compute this photometry would obviously lead to new systematics in the stellar mass computation. In addition to the choice of the IMF, which controls the amount of stars formed as a function of their mass (and therefore in particular the overall mass-to-light ratio), the chosen prescription for stellar mass losses as a function of time and metallicity is important. In the work presented here, the simulated photometry is computed with BC03 SSP templates (see Section~\ref{appendix:imf}), and fitted with a SED library which includes a higher diversity of stellar mass loss prescriptions. Therefore, we do not assume to know \textit{a priori} the stellar mass loss prescriptions of the simulated galaxies, and effectively the best-fit is not always a BC03 template. However, in practice the simulated galaxies have all the same stellar mass loss prescriptions, and these prescriptions are spatially constant within the galaxies: this is unlikely to be the case in the real Universe.
Therefore the simulated galaxy population present less diversity than the observed one, and one could expect the SED-fitting to perform much better on the {\sc Horizon-AGN} Universe than on the real one.
\\
In despite of these limitations, we found that the $z_{\rm phot}$ accuracy of the virtual catalogue is comparable to that of COSMOS2015. 
This suggests that varying IMF and stellar mass loss prescriptions should not dramatically impact redshift reconstruction, which indeed relies on galaxy colours (i.e. the relative values of flux in different bands, and not the absolute value of the flux). Only the redshift errors are lower in the simulated catalogue with respect to the real one. 
However a much larger and systematic impact on stellar mass and SFR is expected, and the reconstruction accuracy  quoted in Section~\ref{sec:physprop} should be understood as an optimistic case. On the other hand, we have shown that with a very good photometry stellar masses are well retrieved (within 0.12 dex) with the COSMOS-like configuration when the IMF is fixed. Therefore, one can expect that all additional systematics will be driven by the uncertainty on the IMF. Comparing the mass from the SED-fitting with an independent measurement (kinematics, small-scale lensing) can therefore be a way to constrain the IMF. 

\section{Impact of absorption on SED}
\label{appendix:photoz}

Dust attenuation is implemented in the virtual photometry as post-processing at the scale of $\sim$1~kpc, following the  prescription detailed in Appendix~\ref{appendix:photometry}. In particular, at equal dust mass, the resulting attenuation of the total spectrum will depend on the geometry of the galaxy and the angle under which it is seen. Conversely, at the SED-fitting stage, dust attenuation  is kept as simple as possible and therefore does not depend on galaxy geometry: all the SSPs are assumed to undergo the same attenuation given a specific attenuation curve. 
In a similar way, IGM absorption is implemented in the {\sc Horizon-AGN} lightcone independently for each galaxy, knowning the foreground distribution of HI. However, at the SED-fitting stage, IGM absorption is accounted for with an average redshift-dependent relation for all galaxies. 
These differences can play a role in driving the scatter and the systematic trends observed in mass and star-formation rate comparison. Further details on this effect are presented now.  
 
\subsection{$z_{\rm phot}$ performance}
  {\sc LePhare} is run with 3 different input photometric catalogues, in order to show the impact of dust and IGM on  $z_\mathrm{phot}$. The main run (presented in Section~\ref{sec:photoz}) includes both inter-galactic and intra-galactic (dust) absorption. An additional run is performed on an input photometry that does not include absorption from either components, and the last one is performed on  an input photometry that includes only inter-galactic absorption. The comparison of the performance of these runs directly tells us about the impact of inhomogenous IGM and spatially varying dust absorption on the SED-fitting performance. \\
{Results of the $z_{\rm phot}$ performance for no absorption, IGM-absorption only and IGM+dust absorption are presented in Table~\ref{tab:dust}. On overall IGM slightly helps to constrain the $z_{\rm phot}$. However, 
as discussed in Section~\ref{sec:photoz}, we note that overestimating the IGM absorption at the SED-fitting stage can make a large fraction of the catastrophic outliers falling in $\left[ z_{\rm sim}>2.5 \right]  \cap \left[ z_{\rm obs}<1.5 \right]$. 
Adding dust to the photometry  reduces slightly the performance. 
Table~\ref{tab:dust} has to be interpreted with caution however. In a given magnitude bin, the dust-free catalogue probes a galaxy population which can overlap a fainter magnitude bin in the dusty catalog. In other words, not only the presence/absence of dust drives the difference in the performance of the $z_{\rm phot}$ in a given magnitude bin, but also the possible differences in the intrinsic SED of the galaxies. }
\begin{table}
\begin{center}
  \caption{Statistical errors (NMAD) and percentage of catastrophic errors ($\eta$) in different $i$ magnitude bins, without absorption, with IGM absorption and with both dust and IGM absorption. }\label{tab:dust}
\begin{tabular}{ c | c c c c c c} \hline
     $i$ band   & \multicolumn{2}{c}{No absorption} &    \multicolumn{2}{c}{IGM} & \multicolumn{2}{c}{IGM+dust}  \\
             mag            &  NMAD & $\eta$ (\%)   &  NMAD & $\eta$  (\%) &  NMAD & $\eta$  (\%)  \\ \hline
$($22,23$\rbrack$ & 0.009 & 0.0  & 0.009 & 0.0 & 0.008 &  0.0 \\
$($23,24$\rbrack$ & 0.013 & 0.0 & 0.013 & 0.0 & 0.014 & 0.0 \\
$($24,25$\rbrack$ & 0.022 & 0.6 & 0.021 & 0.4 & 0.026 & 0.5 \\
$($25,26$\rbrack$ & 0.049 & 8.7 & 0.045 & 7.6 & 0.052 & 9.2 \\
  \hline
\end{tabular}
\end{center}
\end{table}

\begin{figure*}
\LARGE{\textsf{Including dust attenuation, redshifts fixed to $z_\mathrm{sim}$ values:}}
\includegraphics[width=0.99\textwidth]{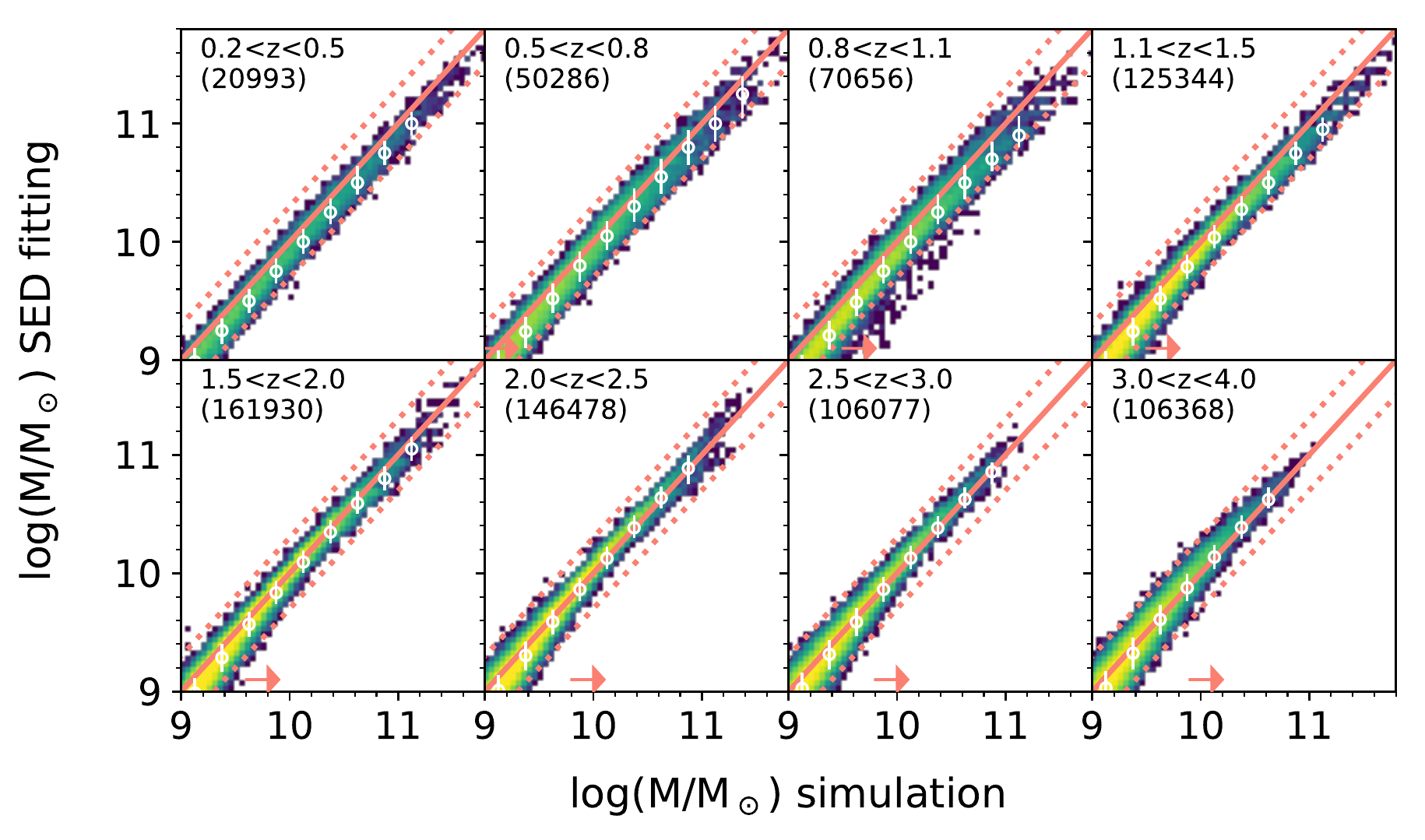}\\\vspace{0.7cm}
\LARGE{\textsf{\textsf{Without dust attenuation, redshifts fixed to $z_\mathrm{sim}$ values:}}}
\includegraphics[width=0.99\textwidth]{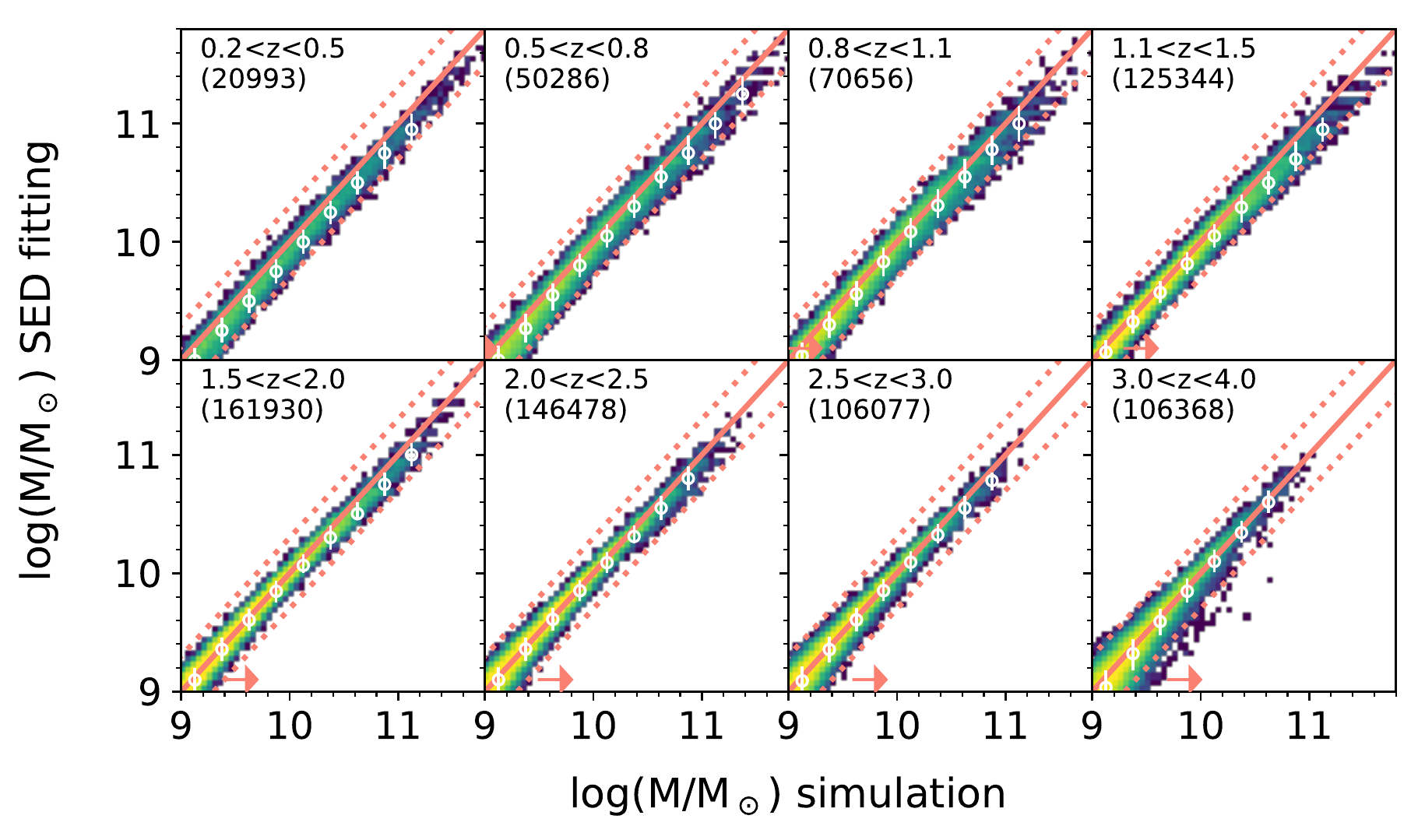}
\caption{Comparison, in different $z_\mathrm{sim}$-bins, between stellar masses estimated through SED fitting and intrinsic values. Dust attenuation is included in the \textit{upper} set of panels, while in the \textit{lower} plots dust-free photometry has been used in the SED fitting.
In both cases, the redshift is fixed to its intrinsic value ($z_\mathrm{sim}$) during the computation.
In each panel, the \textit{solid} line is the 1:1 relation while \textit{dashed} lines show $\pm0.3$\,dex offset from it.}
\label{Fig:masszsim}
\end{figure*}
\begin{figure*}
\LARGE{\textsf{Including dust attenuation, redshifts fixed to $z_\mathrm{sim}$ values:}}
\includegraphics[width=0.99\textwidth]{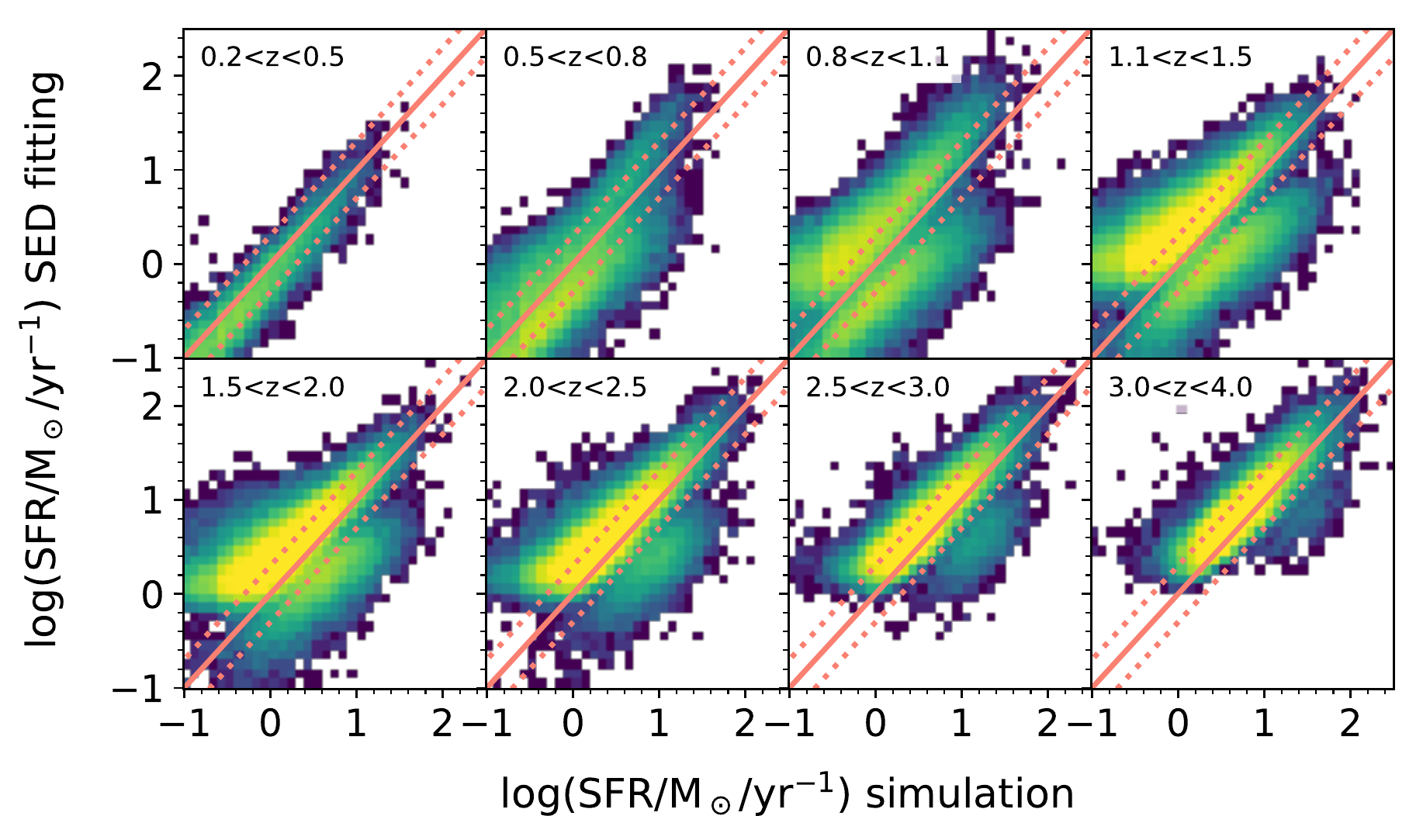}\\\vspace{0.7cm}
\LARGE{\textsf{\textsf{Without dust attenuation, redshifts fixed to $z_\mathrm{sim}$ values:}}}
\includegraphics[width=0.99\textwidth]{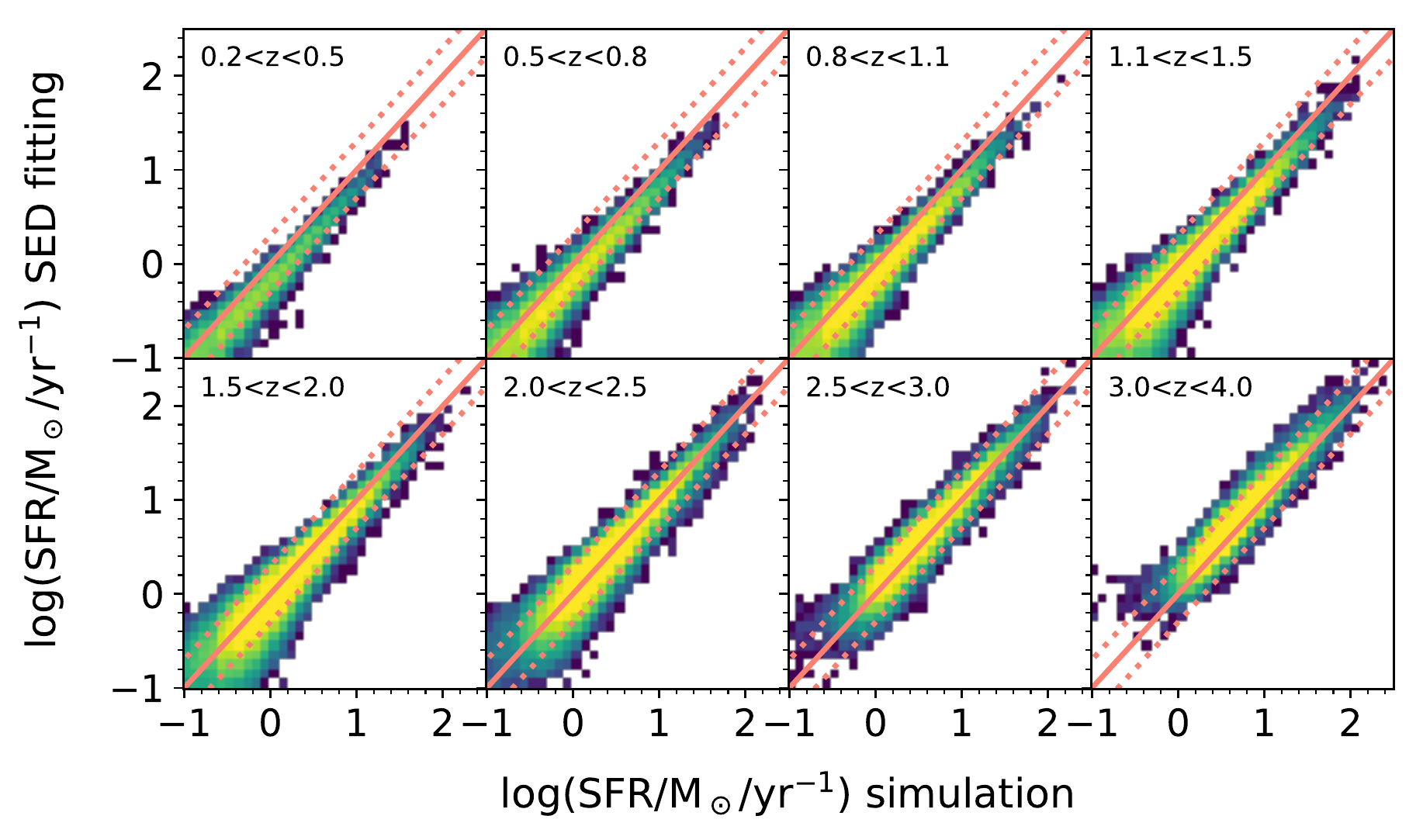}
\caption{Comparison, in different $z_\mathrm{sim}$-bins, between SFR estimated through SED fitting and intrinsic values. Dust attenuation is included in the \textit{upper} set of panels, while in the \textit{lower} plots dust-free photometry has been used in the SED fitting.
In both cases, the redshift is fixed to its intrinsic value ($z_\mathrm{sim}$) during the computation.
In each panel, the \textit{solid} line is the 1:1 relation while \textit{dashed} lines show $\pm0.3$\,dex offset from it.}
\label{Fig:sfrzsim}
\end{figure*}
\subsection{Stellar mass estimation}
As the impact of IGM is limited at the very high redshift, the subsequent analysis focus on the impact of absorption as a whole on stellar mass and SFR estimates. In order to isolate the effect of absorption, the physical parameter estimation is done at the intrinsic redshift of the galaxies (i.e. the mass and SFR uncertainties do not include the $z_{\rm phot}$ uncertainties) on two input photometric catalogues (with and without absorption).  

Fig.~\ref{Fig:masszsim} presents the comparison between the intrinsic and observed stellar mass, for galaxies selected at $K_{\rm s}<24.7$. Horizontal orange arrows indicate the mass completeness in each redshift bin. Without absorption, mass completeness is naturally better as galaxies are brighter. Dust attenuation induces a negligible overestimation of the observed stellar mass especially at high redshift and for massive galaxies. This is expected, as the mass estimates is essentially provided by the NIR bands which are barely affected by the dust. However, when only optical bands are available (e.g. in the LSST-only configuration), dust is expected to have a more dramatic effect, similar to the one observed in the SFR computation (see the discussion below). 

Without any kind of absorption (bottom panel), there is a persistent underestimation of stellar mass by at most 0.1 dex at low redshift, owing to the simplistic SFHs and metallicity distributions in the template library. As noted before (see Section~\ref{sec:physprop}) the effect of dust (overestimation) and SFHs (underestimation) can act in opposite direction and therefore tend to compensate each other.

\subsection{SFR estimation}

\label{Ap:sfr}
\begin{figure}
\includegraphics[scale=0.45,trim={1.5cm 5cm 0.5cm 2.5cm},clip]{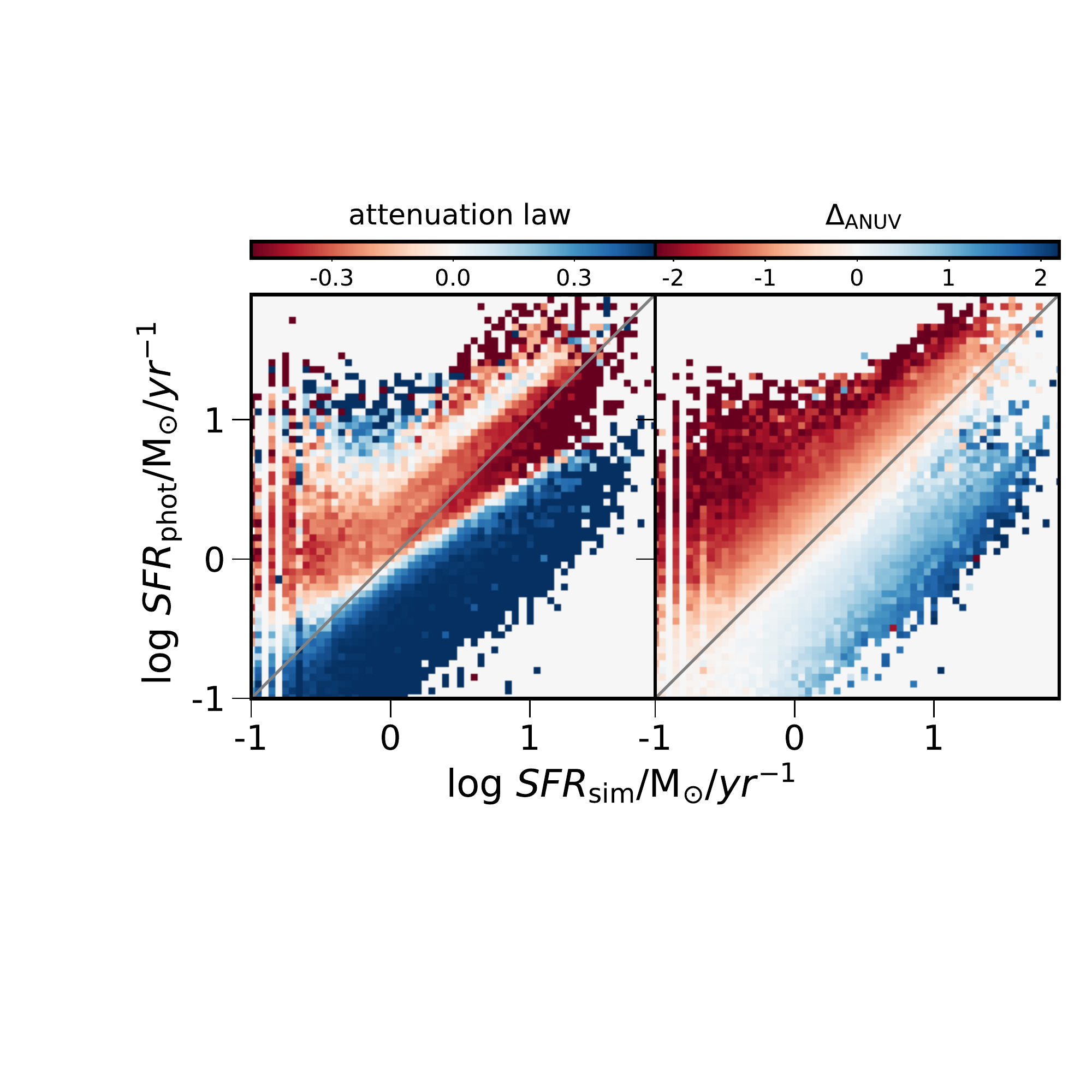}
\caption{Comparison between SFR estimated from SED-fitting and intrinsic values at $1.1<z<1.5$. {\textit{Left}}: The diagram is colour-coded by the attenuation law used in the best-fit template.  Values of 0.5 and -0.5 are given to galaxies fitted with \citet{arnouts13} and the modified \citet{calzetti2000} respectively. \textit{Right}: The diagram is colour-coded by the difference in the attenuation computed in the \textit{NUV} band between the one of the best-fit template, and the intrinsic one: $\Delta_{A_{NUV}}={\rm A}_{NUV}^{\rm sim}-{\rm A}_{NUV}^{\rm phot}$.}
\label{fig:AV}
\end{figure}

Fig.~\ref{Fig:sfrzsim} presents the comparison between the intrinsic and observed SFR, for galaxies selected at $K_{\rm s}<24.7$ and $0.2<z<4$. 
The effect of dust is dramatic: it drives a large scatter and a bimodality, with a systematic overestimation of the SFR for a fraction of the population up to at least  $z\sim 2.5$. This is a consequence of the degeneracy between dust and SFR. 
Fig.~\ref{fig:AV} isolates what in the dust modelling drives the effect. The comparison between SFR estimated from the best-fitted template and the intrinsic  SFR is presented at $1.1<z<1.5$.  The diagram is colour-coded by the  extinction curve used in the fit (\textit{left} panel) and $\Delta_{A_{NUV}}$ (\textit{right} panel), where $\Delta_{{\rm A}_{NUV}}={\rm A}_{NUV}^{\rm sim}-{\rm A}_{NUV}^{\rm phot}$. Qualitatively, when the best-fit template is derived using  \citet{arnouts13}'s extinction curve, $A_{NUV}$ is overestimated and the SFR is consequently overestimated. The reverse happens when the best-fit template is attenuated using 
\citet{calzetti2000}'s extinction curve. 

\section{Zero-point magnitude offsets}
\label{App:Lephare}

It is common in the literature to apply an offset to apparent magnitudes in order to correct for systematics due to calibration discrepancies between different filters \citep[e.g.,][]{ilbert13}.  These offsets are computed by {\sc LePhare} using a spectroscopic sub-sample: the code fits galaxy SEDs after fixing their redshifts at the spectroscopic value, then it compares the magnitude observed in each filter to the one of the best-fit model (i.e., the template minimizing the reduced  $\chi^{2}$).  An offset is added in each filter to reduce the difference between predicted and observed magnitudes. The code iterates the procedure until convergence, finding the final values of the zero-point offsets that will be add by default in the next {\sc LePhare} runs. 

Although such a procedure is quite efficient at improving SED-fitting results, they might also introduce a bias if the used spectroscopic sub-sample is not representative of the whole population. 
Moreover zero-point offsets correct, at least partly, for possible incompleteness in the template library. For this reason some authors prefer not apply them, e.g.\ when deriving  stellar masses, because besides solving calibration issues they may affect the physical interpretation of the SED fitting results \citep[see][]{moutard16b}. 

We compute the zero-point offsets in the  COSMOS filters, not for a spectroscopic-subsample but for the whole galaxy catalogue. All the offsets are found to be smaller than $0.02$ mag, which is of the order of the minimal photometric errors, with the exception of  0.029 and 0.039 mag in the $z^{++}$ and $K_{\rm s}$ band respectively. 
In COSMOS2015 these offsets are generally much larger \cite[see e.g. Table~4 in][]{laigle16},  an indication that in real datasets 
they are mainly coming from calibration issues. Because such issues are not present in the virtual photometry, there is no need to include zero-point  offsets to our virtual magnitudes.

\begin{figure*}
\begin{center}
\includegraphics[scale=0.65,trim={5cm 0.4cm 0.cm 0cm},clip]{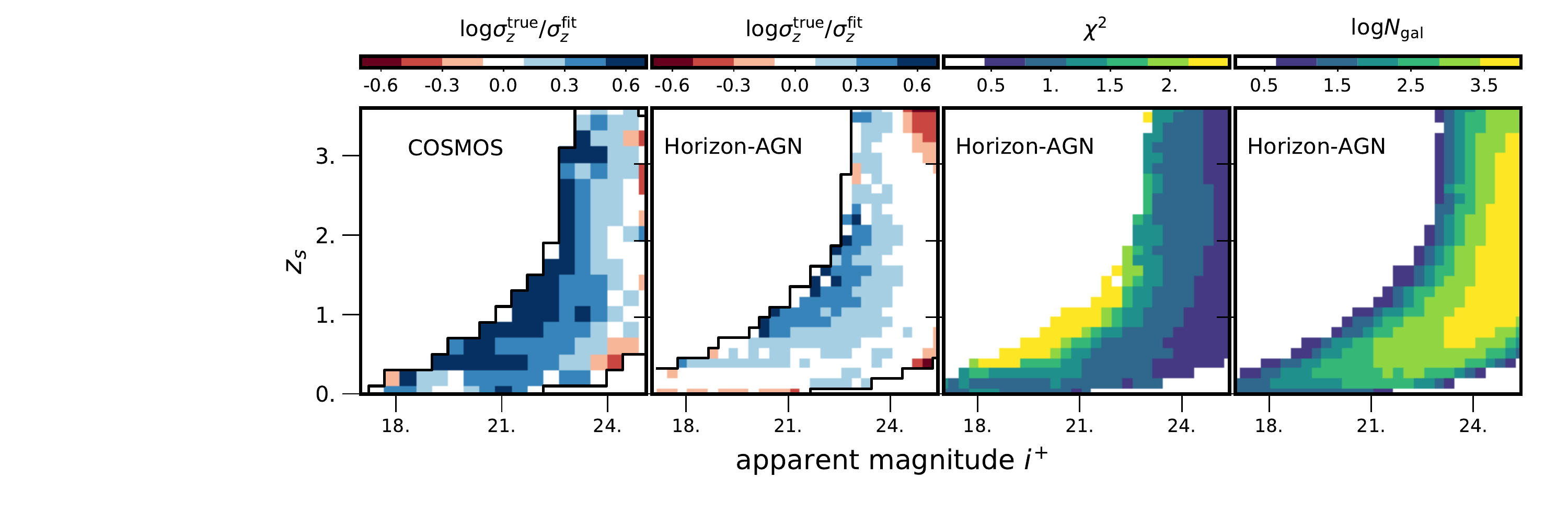}
\caption{The plane $z_{s}$ (either spectroscopic redshift for COSMOS or intrinsic redshift in {\sc Horizon-AGN}) versus apparent magnitude in the $i^{+}$-band color-coded by $\log(\sigma_{z}^{\rm true}/\sigma_{z}^{\rm fit})$ in COSMOS2015 (\textit{extreme-left}) and {\sc Horizon-AGN} (\textit{middle-left}), the reduced $\chi^{2}$ in {\sc Horizon-AGN} (\textit{middle right}) and $\log(N_{\rm gal})$ in {\sc Horizon-AGN}.}    
\end{center}
\label{fig:zerror}
 \end{figure*}

\section{Estimating the redshift errors}
\label{App:zerrors}
{
\subsection{Robustness of $\sigma_{z}^{\rm fit}$}
\label{App:sigz}
By measuring the cumulative distribution of $\vert z_{p}-z_{s}\vert/(1+\sigma_{z}^{\rm fit})$ in bins of magnitudes (figure 13 in L16) for the high-confidence spectroscopic sub-sample in COSMOS, L16 concluded that $\sigma_{z}^{\rm fit}$ derived by {\sc LePhare} underestimates $\sigma_{z}^{\rm true}$, with a trend increasing with fainter magnitudes. In an effort to better quantify this trend, Figure~\ref{fig:zerror} displays $\log \sigma_{z}^{\rm true}/\sigma_{z}^{\rm fit}$ in bins of magnitude and redshift for the COSMOS spectroscopic sample (\textit{extreme left} panel) and the {\sc Horizon-AGN} (\textit{middle left} panel) simulated sample. This plot highlights that {\sc Horizon-AGN}  presents also an underestimation of  $\sigma_{z}^{\rm fit}$ at bright magnitudes in the redshift range $1<z<2.5$. This underestimation might be due either to a remaining underestimation of magnitude errors, or to a lack of representativeness of the template library. The distribution of the reduced $\chi^{2}$ in this plane (\textit{middle right} panel) speaks in favour of this claim, as the regions with higher $\chi^{2}$ broadly match those where  $\sigma_{z}^{\rm fit}$  is underestimated the most. The underestimation is more severe in COSMOS, because the real photometry presents more diversity than the simulated one.}
 
\subsection{Catastrophic outliers}
\label{App:outliers}
\begin{figure}
\includegraphics[scale=0.5]{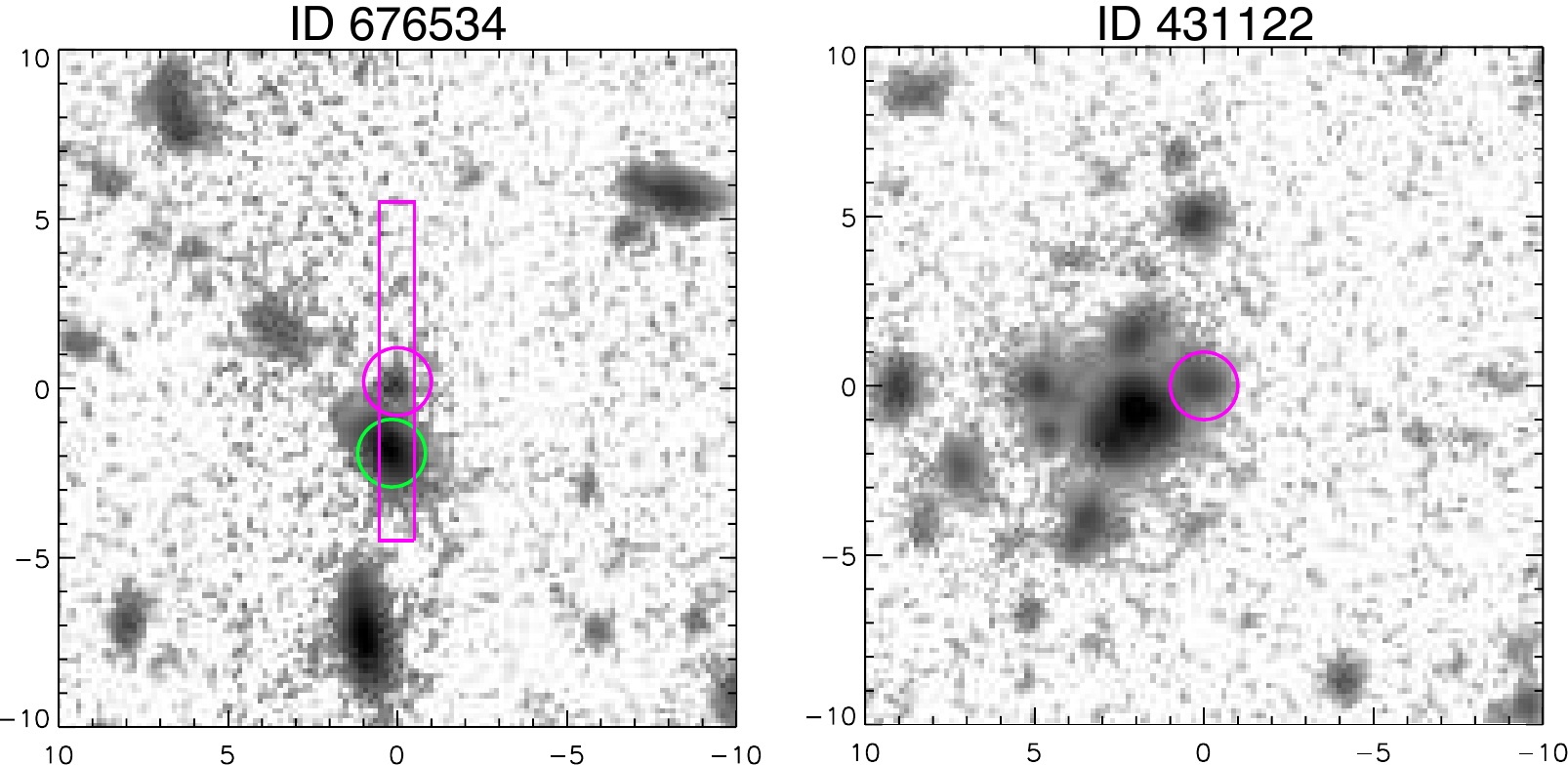}
\caption{Examples of spectroscopic or photometric issues, not reproduced in our simulation, leading to $z_\mathrm{phot}$ catastrophic failures. Images from the Subaru $i$ band are shown here as $20^{\prime\prime}\times20^{\prime\prime}$ postage stamps centred on the target galaxy (purple circle). ID numbers are from COSMOS2015. 
\textit{Left:} The purple frame indicates the $10\arcsec$-long VIMOS slit ($1\arcsec$ width) and the green circle is the on-slit misidentified observed object. \textit{Right:} blending issue where a nearby, brighter galaxy contaminates the photometry of the target}
\label{Fig:Failuresb}
\end{figure}

Let us  now  investigate what causes the higher fraction of catastrophic outliers in the observed zCOSMOS sample \citep{lilly07} with respect to the virtual photometric catalogue, as displayed in Fig.~\ref{fig:zphotzspec_cosmos}. 
It is  important to check if this discrepancy is driven by additional observational limitations (spectroscopic redshift misidentification\footnote{This is a potential issue specific to the zCOSMOS-Deep sample, zCOSMOS-Bright being much more secure.}, crowed photometry, etc.) independent on  photometry modeling, or if the virtual photometric catalogue actually misses some essential components of the real galaxy population, which would make it a poor predicator of the accuracy of SED-fitting performance. 
\\
Let us therefore focus on the failure for the objects marked as red squares on Fig.~\ref{fig:zphotzspec_cosmos}. After individual inspection of the spectroscopy and the photometry 
for each of these objects, we conclude that in the large majority of the cases, the failure arises because of one of the two reasons:

\paragraph*{Uncertain photometry} This case happens for $\sim 35$ per cent of the outliers.  The photometry extraction might be uncertain because of clumpy galaxies which might be over-split or even  identified as two different objects, or on the contrary because of blended objects (which is an issue particularly severe for $NUV$ or IRAC bands given the confusion limit). In the latter case, removing the IRAC bands in the SED fitting improves the match with the spectroscopic redshifts.
This process does not impact the virtual catalogue because the identification of the virtual galaxies is done directly in 3D.
\paragraph*{Spectroscopic misidentification} 
For another $\sim$35 per cent of the outliers, a second object on the VIMOS slit of a length of $>10\arcsec$ is misidentified as the target in the zCOSMOS-Deep observations.  Therefore, the spectroscopic redshift attributed to the original target is erroneous.  In Figure~\ref{Fig:Failuresb}, an example of the misidentification issue is shown.  We found that these $z_{\rm spec}$ often agree well with the photometric objects of the second on-slit objects.   The full description on the zCOSMOS-Deep sample and the redshift evaluation will be given in a future paper (Lilly et al., in prep).
\\
Finally  note that for a few outliers in COSMOS at low redshift, removing the $NUV$ photometry in the SED-fitting improves the match with the spectroscopic redshifts. Recall that the $NUV$ band is not used in the SED fitting of the {\sc Horizon-AGN} COSMOS-like catalogue, which could also be a reason for the lower fraction of outliers.  

\bsp    
\label{lastpage}
\end{document}